\documentclass[conference]{IEEEtran}
\IEEEoverridecommandlockouts

\usepackage{cite}
\usepackage{amsmath,amssymb,amsfonts}
\usepackage{algorithmic}
\usepackage{graphicx}
\usepackage{textcomp}
\usepackage{xcolor}
\def\BibTeX{{\rm B\kern-.05em{\sc i\kern-.025em b}\kern-.08em
    T\kern-.1667em\lower.7ex\hbox{E}\kern-.125emX}}

\usepackage{etoolbox}
\usepackage{multirow}
\usepackage{tabularx}

\newcolumntype{C}{>{\centering\arraybackslash}X}
\usepackage{booktabs}
\usepackage{lipsum}
\usepackage[aboveskip=0pt]{subcaption}
\usepackage{amsthm}
\usepackage{pgfplots}
\pgfplotsset{compat=1.16}
\newtheorem{lem}{Lemma}

\newtheorem{theorem}{Theorem}
\newtheorem{corollary}{Corollary}[theorem]
\usepackage{algorithmic,float}
\usepackage[ruled,linesnumbered,noend]{algorithm2e}
\let\oldnl\nl
\newcommand{\nonl}{\renewcommand{\nl}{\let\nl\oldnl}}
\usepackage{etoolbox}
\makeatletter
\patchcmd\algocf@Vline{\vrule}{\vrule \kern-0.4pt}{}{}
\patchcmd\algocf@Vsline{\vrule}{\vrule \kern-0.4pt}{}{}
\makeatother
\SetKwInput{KwInput}{Input}               
\SetKwInput{KwOutput}{Output}              
\usepackage{tikz}
\usepackage{environ}
\usepackage{mathtools}
\usepackage[normalem]{ulem} 
\usepackage{caption} 

\definecolor{ballblue}{rgb}{0.13, 0.67, 0.8}

\newlength\myindent
\makeatletter
\newcommand{\labeltext}[2]{
	\@bsphack
	\csname phantomsection\endcsname 
	\def\@currentlabel{#1}{\label{#2}}
	\@esphack
}
\makeatother
\newcommand*{\thead}[1]{\multicolumn{1}{|c|}{\bfseries #1}}

\usetikzlibrary{shapes.geometric,arrows,arrows.meta,calc,fit,matrix}
\definecolor{arrowcolor}{rgb}{.27,.45,.77}
\tikzstyle{arrow} = [arrowcolor,opacity=1,thin, {Triangle[angle=60:1.2mm]}-{Triangle[angle=60:1.2mm]}]
\tikzstyle{chart} = [rectangle, minimum width=3cm, minimum height=1cm, text centered,text width =3cm, draw=black, fill=white!30]
\tikzstyle{circlular} = [circle, minimum width=1cm, minimum height=1cm, text centered,text width =1cm, draw=black, fill=white!30]
\tikzstyle{circ} = [circle, minimum width=5mm, minimum height=5mm, text centered,text width =5mm, draw=black, fill=yellow!30]
\tikzstyle{roundrect3} = [rectangle,rounded corners, minimum width=3cm, minimum height=1.6cm, text centered,text width =1.4cm, draw=black, fill=red, text opacity=1]
\tikzstyle{roundrect4} = [rectangle,rounded corners, minimum width=3cm, minimum height=1.6cm, text centered,text width =1.4cm, draw=black, fill=blue, text opacity=1]
\tikzstyle{roundrect6} = [rectangle,rounded corners, minimum width=3cm, minimum height=1.6cm, text centered,text width =1.4cm, draw=black, fill=green, text opacity=1]
\tikzstyle{invisible} = [minimum width=3cm, minimum height=2cm, text centered,text width =1.4cm, draw=black, opacity = 1]
\tikzstyle{tinyarrow} = [thin,->,>=stealth]
\tikzstyle{strippedline} = [thick,dotted,>=stealth]
\BeforeBeginEnvironment{appendices}{\clearpage}

\makeatletter
\newcommand{\removelatexerror}{\let\@latex@error\@gobble}
\makeatother

\newcommand\scalemath[2]{\scalebox{#1}{\mbox{\ensuremath{\displaystyle #2}}}}

\begin{document}

\title{Mining Seasonal Temporal Patterns in Time Series}

\author{\IEEEauthorblockN{Van Long Ho\IEEEauthorrefmark{1},
		Nguyen Ho\IEEEauthorrefmark{1}, Torben Bach Pedersen\IEEEauthorrefmark{1}}
	\IEEEauthorblockA{Department of Computer Science,
		Aalborg University, Denmark\\
		\IEEEauthorrefmark{1}\{vlh, ntth, tbp\}@cs.aau.dk}}

\maketitle

\begin{abstract}
Very large time series are increasingly available from an ever wider range of IoT-enabled sensors, from which significant insights can be obtained through mining temporal patterns from them. A useful type of patterns found in many real-world applications exhibits periodic occurrences, and is thus called \textit{seasonal temporal patterns} (STP). Compared to regular patterns, mining seasonal temporal patterns is more challenging since traditional measures such as \textit{support} and \textit{confidence} do not capture the seasonality characteristics. Further, the anti-monotonicity property does not hold for STPs, and thus, resulting in an exponential search space. This paper presents our Frequent Seasonal Temporal Pattern Mining from Time Series (FreqSTPfTS) solution providing: (1) The first solution for seasonal temporal pattern mining (STPM) from time series that can mine STP at different data granularities. (2) The STPM algorithm that uses efficient data structures and two pruning techniques to reduce the search space and speed up the mining process. (3) An approximate version of STPM that uses mutual information, a measure of data correlation, to prune unpromising time series from the search space. (4) An extensive experimental evaluation showing that STPM outperforms the baseline in runtime and memory consumption, and can scale to big datasets. The approximate STPM is up to an order of magnitude faster and less memory consuming than the baseline, while maintaining high accuracy.
\end{abstract}

\section{Introduction}
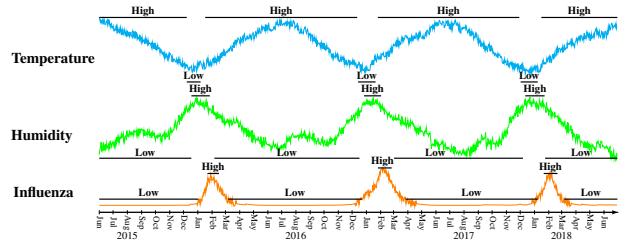
\begin{figure}[!t]
	\centering
	\resizebox{0.95\columnwidth}{!}{
		\begin{tikzpicture}
	
	\newcommand{\DrawEvent}[4]{
		\draw [black,thick] (axis cs: #1, #3+0.06) -- (axis cs: #2, #3+0.06);
		\node [black, scale=0.9] at (axis cs: #1/2+#2/2, #3+0.15) {\footnotesize \textbf{#4}};
	}
	
	\newcommand{\DrawEventShiftLeftText}[5]{
		\draw [black,thick] (axis cs: #1, #3+0.06) -- (axis cs: #2, #3+0.06);
		\node [black, scale=0.9] at (axis cs: #1/2+#2/2 - #5, #3+0.15) {\footnotesize \textbf{#4}};
	}
	
	\begin{axis}[hide axis,
		axis line style={draw=none},
		tick style={draw=none},
		ymin = -.6, ymax = 2.9,
		xmin = -210, xmax = 1250,
		xticklabels={,,},
		yticklabels={,,},
		xscale=2.3,
		]
		\addplot+ [mark=none, color=cyan] table [x=index, y=temp, col sep=comma] {data/data3att_rerange_smooth_noise_Using.csv};
		\addplot+ [mark=none, color=orange] table [x=index, y=TypeA, col sep=comma] {data/data3att_rerange_smooth_noise_Using.csv};
		\addplot+ [mark=none, color=green] table [x=index, y=humidity, col sep=comma] {data/data3att_rerange_smooth_noise_Using.csv};
		
		\node at (axis cs: -100, 1) {\normalsize \textbf{Humidity\phantom{aaa}}};
		\node at (axis cs: -100, 2.1) {\normalsize \textbf{Temperature}};
		\node at (axis cs: -100, 0.2) {\normalsize \textbf{Influenza\phantom{aai}}};
		\draw [black, thin, -latex] (axis cs: 0, -.1) -- (axis cs: 1127, -.1) node {};
		
		\draw [black, thin] (axis cs: 0, -.1) -- (axis cs: 1127, -.1) node {}; 
		
		\pgfplotsinvokeforeach{0, 519, 1035, 396, 274, 914, 792, 153, 30, 670, 549, 1065, 427, 305, 945, 183, 823, 700, 61, 580, 1096, 458, 335, 976, 853, 214, 731, 92, 611, 488, 1004, 366, 884, 245, 761, 122, 639}
		{\draw [black, thin] (axis cs: #1, -.13) -- (axis cs: #1, -.07) node {};}
		
		\pgfplotsinvokeforeach{945, 580, 214}
		{\node[scale=0.6, rotate=-90] at (axis cs:#1, -.3) {Jan\phantom{lg}};}
		\pgfplotsinvokeforeach{976, 611, 245}
		{\node[scale=0.6, rotate=-90] at (axis cs:#1, -.3) {Feb\phantom{lg}};}
		\pgfplotsinvokeforeach{274, 1004, 639}
		{\node[scale=0.6, rotate=-90] at (axis cs:#1, -.3) {Mar\phantom{lg}};}
		\pgfplotsinvokeforeach{305, 1035, 670}
		{\node[scale=0.6, rotate=-90] at (axis cs:#1, -.3) {Apr\phantom{lg}};}
		\pgfplotsinvokeforeach{1065, 700, 335}
		{\node[scale=0.6, rotate=-90] at (axis cs:#1, -.3) {May\phantom{lg}};}
		\pgfplotsinvokeforeach{0, 1096, 731, 366}
		{\node[scale=0.6, rotate=-90] at (axis cs:#1, -.3) {Jun\phantom{lg}};}
		\pgfplotsinvokeforeach{761, 396, 30}
		{\node[scale=0.6, rotate=-90] at (axis cs:#1, -.3) {Jul\phantom{lg}};}
		\pgfplotsinvokeforeach{792, 427, 61}
		{\node[scale=0.6, rotate=-90] at (axis cs:#1, -.3) {Aug\phantom{lg}};}
		\pgfplotsinvokeforeach{458, 92, 823}
		{\node[scale=0.6, rotate=-90] at (axis cs:#1, -.3) {Sep\phantom{lg}};}
		\pgfplotsinvokeforeach{488, 122, 853}
		{\node[scale=0.6, rotate=-90] at (axis cs:#1, -.3) {Oct\phantom{lg}};}
		\pgfplotsinvokeforeach{153, 884, 519}
		{\node[scale=0.6, rotate=-90] at (axis cs:#1, -.3) {Nov\phantom{lg}};}
		\pgfplotsinvokeforeach{914, 549, 183}
		{\node[scale=0.6, rotate=-90] at (axis cs:#1, -.3) {Dec\phantom{lg}};}
		
		\node[scale=0.7] at (axis cs: 61, -.43) {2015};
		\node[scale=0.7] at (axis cs: 427, -.43) {2016};
		\node[scale=0.7] at (axis cs: 792, -.43) {2017};
		\node[scale=0.7] at (axis cs: 1004, -.43) {2018};

		\DrawEvent{0}{215}{0.03}{Low};
		\DrawEvent{235}{260}{0.4}{High};
		\DrawEvent{280}{570}{0.03}{Low};
		\DrawEventShiftLeftText{605}{635}{0.48}{High}{5};
		\DrawEvent{665}{952}{0.03}{Low};
		\DrawEvent{965}{995}{0.4}{High};
		\DrawEvent{1010}{1125}{0.03}{Low};

		\DrawEvent{0}{200}{0.62}{Low};
		\DrawEvent{200}{240}{1.52}{High};
		\DrawEvent{250}{565}{0.62}{Low}{20};
		\DrawEvent{575}{612}{1.52}{High};
		\DrawEventShiftLeftText{640}{920}{0.62}{Low}{50};
		\DrawEvent{925}{967}{1.52}{High};
		\DrawEvent{968}{1125}{0.62}{Low};

		\DrawEvent{0}{190}{2.65}{High};
		\DrawEvent{190}{220}{1.72}{Low};
		\DrawEvent{230}{560}{2.65}{High};
		\DrawEvent{562}{600}{1.72}{Low};
		\DrawEvent{605}{912}{2.65}{High};
		\DrawEvent{915}{952}{1.72}{Low};
		\DrawEvent{960}{1125}{2.65}{High};
		
	\end{axis}

\end{tikzpicture}
	}
	\vspace{-0.05in}
	\caption{Weather and Influenza time series}
	\label{fig:TimeSeries}
\end{figure}

\vspace{-0.01in}
The widespread of IoT systems enables the collection of big time series from domains such as energy, transportation, climate, and healthcare. Mining such time series can discover hidden patterns and offer new insights into the application domains to support evidence-based decision making and planning. Often, pattern mining methods such as sequential pattern mining (SPM) \cite{sequential_lam2014mining, sequential_huang2008general} 
and temporal pattern mining (TPM) \cite{ho2020efficient, lee2020z} are used to extract frequent (temporal) relations between events. In SPM, events occur in sequential order, whereas in TPM, events carry additional temporal information such as occurrence time, making relations between temporal events are more expressive and comprehensive. 
A useful type of temporal patterns found in many real-world applications are those that exhibit periodic occurrences. Such patterns occur concentrated within a particular time period, and then repeat that concentrated occurrence periodically. They are thus called \textit{seasonal temporal patterns}. Here, the term \textit{seasonal} indicates the periodic re-occurrence, while the term \textit{temporal pattern} indicates patterns that are formed by the temporal relations between events, such as follows, contains, overlaps.
Seasonal temporal patterns are useful in revealing seasonal information of temporal events and their relations. For example, in healthcare, health experts might be interested in finding seasonal diseases in a geographical location, as exemplified in Fig. \ref{fig:TimeSeries} using the real-world data from Kawasaki, Japan between 2015 - 2018 \cite{diseasedata}, \cite{openweather}. 
Here, a seasonal temporal pattern involving weather and epidemic events can be found: \{Low Temperature \textit{overlaps} High Humidity \textit{followed by} High Influenza Cases\}. This pattern occurs yearly and is concentrated in January, February. Detecting such seasonal diseases will support health experts in prevention and planning. In market analysis, knowing the periodic rise of certain stocks and their relations to other impact factors can be of interests for traders to plan better trading strategies. In marketing, identifying the order of search keywords that appear seasonally in the search engine can be useful to better understand customer needs and thereby improve the marketing plans. 

\textbf{Challenges.} Although seasonal temporal patterns are useful, mining them is a challenging task for several reasons. First, the \textit{support} measure used by TPM is not sufficient to mine seasonal patterns, since the traditional \textit{support} represents the frequency of a pattern across the entire dataset, and thus, cannot capture the seasonality characteristic of seasonal patterns. Second, since temporal patterns are constructed based on temporal events, the complex relations between temporal events create an exponential and large search space of size $O(n^h 3^{h^2})$ ($n$ is the number of events and $h$ is the length of temporal patterns).  
Finally, since seasonal temporal patterns do not uphold the anti-monotonicity property, i.e., the non-empty subsets of a seasonal temporal pattern may not be seasonal, mining seasonal temporal patterns is more computationally expensive as the typical pruning technique based on anti-monotonicity property cannot be applied. This raises the need for an efficient seasonal temporal pattern mining approach with effective prunings to tackle the exponential search space. Existing work such as \cite{kiran2015discovering, periodic_kiran2019discovering} proposes solutions to mine seasonal itemsets. However, they do not consider the temporal aspect of items/ events, thus, addressing the exponential search space of seasonal temporal patterns is still an open problem. 

\textbf{Contributions.} In the present paper, we present our Frequent Seasonal Temporal Pattern Mining from Time Series (FreqSTPfTS) solution that addresses all the above challenges. Specifically, our key contributions are as follows. (1) We propose the first solution to mine \textit{seasonal temporal patterns} from time series. Within the process, we introduce several measures to assess the seasonality characteristics, and use these to formally define the concept of \textit{seasonal temporal patterns} in time series. The formulation allows to flexibly mine seasonal temporal patterns at different granularities. 
(2) Our Seasonal Temporal Pattern Mining (STPM) algorithm is efficient and has several important novelties. First, STPM employs efficient data structures, i.e., the hierarchical hash tables, to enable fast retrieval of candidate events and patterns during the mining process. Second, we define a new measure \textit{maxSeason} that upholds the anti-monotonicity property, and design two efficient pruning techniques: Apriori-like pruning and transitivity pruning. (3) Based on mutual information, we propose a novel approximate version of STPM to prune unpromising time series and significantly reduce the search space, while maintaining highly accurate results. The approximate STPM can scale on big datasets, i.e., many time series and many sequences. (4) We perform extensive experimental evaluation on synthetic and real-world datasets from various domains showing that STPM outperforms the baseline in both runtime and memory usage. The approximate STPM achieves up to an order of magnitude speedup w.r.t. the baseline, while obtaining high accuracy compared to the exact STPM. 
\section{Related work}\label{sec:relatedwork}\vspace{-0.02in} 
Finding seasonal patterns that represent temporal periodicity in time series is an important research topic, and has received substantial attention in the last decades. By considering seasonality as periodic occurrences, different techniques have been proposed to find periodic sub-sequences in time series data. Such techniques, first introduced by Han et al. in \cite{han1998mining, han1999efficient}, and later extended by \cite{assfalg2009periodic, motif_zhang2022efficient, motif_liu2016fast, motif_mohammad2013approximately,kegel2021season}, are called \textit{motif} discovery techniques. 
However, since motifs are defined as similar time series sub-sequences, motif discovery can only find recurrent sub-sequences rather than periodic temporal patterns. 

Another research direction in this area concerns periodic association rules \cite{tanbeer2009discovering, uday2010towards, kiran2015discovering, amphawan2009mining, cappiello2015co, ho2015data, barkat2014open, ho2019amic, ho2016adaptive, ho2016characterizing, gribaudo2014analysis, ho2019efficient, ho2017improving, ho2021efficient, ho2021efficientvldb, ho2022unified, ho2020efficientfull, ho2017towards, ho2019amicfull, ho2013activity, ho2020efficienttimedelay,  periodic_fournier2022tspin, periodic_kiran2020discovering, periodic_kiran2019discovering, javed2021hova}. Such techniques can identify seasonal associations between itemsets, for example, market-basket analysis to reveal the seasonal occurrence of the association \{Glove $\Rightarrow$ Winter Hat\} during the winter season. To mine such seasonal itemset patterns in transactional databases, Tanbeer et al. in \cite{tanbeer2009discovering} proposed the PFP-growth algorithm using \textit{minSup} and \textit{maxPer} as seasonality measures. In their method, a tree structure called PF-tree is used as a compact representation of periodic frequent itemsets, with \textit{maxPer} imposing the periodic constraint, and \textit{minSup} imposing the frequency constraint on the pattern occurrences. Although PFP-growth can capture seasonality characteristic through the \textit{maxPer} measure, the use of \textit{minSup} means that it cannot identify rare seasonal patterns. Follow-up work such as \cite{amphawan2009mining, uday2010towards} improves different aspects of PFP-growth, for example, Amphawan et al. \cite{amphawan2009mining} propose \textit{period summary} to approximate the pattern periodicity to reduce the memory cost, Uday et al. \cite{uday2010towards} use the concept of \textit{item-specific support} to address the rare pattern problem. Recently, Javed et al. \cite{javed2021hova} propose hashed occurrence vectors and Apriori-based approach to speed up periodic itemsets mining.

In a more recent work \cite{kiran2015discovering}, Uday et al. propose the RP-growth algorithm to discover recurring itemset patterns in transactional databases. RP-growth uses an RP-tree to maintain frequent itemsets, and recursively mines the RP-tree to discover recurring ones. In their follow-up work, the same authors introduce several improvements of \cite{kiran2015discovering}. In \cite{kiran2016efficient}, they propose the Periodic-Frequent Pattern-growth++ (PFP-growth++) algorithm that employs two new concepts, \textit{local-periodicity} and \textit{periodicity}, to capture locally optimal and globally optimal solutions of recurring patterns. This enables 2-phase pruning to improve the runtime efficiency. In \cite{periodic_kiran2019discovering},  the authors extend PFP-growth++ to find periodic spatial patterns in spatio-temporal databases. In \cite{periodic_kiran2020discovering}, PFP-growth++ is extended to find maximal periodic frequent patterns. In \cite{kiran2019finding}, they further improve PFP-growth++ to be memory efficient by proposing a concept called \textit{period summary} to effectively summarize the temporal occurrence information of an itemset in a Periodic Summary-tree (PS-tree), and designing Periodic Summary Pattern Growth algorithm (PS-growth) to find all periodic-frequent itemset patterns from PS-tree. Nevertheless, all the mentioned work can only discover seasonal patterns between itemsets.  
To the best of our knowledge, no existing work addresses the seasonal temporal pattern mining that finds seasonal occurrences of temporal patterns. In Section \ref{sec:experiment}, we adapt the state-of-the-art method for periodic itemset mining \textit{PS-growth} to mine seasonal temporal patterns, and use it as an experimental baseline.
\section{Preliminaries}\label{sec:preliminary}
\subsection{Time Granularity}\vspace{-0.02in}
\hspace{-0.15in}\textbf{Definition 3.1} (Time domain) A time domain $\mathcal{T}$ consists of an ordered set of time instants that are isomorphic to the natural numbers. The time instants in $\mathcal{T}$ have a time unit, presenting how they are measured. 

\hspace{-0.15in}\textbf{Definition 3.2} (Time granularity) Given a time domain $\mathcal{T}$, a \textit{time granularity} $G$ is a \textit{complete and non-overlapping equal partitioning} of $\mathcal{T}$, i.e., $\mathcal{T}$ is divided into non-overlapping equal partitions. Each non-empty partition $G_i \in G$ is called a (time) \textit{granule}. The position of a granule $G_i$ in $G$, denoted as $p(G_i)$, is identified by counting the number of granules which appear before and up to (including) $G_i$. 
The \textit{period} between two granules $G_i$ and $G_j$ in granularity $G$ measures the time duration between $G_i$ and $G_j$, and is computed as: $pr_{\textit{ij}} = |p(G_i) - p(G_j)|$, where $p(G_i)$ and $p(G_j)$ are the positions of $G_i$ and $G_j$, respectively.

As an example, consider a time domain $\mathcal{T}$ consisting of an ordered set of minutes. The time instants minute$_1$, minute$_2$, etc. are isomorphically mapped to the natural numbers, and are measured in the \textit{Minute} time unit.  
Here, $\mathcal{T}$ can have different time granularities such as Minute, 5-Minutes, or even Hour, Day, Year. The position of granule Minute$_2$ in the Minute granularity is $p(\text{Minute}_2)=2$. 
The period between the Minute$_1$ and Minute$_6$ granules is: $|p(\text{Minute}_6) - p(\text{Minute}_1)| = 5$, indicating that the time duration between them is $5$ minutes. We note that the period is only defined between granules of the same granularity.

\hspace{-0.15in}\textbf{Definition 3.3} (Finer time granularity) 
A time granularity $G$ is \textit{finer} than a time granularity $H$ if and only if for every granule $H_j \in H$, there exists $m$ adjacent granules $G_{i+1},...,G_{i+m}$ $\in G$ such that $H_j = G_{i+1} \cup ... \cup G_{i+m}$ where $m \geq 1$. We call $G$ is \textit{m-Finer} than $H$, denoted as $G \trianglelefteq_m H$. 

In the previous example, we have the Minute granularity is 60-Finer than the Hour granularity.

\hspace{-0.15in}\textbf{Definition 3.4} (Time granularity hierarchy) Given a time domain $\mathcal{T}$, the different time granularities of $\mathcal{T}$ form a \textit{time granularity hierarchy} $\mathcal{H}$ where each level in $\mathcal{H}$ represents one specific granularity, with the lower levels in the hierarchy having finer granularity than the higher levels.

\begin{figure}[!t]
	\centering
	\includegraphics[width=1\linewidth, height=1.8cm]{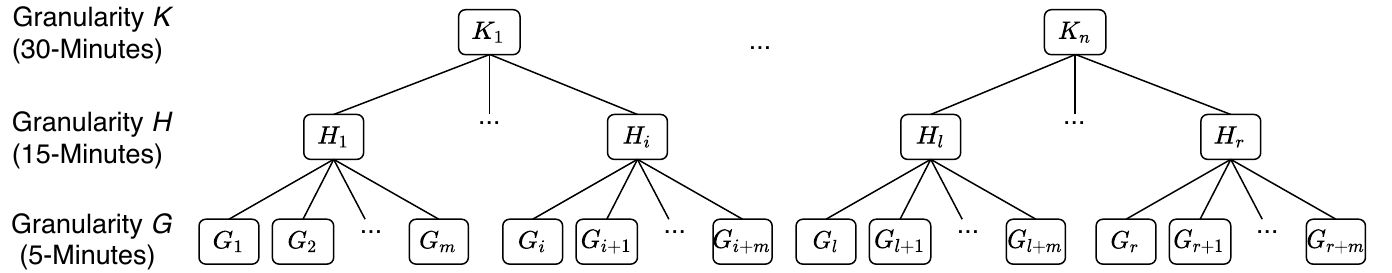}
	\vspace{-0.2in}
	\caption{Time granularity hierarchy $\mathcal{H}$}
	\label{fig:TimeHierarchy}
	\vspace{-0.1in}
\end{figure}

Fig. \ref{fig:TimeHierarchy} shows an example of the time granularity hierarchy. Here, to be consistent with examples in the following sections, we assume granularity $G$ is 5-Minutes and is the finest, whereas granularity $H$ is 15-Minutes and $G \trianglelefteq_3 H$.

\begin{table}[!t]
	\begin{minipage}{\columnwidth}
		\caption{Frequently Used Notations}
		\vspace{-0.08in}
		\resizebox{\columnwidth}{!}{
			\begin{tabular}{ |l|l|} 
				\hline {\bfseries Notation} & {\bfseries Description} \\ 
				\hline
				$\mathcal{T}$, $\mathcal{H}$  &  time domain $\mathcal{T}$ and time granularity hierarchy $\mathcal{H}$ \\
				$p(G_i)$ & the position of the granule $G_i$\\
				$G \trianglelefteq_m H$ & granularity $G$ is \textit{m-Finer} than granularity $H$\\
				$X$, $X_S$ & time series $X$ and symbolic time series $X_S$\\
				$E_{\triangleright e}$ & temporal event $E$ has an event instance $e$\\
				$g$$: X_S \rightarrow_m H$ & sequence mapping from $X_S$ to granularity $H$\\
				$Seq_i$ = $<$$e_1$,...,$e_n$$>$& a temporal sequence of $n$ event instances\\
				$\mathcal{D_{\text{SYB}}}$, $\mathcal{D}_{\text{SEQ}}$ & symbolic database and temporal sequence database \\
				$H_i^E$, $H_i^P$ & event $E$ (pattern $P$) occurs at granularity $H_i$\\			
				$\text{SUP}^E$, $\text{SUP}^{P}$ & support set of event $E$ (pattern $P$)\\
				$\text{NearSUP}_i^P$ & near support set $i$ of pattern $P$\\
				$\textit{den}(\text{NearSUP}_i^P)$ &  density of the near support set\\
				$\textit{dist}(\text{NearSUP}_i^P, \text{NearSUP}_j^P)$ & distance between two near support sets\\
				$\textit{seasons}(P)$ & number of seasons of pattern $P$\\
				\hline 
			\end{tabular} 
		}
		\label{tbl:notation}
	\end{minipage}
\end{table}

\subsection{Symbolic Representation of Time Series}\vspace{-0.02in}
Consider the time domain $\mathcal{T}$. Let $\mathcal{H}$ be the time granularity hierarchy of $\mathcal{T}$, and $G$ be the finest granularity in $\mathcal{H}$. 

\hspace{-0.15in}\textbf{Definition 3.5} (Time series) A \textit{time series} $X = x_1, x_2, ..., x_n$ in the time domain $\mathcal{T}$ is a sequence of data values that measure the same phenomenon during an observation time period in $\mathcal{T}$, and are chronologically ordered. 
We say that $X$ has granularity $G$ if $X$ is sampled at every time instant $t_i$ in $\mathcal{T}$. 

A \textit{symbolic time series} $X_S$ of $X$ encodes the raw values of $X$ into a sequence of symbols using a mapping function $f$$: X$$\rightarrow$$\Sigma_{X}$ that maps each value $x_i \in X$ into a symbol $\omega \in \Sigma_{X}$. The finite set of permitted symbols used to encode $X$ is called the \textit{symbol alphabet} of $X$, denoted as $\Sigma_X$. Since the mapping function $f$ performs the 1-to-1 mapping from $X$ to $X_S$, $X_S$ has the same granularity $G$ as $X$.

For example, let $X$ = 1.82, 1.25, 0.46, 0.0 be a time series representing the energy usage of an electrical device recorded every 5 minutes. By using $\Sigma_X$ = \{1, 0\} (1: ON, 0: OFF), we obtain $X_S$ = 1, 1, 1, 0. The mapping function $f$ can be defined using time series representation techniques such as SAX \cite{lin2003symbolic}. 

\hspace{-0.15in}\textbf{Definition 3.6} (Symbolic database)
Given a set of time series $\mathcal{X}=\{X_1,...,X_n\}$, the set of symbolic representations of the time series in $\mathcal{X}$ forms a \textit{symbolic database} $\mathcal{D_{\text{SYB}}}$.

Table \ref{tbl:SymbolDatabase} shows an example of the symbolic database $\mathcal{D}_{\text{SYB}}$ using $\Sigma$ = \{0, 1\}. There are 5 time series: \{C, D, F, M, N\} (C: Cooker, D: Dish Washer, F: Food Processor, M: Microwave, N: Nespresso Coffee) representing the energy usage of electrical devices at 5-Minutes granularity. 

\begin{table*}
	\begin{minipage}{1\textwidth}
		\captionsetup{justification=centering, font=small}
		\caption{\small A Symbolic Database $\mathcal{D}_{\text{SYB}}$ (G: 5-Minutes granularity)}
		\vspace{-0.08in}
		\resizebox{\textwidth}{1cm}{
			\setlength{\tabcolsep}{0.5mm} 
			\renewcommand{\arraystretch}{1.5}
			\begin{tabular}{|c|c|ccc|ccc|ccc|ccc|ccc|ccc|ccc|ccc|ccc|ccc|ccc|ccc|ccc|ccc|}
				\hline
				\multicolumn{2}{|c|}{\bfseries Granules in $\textbf{\textit{G}}$} & \;$\textbf{\textit{G}}_1$\; & $\textbf{\textit{G}}_2$ & \;$\textbf{\textit{G}}_3$\; & \;$\textbf{\textit{G}}_4$\; & $\textbf{\textit{G}}_5$ & \;$\textbf{\textit{G}}_6$\;  & \;$\textbf{\textit{G}}_7$\; & $\textbf{\textit{G}}_8$ & \;$\textbf{\textit{G}}_9$\;  & $\textbf{\textit{G}}_{10}$ & $\textbf{\textit{G}}_{11}$ & $\textbf{\textit{G}}_{12}$ & $\textbf{\textit{G}}_{13}$ & $\textbf{\textit{G}}_{14}$ & $\textbf{\textit{G}}_{15}$ & $\textbf{\textit{G}}_{16}$ & $\textbf{\textit{G}}_{17}$ & $\textbf{\textit{G}}_{18}$  & $\textbf{\textit{G}}_{19}$ & $\textbf{\textit{G}}_{20}$ & $\textbf{\textit{G}}_{21}$  & $\textbf{\textit{G}}_{22}$ & $\textbf{\textit{G}}_{23}$ & $\textbf{\textit{G}}_{24}$  & $\textbf{\textit{G}}_{25}$ & $\textbf{\textit{G}}_{26}$ & $\textbf{\textit{G}}_{27}$ & $\textbf{\textit{G}}_{28}$ & $\textbf{\textit{G}}_{29}$ & $\textbf{\textit{G}}_{30}$ & $\textbf{\textit{G}}_{31}$ & $\textbf{\textit{G}}_{32}$ & $\textbf{\textit{G}}_{33}$ & $\textbf{\textit{G}}_{34}$ & $\textbf{\textit{G}}_{35}$ & $\textbf{\textit{G}}_{36}$ & $\textbf{\textit{G}}_{37}$ & $\textbf{\textit{G}}_{38}$ & $\textbf{\textit{G}}_{39}$ & $\textbf{\textit{G}}_{40}$ & $\textbf{\textit{G}}_{41}$ & $\textbf{\textit{G}}_{42}$\\ 
				\hline
				\multicolumn{2}{|c|}{\bfseries Position} & \textbf{\;1\;} & \textbf{\;2\;} & \textbf{\;3\;} & \textbf{\;4\;} & \textbf{\;5\;} & \textbf{\;6\;} & \textbf{\;7\;} & \textbf{\;8\;} & \textbf{\;9\;} & \textbf{10} & \textbf{11} & \textbf{12} & \textbf{13} & \textbf{14} & \textbf{15} & \textbf{16} & \textbf{17} & \textbf{18} & \textbf{19} & \textbf{20} & \textbf{21} & \textbf{22} & \textbf{23} & \textbf{24} & \textbf{25} & \textbf{26} & \textbf{27} & \textbf{28} & \textbf{29} & \textbf{30} & \textbf{31} & \textbf{32} & \textbf{33} & \textbf{34} & \textbf{35} & \textbf{36} & \textbf{37} & \textbf{38} & \textbf{39} & \textbf{40} & \textbf{41} & \textbf{42}\\ 
				\hline
				\multirow{5}{1cm}{\bfseries Time series}  & \textbf{C} & \normalsize 1 & \normalsize 1 & \normalsize 0 & \normalsize 1 & \normalsize 0 & \normalsize 0 & \normalsize 1 & \normalsize 1 & \normalsize 0 & \normalsize 0 & \normalsize 0 & \normalsize 0 & \normalsize 0 & \normalsize 0 & \normalsize 0 & \normalsize 0 & \normalsize 0 & \normalsize 0 & \normalsize 1 & \normalsize 1 & \normalsize 1 & \normalsize 1 & \normalsize 1 & \normalsize 1 & \normalsize 0 & \normalsize 0 & \normalsize 0 & \normalsize 0 & \normalsize 0 & \normalsize 0 & \normalsize 1 & \normalsize 0 & \normalsize 0 & \normalsize 1 & \normalsize 1 & \normalsize 0 & \normalsize 0 & \normalsize 0 & \normalsize 0 & \normalsize 1 & \normalsize 1 & \normalsize 0\\ 
				\cline{2-44}  
				  & \textbf{D} & \normalsize 1 & \normalsize 0 & \normalsize 0 & \normalsize 1 & \normalsize 0 & \normalsize 0 & \normalsize 1 & \normalsize 1 & \normalsize 0 & \normalsize 1 & \normalsize 1 & \normalsize 0 & \normalsize 0 & \normalsize 0 & \normalsize 0 & \normalsize 0 & \normalsize 0 & \normalsize 0 & \normalsize 1 & \normalsize 1 & \normalsize 1 & \normalsize 1 & \normalsize 1 & \normalsize 1 & \normalsize 0 & \normalsize 0 & \normalsize 0 & \normalsize 0 & \normalsize 0 & \normalsize 0 & \normalsize 1 & \normalsize 0 & \normalsize 0 & \normalsize 1 & \normalsize 0 & \normalsize 0 & \normalsize 1 & \normalsize 1 & \normalsize 0 & \normalsize 1 & \normalsize 1 & \normalsize 0\\ 
				  \cline{2-44}  
				  & \textbf{F} & \normalsize 0 & \normalsize 0 & \normalsize 1 & \normalsize 0 & \normalsize 1 & \normalsize 1 & \normalsize 0 & \normalsize 0 & \normalsize 1 & \normalsize 0 & \normalsize 0 & \normalsize 1 & \normalsize 1 & \normalsize 1 & \normalsize 1 & \normalsize 0 & \normalsize 0 & \normalsize 0 & \normalsize 0 & \normalsize 0 & \normalsize 0 & \normalsize 0 & \normalsize 0 & \normalsize 0 & \normalsize 1 & \normalsize 1 & \normalsize 1 & \normalsize 1 & \normalsize 1 & \normalsize 1 & \normalsize 0 & \normalsize 0 & \normalsize 1 & \normalsize 0 & \normalsize 0 & \normalsize 1 & \normalsize 0 & \normalsize 0 & \normalsize 1 & \normalsize 0 & \normalsize 0 & \normalsize 1\\ 
				  \cline{2-44} 				 				  
				  & \textbf{M} & \normalsize 1 & \normalsize 1 & \normalsize 1 & \normalsize 1 & \normalsize 0 & \normalsize 0 & \normalsize 1 & \normalsize 1 & \normalsize 1 & \normalsize 1 & \normalsize 1 & \normalsize 0 & \normalsize 1 & \normalsize 1 & \normalsize 1 & \normalsize 1 & \normalsize 1 & \normalsize 1 & \normalsize 0 & \normalsize 0 & \normalsize 0 & \normalsize 1 & \normalsize 1 & \normalsize 1 & \normalsize 1 & \normalsize 1 & \normalsize 1 & \normalsize 1 & \normalsize 1 & \normalsize 1 & \normalsize 1 & \normalsize 1 & \normalsize 1 & \normalsize 0 & \normalsize 0 & \normalsize 0 & \normalsize 1 & \normalsize 1 & \normalsize 1 & \normalsize 0 & \normalsize 0 & \normalsize 0\\ 
				  \cline{2-44}  				  
				  & \textbf{N} & \normalsize 1 & \normalsize 1 & \normalsize 0 & \normalsize 1 & \normalsize 1 & \normalsize 1 & \normalsize 1 & \normalsize 1 & \normalsize 1 & \normalsize 1 & \normalsize 1 & \normalsize 0 & \normalsize 1 & \normalsize 1 & \normalsize 1 & \normalsize 1 & \normalsize 1 & \normalsize 1 & \normalsize 0 & \normalsize 0 & \normalsize 0 & \normalsize 0 & \normalsize 0 & \normalsize 0 & \normalsize 1 & \normalsize 1 & \normalsize 1 & \normalsize 1 & \normalsize 1 & \normalsize 1 & \normalsize 1 & \normalsize 1 & \normalsize 1 & \normalsize 1 & \normalsize 1 & \normalsize 1 & \normalsize 1 & \normalsize 1 & \normalsize 1 & \normalsize 0 & \normalsize 0 & \normalsize 0\\
				\hline 
			\end{tabular}
		}
		\label{tbl:SymbolDatabase}
		\vspace{.3cm}
	\end{minipage}
	
	\begin{minipage}{.4\textwidth}
		\captionsetup{justification=centering, font=small}
		\caption{\small Temporal Relations between Events}
		\vspace{-0.08in}
		\scriptsize
		\resizebox{\textwidth}{2.7cm}{
			\begin{tabular}{|m{.28\columnwidth}| m{.94\columnwidth}|}
				\hline
				Follows: \hspace{0.2cm}$E_{i_{\triangleright e_i}} \rightarrow E_{j_{\triangleright e_j}}$ & \begin{tikzpicture}
					
					\draw (0,0) -- node[below]{\textbf{e$_{i}$}} ++(0.7,0);
					\filldraw (0,0) circle (2pt)node[above]{\small t$_{s_i}$} ;
					\filldraw (0.7,0) circle (2pt)node[above ]{\small t$_{e_i}{\pm \epsilon}$};
					
					\filldraw (0.9,0) circle (2pt)node[below right]{\small t$_{s_j}$};
					\filldraw (2.3,0) circle (2pt)node[below]{\small t$_{e_j}$};
					\draw (0.9,0) -- node[above]{\textbf{e$_{j}$}} ++(1.4,0);
					
					\draw (3.5,0) -- node[below]{\textbf{e$_{i}$}} ++(0.7,0);
					\filldraw (3.5,0) circle (2pt)node[above]{\small t$_{s_i}$} ;
					\filldraw (4.2,0) circle (2pt)node[above]{\small t$_{e_i}{\pm \epsilon}$};
					
					\filldraw (4.7,0) circle (2pt)node[below right]{\small t$_{s_j}$};
					\filldraw (6.1,0) circle (2pt)node[below]{\small t$_{e_j}$};
					\draw (4.7,0) -- node[above]{\textbf{e$_{j}$}} ++(1.4,0);
					
					\node [align=center] at (3.5,-0.7) {\small t$_{e_i}{\pm \epsilon}$ $\le$ t$_{s_j}$};
					
				\end{tikzpicture}  \\ \hline
				
				Contains: \hspace{0.2cm}$E_{i_{\triangleright e_i}} \succcurlyeq E_{j_{\triangleright e_j}}$ & 	\begin{tikzpicture}
					\draw (0,0) -- node[above]{\textbf{e$_i$}} ++(2.0,0);
					\filldraw (0,0) circle (2pt)node[above]{\small t$_{s_i}$} ;
					\filldraw (2,0) circle (2pt)node[above]{\small t$_{e_i}\pm \epsilon$};
					
					\draw (0,-0.5) -- node[above]{\textbf{e$_{j}$}} ++(2.0,0);
					\filldraw (0,-0.5) circle (2pt)node[below]{\small t$_{s_j}$} ;
					\filldraw (2,-0.5) circle (2pt)node[below]{\small t$_{e_j}$};
					
					\draw (3.5,0) -- node[above]{\textbf{e$_i$}} ++(2.1,0);
					\filldraw (3.5,0) circle (2pt)node[above]{\small t$_{s_i}$} ;
					\filldraw (5.6,0) circle (2pt)node[above]{\small t$_{e_i}\pm \epsilon$};	
					
					\draw (3.9,-0.5) -- node[above]{\textbf{e$_j$}} ++(1.4,0);
					\filldraw (3.9,-0.5) circle (2pt)node[below]{\small t$_{s_j}$} ;
					\filldraw (5.3,-0.5) circle (2pt)node[below]{\small t$_{e_j}$};	
					
					\draw (0,-1.5) -- node[above]{\textbf{e$_i$}} ++(2.0,0);
					\filldraw (0,-1.5) circle (2pt)node[above]{\small t$_{s_i}$} ;
					\filldraw (2.0,-1.5) circle (2pt)node[above]{\small t$_{e_i}\pm \epsilon$};
					
					\draw (-0,-2) -- node[above]{\textbf{e$_{j}$}} ++(1.5,0);
					\filldraw (0,-2) circle (2pt)node[below]{\small t$_{s_j}$} ;
					\filldraw (1.5,-2) circle (2pt)node[below]{\small t$_{e_j}$};

					\draw (3.6,-1.5) -- node[above]{\textbf{e$_i$}} ++(2.0,0);
					\filldraw (3.6,-1.5) circle (2pt)node[above]{t$_{s_i}$} ;
					\filldraw (5.6,-1.5) circle (2pt)node[above]{t$_{e_i}\pm \epsilon$};
					
					\draw (4.1,-2) -- node[above]{\textbf{e$_{j}$}} ++(1.5,0);
					\filldraw (4.1,-2) circle (2pt)node[below]{t$_{s_j}$} ;
					\filldraw (5.6,-2) circle (2pt)node[below]{t$_{e_j}$};
					\node [align=center] at (3.0,-2.7) {\small (t$_{s_{i}} \le$ t$_{s_j}$)  $\wedge$ (t$_{e_i}{\pm \epsilon}$ $\ge$ t$_{e_j}$)};
					
				\end{tikzpicture} \\ \hline
				
				Overlaps:\hspace{0.2cm} $E_{i_{\triangleright e_i}} \between E_{j_{\triangleright e_j}}$ & \begin{tikzpicture}
					
					\draw (0,0) -- node[above]{\textbf{e$_i$}} ++(2,0);
					\filldraw (0,0) circle (2pt)node[above]{\small t$_{s_i}$} ;
					\filldraw (2,0) circle (2pt)node[above]{\small t$_{e_i}\pm \epsilon$};
					
					\draw (1,-0.75) -- node[above right]{\textbf{e$_{j}$}} ++(2.0,0);
					\filldraw (1,-0.75) circle (2pt)node[below]{\small t$_{s_j}$} ;
					\filldraw (3,-0.75) circle (2pt)node[below]{\small t$_{e_j}$};
					
					\draw[dashed] (1,0) -- (1,-0.75);
					\draw[dashed] (2,0) -- (2,-0.75);
					
					\draw[dashed,>=latex,thin,<->] (1,-0.325) -- node[above]{d$_{o}$} ++(1,0);
					
					\node [align=center] at (1.75,-1.45) {\small (t$_{s_i}<$ t$_{s_j}$) $\wedge$ (t$_{e_i}{\pm \epsilon}$ $<$ t$_{e_j}$) $\wedge$ 
						\small (t$_{e_i}$ $-$ t$_{s_j}$ $\ge$ d${_o}{\pm \epsilon}$)};
					
				\end{tikzpicture} \\
				\hline
		\end{tabular} }
		\label{tbl:relations}
	\end{minipage}
	\begin{minipage}{.6\textwidth}
		\captionsetup{justification=centering, font=small}
		\caption{\mbox{\small A Temporal Sequence Database $\mathcal{D}_{\text{SEQ}}$ (H: 15-Minutes granularity)}}
		\vspace{-0.08in}
		\label{tbl:SequenceDatabase}
		\resizebox{\textwidth}{2.7cm}{
			\begin{tabular}{ |c|c|p{11cm}| }
				\hline  {\bfseries Granules} & {\bfseries Position} & {\bfseries \;\;\;\;\;\;\;\;\;\;\;\;\;\;\;\;\;\;\;\;\;\;\;\;\;\;\;\;\;\;\;\;\;\;\;\;\;\;\;\;\;\;\;\;\;\;\;\;\;\;\;Temporal sequences} \\
				\hline  
				$\textbf{\textit{H}}_1$=$\{\textbf{\textit{G}}_1$,$\textbf{\textit{G}}_2$,$\textbf{\textit{G}}_3\}$ & 1   &  (C:1,[$G_1,G_2$]), (C:0,[$G_3,G_3$]), (D:1,[$G_1,G_1$]), (D:0,[$G_2,G_3$]), (F:0,[$G_1,G_2$]), (F:1,[$G_3,G_3$]), (M:1,[$G_1,G_3$]), (N:1,[$G_1,G_2$]), (N:0,[$G_3,G_3$])
				\\
				\hline 
				$\textbf{\textit{H}}_2$=$\{\textbf{\textit{G}}_4$,$\textbf{\textit{G}}_5$,$\textbf{\textit{G}}_6\}$ & 2   &  (C:1,[$G_4,G_4$]), (C:0,[$G_5,G_6$]), (D:1,[$G_4,G_4$]), (D:0,[$G_5,G_6$]), (F:0,[$G_4,G_4$]), (F:1,[$G_5,G_6$]), (M:1,[$G_4,G_4$]), (M:0,[$G_5,G_6$]), (N:1,[$G_4,G_6$])
				\\ 
				\hline 
				$\textbf{\textit{H}}_3$=$\{\textbf{\textit{G}}_7$,$\textbf{\textit{G}}_8$,$\textbf{\textit{G}}_9\}$ & 3   &  (C:1,[$G_7,G_8$]), (C:0,[$G_9,G_9$]), (D:1,[$G_7,G_8$]), (D:0,[$G_9,G_9$]), (F:0,[$G_7,G_8$]), (F:1,[$G_9,G_9$]), (M:1,[$G_7,G_9$]), (N:1,[$G_7,G_9$])
				\\
				\hline 
				$\textbf{\textit{H}}_4$=$\{\textbf{\textit{G}}_{10}$,$\textbf{\textit{G}}_{11}$,$\textbf{\textit{G}}_{12}\}$ & 4   &  (C:0,[$G_{10},G_{12}$]), (D:1,[$G_{10},G_{11}$]), (D:0,[$G_{12},G_{12}$]), (F:0,[$G_{10},G_{11}$]),  (F:1,[$G_{12},G_{12}$]), (M:1,[$G_{10},G_{11}$]), (M:0,[$G_{12},G_{12}$]), (N:1,[$G_{10},G_{11}$]), (N:0,[$G_{12},G_{12}$])
				\\
				\hline
				$\textbf{\textit{H}}_5$=$\{\textbf{\textit{G}}_{13}$,$\textbf{\textit{G}}_{14}$,$\textbf{\textit{G}}_{15}\}$ & 5   &  (C:0,[$G_{13},G_{15}$]), (D:0,[$G_{13},G_{15}$]), (F:1,[$G_{13},G_{15}$]), (M:1,[$G_{13},G_{15}$]), (N:1,[$G_{13},G_{15}$])
				\\
				\hline 
				$\textbf{\textit{H}}_6$=$\{\textbf{\textit{G}}_{16}$,$\textbf{\textit{G}}_{17}$,$\textbf{\textit{G}}_{18}\}$ & 6   &  (C:0,[$G_{16},G_{18}$]), (D:0,[$G_{16},G_{18}$]), (F:0,[$G_{16},G_{18}$]), (M:1,[$G_{16},G_{18}$]), (N:1,[$G_{16},G_{18}$])
				\\ 
				\hline 
				$\textbf{\textit{H}}_7$=$\{\textbf{\textit{G}}_{19}$,$\textbf{\textit{G}}_{20}$,$\textbf{\textit{G}}_{21}\}$ & 7   &  (C:1,[$G_{19},G_{21}$]), (D:1,[$G_{19},G_{21}$]), (F:0,[$G_{19},G_{21}$]), (M:0,[$G_{19},G_{21}$]), (N:0,[$G_{19},G_{21}$])
				\\
				\hline 
				$\textbf{\textit{H}}_8$=$\{\textbf{\textit{G}}_{22}$,$\textbf{\textit{G}}_{23}$,$\textbf{\textit{G}}_{24}\}$ & 8   &  (C:1,[$G_{22},G_{24}$]), (D:1,[$G_{22},G_{24}$]), (F:0,[$G_{22},G_{24}$]), (M:1,[$G_{22},G_{24}$]), (N:0,[$G_{22},G_{24}$])
				\\
				\hline
				$\textbf{\textit{H}}_9$=$\{\textbf{\textit{G}}_{25}$,$\textbf{\textit{G}}_{26}$,$\textbf{\textit{G}}_{27}\}$ & 9   &  (C:0,[$G_{25},G_{27}$]), (D:0,[$G_{25},G_{27}$]), (F:1,[$G_{25},G_{27}$]), (M:1,[$G_{25},G_{27}$]), (N:1,[$G_{25},G_{27}$])
				\\
				\hline 
				$\textbf{\textit{H}}_{10}$=$\{\textbf{\textit{G}}_{28}$,$\textbf{\textit{G}}_{29}$,$\textbf{\textit{G}}_{30}\}$ & 10   &  (C:0,[$G_{28},G_{30}$]), (D:0,[$G_{28},G_{30}$]), (F:1,[$G_{28},G_{30}$]), (M:1,[$G_{28},G_{30}$]), (N:1,[$G_{28},G_{30}$])
				\\ 
				\hline 
				$\textbf{\textit{H}}_{11}$=$\{\textbf{\textit{G}}_{31}$,$\textbf{\textit{G}}_{32}$,$\textbf{\textit{G}}_{33}\}$ & 11   &  (C:1,[$G_{31},G_{31}$]), (C:0,[$G_{32},G_{33}$]), (D:1,[$G_{31},G_{31}$]), (D:0,[$G_{32},G_{33}$]), (F:0,[$G_{31},G_{32}$]), (F:1,[$G_{33},G_{33}$]), (M:1,[$G_{31},G_{33}$]), (N:1,[$G_{31},G_{33}$])
				\\
				\hline 
				$\textbf{\textit{H}}_{12}$=$\{\textbf{\textit{G}}_{34}$,$\textbf{\textit{G}}_{35}$,$\textbf{\textit{G}}_{36}\}$ & 12   &  (C:1,[$G_{34},G_{35}$]), (C:0,[$G_{36},G_{36}$]), (D:1,[$G_{34},G_{34}$]), (D:0,[$G_{35},G_{36}$]), (F:0,[$G_{34},G_{35}$]), (F:1,[$G_{36},G_{36}$]), (M:0,[$G_{34},G_{36}$]),  (N:1,[$G_{34},G_{36}$])
				\\
				\hline
				$\textbf{\textit{H}}_{13}$=$\{\textbf{\textit{G}}_{37}$,$\textbf{\textit{G}}_{38}$,$\textbf{\textit{G}}_{39}\}$ & 13   &  (C:0,[$G_{37},G_{39}$]), (D:1,[$G_{37},G_{38}$]), (D:0,[$G_{39},G_{39}$]), (F:0,[$G_{37},G_{38}$]), (F:1,[$G_{39},G_{39}$]), (M:1,[$G_{37},G_{39}$]), (N:1,[$G_{37},G_{39}$])
				\\
				\hline 
				$\textbf{\textit{H}}_{14}$=$\{\textbf{\textit{G}}_{40}$,$\textbf{\textit{G}}_{41}$,$\textbf{\textit{G}}_{42}\}$ & 14   &  (C:1,[$G_{40},G_{41}$]), (C:0,[$G_{42},G_{42}$]), (D:1,[$G_{40},G_{41}$]), (D:0,[$G_{42},G_{42}$]), (F:0,[$G_{40},G_{41}$]), (F:1,[$G_{42},G_{42}$]), (M:0,[$G_{40},G_{42}$]), (N:0,[$G_{40},G_{42}$])
				\\
				\hline
			\end{tabular} 
		}
	\end{minipage}	
\end{table*}
\vspace{-0.05in}
\subsection{Temporal Event and Temporal Relation}
\hspace{-0.15in}\textbf{Definition 3.7} (Temporal event) Consider a symbolic time series $X_S$. A \textit{temporal event} $E$ in $X_S$ is a tuple $E = (\omega, T)$ where $\omega \in \Sigma_X$ is a symbol, and $T=\{[t_{s_i}, t_{e_i}]\}$ is the set of time intervals during which $X_S$ has the value $\omega$. Each time interval has $t_{s_i}$ as the start time, and $t_{e_i}$ as the end time. 

\textbf{Instance of a temporal event:} The tuple $e = (\omega, [t_{s_i}, t_{e_i}])$ is called an \textit{instance} of the temporal event $E= (\omega, T)$, representing a single occurrence of $E$ during $[t_{s_i}, t_{e_i}]$. We use the notation $E_{\triangleright e}$ to denote that the event $E$ has an instance $e$. 

Consider the symbolic time series C in Table \ref{tbl:SymbolDatabase}. Then $E=(\text{C:1}, \{[G_1,G_2],[G_4,G_4], [G_7,G_8], [G_{19},G_{24}],$ $[G_{31},G_{31}], [G_{34},G_{35}], [G_{40},G_{41}]\})$ is an event of C, representing the time intervals during which C is associated with the symbol 1. The tuple $(\text{C:1}, [G_1,G_2])$ is an instance of $E$. Note that for simplicity, we use the granules to represent the start and end times of the time intervals, as we can trace back the timestamp associated to each granule. 

\textbf{Relations between temporal events:} 
Let $E_i$ and $E_j$ be two temporal events, and $e_i=(\omega_i,[t_{s_i}, t_{e_i}])$, $e_j=(\omega_j,[t_{s_j}, t_{e_j}])$ be their corresponding instances. We rely on the popular Allens relation model \cite{allen} to define 3 basic temporal relations: \textit{Follows}, \textit{Contains}, \textit{Overlaps} between $E_i$ and $E_j$ through $e_i$ and $e_j$. We avoid the exact time mapping problem in Allens relations by adding a tolerance \textit{buffer} $\epsilon$ to the relation's endpoints, while ensuring the relations are \textit{mutually exclusive} (proof in the technical report \cite{ho2022seasonal}). Table \ref{tbl:relations} illustrates the three relations and their conditions, with $\epsilon \ge 0$ being the buffer size, and $d_o$ representing the minimal overlapping duration between two event instances in an Overlaps relation.

\hspace{-0.15in}\textbf{Definition 3.8} (Temporal pattern) Let $\Re$$=$\{Follows, Contains, Overlaps\} be the set of temporal relations. A \textit{temporal pattern} $P$ $=$ $<$$(r_{12}, E_{1},$ $E_{2})$,...,$(r_{(n-1)(n)},E_{n-1},E_{n})$$>$ is a list of triples $(r_{\textit{ij}},E_{i},E_{j})$, each representing a relation $r_{\textit{ij}} \in \Re$ between two events $E_i$ and $E_j$.

Note that each relation $r_{\textit{ij}}$ is formed using the specific instances of $E_i$ and $E_j$. A temporal pattern of $n$ events is called an $n$-event pattern. We use $E_i \in P$ to denote that the event $E_i$ occurs in $P$, and $P_1 \subseteq P$ to say that a pattern $P_1$ is a sub-pattern of $P$. 
An example temporal pattern is shown in Fig. \ref{fig:TimeSeries}: P = $<$(Overlaps, Low Temperature, High Humidity), (Follows, Low Temperature, High Influenza Cases), (Follows, High Humidity, High Influenza Cases)$>$. Here, P is a 3-event pattern, containing pairwise temporal relations between Low Temperature, High Humidity, and High Influenza Cases.

\subsection{Temporal Sequence Database}\vspace{-0.02in}
\hspace{-0.15in}\textbf{Definition 3.9} (Sequence mapping)
Consider a symbolic time series $X_S$ of granularity $G$. Let $H$ be a granularity in $\mathcal{H}$ such that $G \trianglelefteq_m H$. A sequence mapping $g$$: X_S \rightarrow_m H$ maps $m$ adjacent symbols in $X_S$ into a single granule $H_i \in H$. 

For example, consider the symbolic time series C in Table \ref{tbl:SymbolDatabase}. Using $G \trianglelefteq_3 H$, a sequence mapping $g$$: C \rightarrow_3 H$ creates granularity $H$ where the granules are: $H_1$: $<$C:1, C:1, C:0$>$, $H_2$: $<$C:1, C:0, C:0$>$, $H_3$: $<$C:1, C:1, C:0$>$, and so on. 

\hspace{-0.15in}\textbf{Definition 3.10} (Temporal sequence of a symbolic time series) Consider a symbolic time series $X_S$ of granularity $G$. Let $<$$\omega_{1},..., \omega_m$$>$ be a symbolic sequence at granule $H_i$ in $H$, obtained by performing a sequence mapping $g$$:$ $X_S \rightarrow_m H$. A \textit{temporal sequence} $Seq_i=<e_{1},..., e_n>$ is a list of $n$ event instances, each is obtained by grouping consecutive and identical symbols $\omega$ in $H_i$ into an event instance $e = (\omega, [t_{s}, t_{e}])$. 

In the previous example, the temporal sequences of the granules in $H$ are: $Seq_1$ = $<$(C:1, [$G_1, G_2$]), (C:0, [$G_3, G_3$])$>$ at $H_1$, $Seq_2$ = $<$(C:1, [$G_4, G_4$]), (C:0, [$G_5, G_6$])$>$ at $H_2$, $Seq_3$ = $<$(C:1, [$G_7, G_8$]), (C:0, [$G_9, G_9$])$>$ at $H_3$, and so on.

\hspace{-0.15in}\textbf{Definition 3.11} (Temporal sequence database) 
Consider a symbolic database $\mathcal{D_{\text{SYB}}}$ of granularity $G$ (defined in Def 3.6) which contains a collection of symbolic time series $\{X_S\}$, and a granularity $H \in \mathcal{H}$. Let $g: X_S \rightarrow_m H$ be a sequence mapping applied to each symbolic time series $X_S$ in $\mathcal{D_{\text{SYB}}}$. The temporal sequences obtained from the mapping $g$ form a \textit{temporal sequence database} $\mathcal{D}_{\text{SEQ}}$ where each row $i$ is a set of sequences $\{Seq_i\}$ of the same granule $H_i \in H$. 
Furthermore, the temporal sequence database $\mathcal{D}_{\text{SEQ}}$ has granularity $H$.  

Table \ref{tbl:SequenceDatabase} shows an example of $\mathcal{D}_{\text{SEQ}}$, obtained from $\mathcal{D}_{\text{SYB}}$ in Table \ref{tbl:SymbolDatabase} using the mapping $g: X_S \rightarrow_3 H$ on the five symbolic time series \{C, D, F, M, N\}. 

Given a symbolic database $\mathcal{D_{\text{SYB}}}$ of granularity $G$ and a granularity hierarchy $\mathcal{H}$, we can construct different temporal sequence databases $\mathcal{D_{\text{SEQ}}}$ of different granularities $H \in \mathcal{H}$ by using different sequence mappings $g: X_S \rightarrow_m H$. For instance, in the previous example, using $g: X_S \rightarrow_3 H$, we obtain $\mathcal{D_{\text{SEQ}}}$ at 15-Minutes granularity. Using $g: X_S \rightarrow_{12} H$, we obtain $\mathcal{D_{\text{SEQ}}}$ at 1-Hour granularity.
\subsection{Frequent Seasonal Temporal Pattern}\vspace{-0.02in}
\hspace{-0.15in}\textbf{Definition 3.12} (Support set of a temporal event) Consider a temporal sequence database $\mathcal{D}_{\text{SEQ}}$ of granularity $H$, and a temporal event $E$. The set of granules $H_i$ in $\mathcal{D}_{\text{SEQ}}$ where $E$ occurs, arranged in an increasing order, is called the \textit{support set} of event $E$ and is denoted as $\text{SUP}^E = \{H_l^E, ..., H_r^E \}$, where $1 \leq l \leq r \leq |\mathcal{D}_{\text{SEQ}}|$. The granule $H_i$ at which event $E$ occurs is denoted as $H_i^E$. The support set of a group of events, denoted as $\text{SUP}^{(E_i, ..., E_k)}$, and the support set of a temporal pattern, denoted as $\text{SUP}^P = \{H_l^P, ..., H_r^P \}$, are defined similarly to that of a temporal event.

\hspace{-0.15in}\textbf{Definition 3.13} (Near support set of a temporal pattern) 
Consider a pattern $P$ with the support set $\text{SUP}^P = \{H_l^P, ..., H_r^P \}$. Let \textit{maxPeriod} be the \textit{maximum period threshold}, representing the predefined maximal period between any two consecutive granules in $\text{SUP}^P$.  The set $\text{SUP}^P$ is called a \textit{near support set} of $P$ if $\forall (H_o^P, H_p^P) \in \text{SUP}^P$$:$ $(H_o^P \text{and } H_p^P \text{are consecutive})$ $\wedge$ $|p(H_o^P)-p(H_p^P)| \le \textit{maxPeriod}$, where $p(H_o^P)$ and $p(H_p^P)$ are the positions of $H_o^P$ and $H_p^P$ in granularity $H$. We denote the near support set of pattern $P$ as $\text{NearSUP}^P$. 

Intuitively, the near support set of $P$ is a support set where $P$'s occurrences are close in time. Moreover, $\text{NearSUP}^P$ is called a \textit{maximal near support set} if $\text{NearSUP}^P$ has no other superset beside itself which is also a near support set. The near support set of an event is defined similarly to that of a pattern.

\begin{figure}[!t]
	\centering
	\includegraphics[width=1\linewidth]{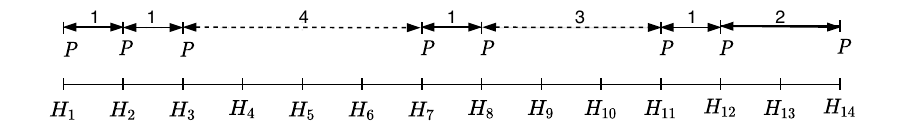}
	\vspace{-0.2in}
	\caption{Near support sets of pattern $P$ = (C:1 $\succcurlyeq$ D:1)}
	\label{fig:seasonalpattern}
	\vspace{-0.14in}
\end{figure}

As an example, consider the pattern $P$ = (Contains, C:1, D:1) (or C:1 $\succcurlyeq$ D:1)
in Table \ref{tbl:SequenceDatabase}, and let $\textit{maxPeriod}=2$. Here, the support set of $P$ is $\text{SUP}^{P}$ = $\{H_1, H_2, H_3, H_7, H_8, H_{11}, H_{12}, H_{14}\}$. Hence, $P$ has three maximal near support sets: $\text{NearSUP}_1^{P} = \{H_1,H_2,H_3\}$, $\text{NearSUP}_2^{P} = \{H_7,H_8\}$, and $\text{NearSUP}_3^{P} = \{H_{11},H_{12},H_{14}\}$. 
 Fig. \ref{fig:seasonalpattern} illustrates the three near support sets of $P$.

\hspace{-0.15in}\textbf{Definition 3.14} (Season of a temporal pattern) Let $\text{NearSUP}^P$ be a near support set of a pattern $P$. Then $\text{NearSUP}^P$ is called a \textit{season} of $P$ if $\textit{den}(\text{NearSUP}^P)$ $=$ $\mid$$\text{NearSUP}^P$$\mid$ $\ge \textit{minDensity}$, where $\scalemath{0.95}{\textit{den}(\text{NearSUP}^P)}$ counts the number of granules in $\text{NearSUP}^P$ called the \textit{density of}  $\scalemath{0.95}{\text{NearSUP}^P}$, and $\textit{minDensity}$ is a predefined minimum density threshold.

For instance, in the previous example, we have $\textit{den}(\text{NearSUP}_1^{P}) = |\text{NearSUP}_1^{P}|=3$. Similarly, $\textit{den}(\text{NearSUP}_2^{P}) = 2$, $\textit{den}(\text{NearSUP}_3^{P}) = 3$. If the occurrences of a pattern $P$ are dense enough, the near support set becomes a season of $P$. 
Intuitively, a {\em season} of a temporal pattern is a {\em concentrated occurrence period}, separated by a long {\em gap period} of no/few occurrences, before the next season starts. The season of an event is defined similarly as for a pattern.

The \textit{distance} between two seasons $\text{NearSUP}_i^P$ = $\{H_k^P, ..., H_n^P\}$ and $\text{NearSUP}_j^P$ = $\{H_r^P, ..., H_u^P\}$ is computed as:
$\textit{dist}(\text{NearSUP}_i^P, \text{NearSUP}_j^P)$ = $\mid$$p(H_n^P)-p(H_r^P)$$\mid$.

Based on the season concept and the distance measure, we define frequent seasonal temporal patterns as follows.

\hspace{-0.15in}\textbf{Definition 3.15} (Frequent seasonal temporal pattern) 
Let $\mathcal{PS}$ = $\left\{\text{NearSUP}^{P}\right\}$ be the set of seasons of a temporal pattern $P$, and $\textit{minSeason}$ be the \textit{minimum seasonal occurrence} threshold, $\textit{distInterval}$ = $[\text{dist}_{\min}, \text{dist}_{\max}]$ be the \textit{distance interval} where $\text{dist}_{\min}$ is the minimum distance and $\text{dist}_{\max}$ is the maximum distance. A temporal pattern $P$ is called a \textit{frequent seasonal temporal pattern} iff $\textit{seasons}(P)$ $=$ $\mid$$\mathcal{PS}$$\mid$ $\ge$ $\textit{minSeason}$ $\wedge$ $\forall (\text{NearSUP}_i^P$, $\text{NearSUP}_j^P)$ $\in$ $\mathcal{PS}$: they are consecutive and $\text{dist}_{\min} \leq \textit{dist}(\text{NearSUP}_i^P, \text{NearSUP}_j^P) \leq \text{dist}_{\max}$. 

Intuitively, a pattern $P$ is \textit{seasonal} if the distance between two consecutive seasons is within the predefined distance interval. Moreover, a seasonal temporal pattern is \textit{frequent} if it occurs more often than a predefined \textit{minimum seasonal occurrence} threshold. The number of seasons of a pattern $P$ is the size of $\mathcal{PS}$, and is computed as $\textit{seasons}(P) = \mid \mathcal{PS} \mid$. 

\textbf{Mining Frequent Seasonal Temporal Patterns from Time Series} \textbf{(FreqSTPfTS).} Given a set of $n$ time series $\mathcal{X}=\{X_1,...,X_n\}$ of granularity $G$, let $\mathcal{D}_{\text{SEQ}}$ be the temporal sequence database of granularity $H \in \mathcal{H}$ obtained from ${\mathcal{X}}$, and $\textit{maxPeriod}$, $\textit{minDensity}$, $\textit{distInterval}$, and $\textit{minSeason}$ be the maximum period, minimum density, distance interval, and minimum seasonal occurrence thresholds, respectively. The FreqSTPfTS problem aims to find all frequent seasonal temporal patterns $P$ in $\mathcal{D}_{\text{SEQ}}$ that satisfy the $\textit{maxPeriod}$, $\textit{minDensity}$, $\textit{distInterval}$, and $\textit{minSeason}$ constraints. 

In Section \ref{sec:ExperimentalSetup}, we provide the guidelines on how to set the values of the four constraints in real-life settings.
\section{Frequent Seasonal Temporal Pattern Mining} \label{sec:FTPMfTSMining}
\vspace{-0.02in} \subsection{Overview of FreqSTPfTS Mining Process} \label{sec:sesonalmining} \vspace{-0.02in}
The FreqSTPfTS mining process consists of two phases. \textbf{Phase 1}, \textit{Data Transformation}, converts a set of time series $\mathcal{X}$ into a symbolic database $\mathcal{D}_{\text{SYB}}$ by using the  mapping function defined in Def. 3.5, and then converts $\mathcal{D}_{\text{SYB}}$ into a temporal sequence database $\mathcal{D}_{\text{SEQ}}$ by applying the sequence mapping defined in Def. 3.9. 
\textbf{Phase 2}, \textit{Seasonal Temporal Pattern Mining (STPM)}, consists of two steps to mine frequent seasonal temporal patterns: \textit{Seasonal Single Event Mining} and \textit{Seasonal k-Event Pattern Mining} (\text{k} $\geq$ $2$). 

Before introducing the STPM algorithm in detail, we first present \textit{candidate seasonal pattern}, a concept designed to support Apriori-like pruning in STPM.

\vspace{-0.05in}
\subsection{Candidate Seasonal Pattern}\label{sec:Candidate}\vspace{-0.02in}	
Pattern mining methods often use the \textit{anti-monotonicity} property of the \textit{support} measure to reduce the search space \cite{hdfs}. This property ensures that an infrequent event $E_i$ cannot form a frequent 2-event pattern $P$, since support($E_i$) $\ge$ support($P$). Hence, if $E_i$ is infrequent, we can safely remove $E_i$ and any of its combinations from the search space, and still guarantee the algorithm completeness.
However, seasonal temporal patterns constrained by the $\textit{maxPeriod}$, $\textit{minDensity}$, $\textit{distInterval}$ and $\textit{minSeason}$ thresholds do \textit{not} uphold this property, as illustrated below. 

Consider an event $E$ = M:1 and a 2-event pattern $P$ = \text{M:1} $\succcurlyeq$ \text{N:1} in Table \ref{tbl:SequenceDatabase}. Let \textit{maxPeriod} = 2, \textit{minDensity} = 3, \textit{distInterval} = [4, 10], and \textit{minSeason} = 2. From the constraints, we can identify the seasons of $E$ and $P$ as: $\mathcal{PS}^{E}$ = $\{\text{NearSUP}_1^{E}\}$ = $\{H_1,H_2,H_3,H_4,H_5,H_6,H_8,H_9,H_{10},H_{11},H_{13}\}$, and $\mathcal{PS}^{P}$ = $\{\{\text{NearSUP}_1^{P}\}$ = $\{H_1,H_3,H_4,H_5,H_6\}$, $\{\text{NearSUP}_2^{P}\}$ = $\{H_{10},H_{11},H_{13}\}\}$. Here, for the pattern $P$, $H_2$ is not present in $\{\text{NearSUP}_1^{P}\}$ since $P$ does not occur in $H_2$, and $H_9$ is not present in $\{\text{NearSUP}_2^{P}\}$ because of the constraint $\text{dist}_{\min}$ = 4. Hence, we have: $\mid$$\mathcal{PS}^{E}$$\mid$$=$$1$ and $\mid$$\mathcal{PS}^{P}$$\mid$$=$$2$. Due to the $\textit{minSeason}$ constraint, $E$ is not a frequent seasonal event, whereas $P$ is. This shows that seasonal temporal patterns do not adhere to the anti-monotonic property. 

To improve STPM performance, we propose the novel \textit{maximum seasonal occurrence} measure, called $\textit{maxSeason}$, that upholds the anti-monotonicity property to prune infrequent patterns and reduce STPM search space. Indeed, $\textit{maxSeason}$ is an upper bound on the number of seasons of a pattern.
	
\hspace{-0.15in}\textbf{Maximum seasonal occurrence of a temporal pattern $P$:} is the ratio between the number of granules in the support set SUP$^P$ of $P$, and the $\textit{minDensity}$ threshold:
\vspace{-0.1in}
	\begin{equation}
		\small
		\textit{maxSeason}({P}) = \frac { |SUP^P| }{\textit{minDensity}} 
		\label{eq:maxSeasonPattern}
	\end{equation}

Eq. \eqref{eq:maxSeasonPattern} divides the number of granules containing $P$ by the minimum density of a season. Thus, it computes the maximum seasons a pattern $P$ can have. The maximum seasonal occurrence of a single event $E$, and of a group of events $(E_i, ..., E_k)$, are defined in a similar way. 
Below, we show how $\textit{maxSeason}$ upholds the anti-monotonicity property.

\begin{lem} \label{lempattern1} \vspace{-0.05in}
		Let $P$ and $P^{'}$ be two temporal patterns such that $P^{'} \subseteq P$. Then $\textit{maxSeason}(P^{'}) \geq \textit{maxSeason}(P)$. 
\end{lem} 
\begin{proof}
	We have: 
	
	$\textit{maxSeason}(P^{'}) = \frac { |SUP^{P^{'}}|  }{\textit{minDensity}}$, $ \textit{maxSeason}(P) = \frac { |SUP^P|  }{\textit{minDensity}}$
	
	Since: 
	$|SUP^{P^{'}}| \geq |SUP^P| \text{ (Derived from Def. 3.12)} $
	
	Hence: $\textit{maxSeason}(P^{'}) \geq \textit{maxSeason}(P)$
\end{proof}
\begin{lem}\label{lemeventpattern}\vspace{-0.05in}
	Let $P$ be a k-event temporal pattern formed by a k-event group $(E_1,...,E_k)$. Then, $\textit{maxSeason}(P) \leq \textit{maxSeason}{(E_1,...,E_k)}$. 
\end{lem} 
\begin{proof}
	Derived directly from Def. 3.12, and Eq. \eqref{eq:maxSeasonPattern}. 
\end{proof}

From Lemmas \ref{lempattern1} and \ref{lemeventpattern}, the $\textit{maxSeason}$ of a pattern $P$ is always at most the $\textit{maxSeason}$ of its sub-pattern $P^{'}$, and of its events $(E_1,...,E_k)$. Thus, $\textit{maxSeason}$ upholds the anti-monotonicity property, and can be used to reduce the STPM search space. Below, we define the \textit{candidate pattern} concept that uses $\textit{maxSeason}$ as a gatekeeper to identify frequent/ infrequent seasonal patterns.
	
\hspace{-0.15in}\textbf{Candidate seasonal pattern:} A temporal pattern $P$ is a \textit{candidate seasonal pattern} if $\textit{maxSeason}(P) \geq \textit{minSeason}$.

Similarly, a group of k events $G_E= (E_1,..., E_k)$ ($k \ge 1$) is a \textit{candidate seasonal k-event group} if $\textit{maxSeason}(G_E) \geq \textit{minSeason}$. 
Intuitively, a pattern $P$ (or k-event group $G_E$) is infrequent if its $\textit{maxSeason}$ is less than $\textit{minSeason}$. Hence, $P$ (or $G_E$) can be safely removed from the search space.

Next, we present our STPM algorithm and detail the two mining steps. Algorithm \ref{algorithmSFTPM} provides the pseudo-code of STPM. 

\vspace{-0.05in} \subsection{Mining Seasonal Single Events} \label{sec:1patternmining}
\begin{figure*}
	\setlength{\tabcolsep}{0pt}
	\begin{tabularx}{\linewidth}{ll}
		\begin{minipage}{.28\linewidth}
			\begin{minipage}{\linewidth}
				\captionsetup{justification=centering, font=small}
				\includegraphics[width=1\textwidth,height=3.1cm]{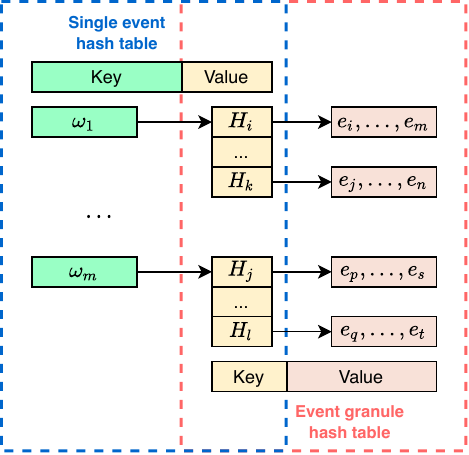}
				\vspace{-0.2in}
				\caption{The $HLH_1$ structure}
				\label{fig:hlh1}
			\end{minipage}
		\end{minipage}&
		\hspace{0.2in}
		\begin{minipage}{.68\linewidth}
			\begin{minipage}{\linewidth}
				\centering
				\captionsetup{justification=centering, font=small}
				\includegraphics[width=1\textwidth,height=3.1cm]{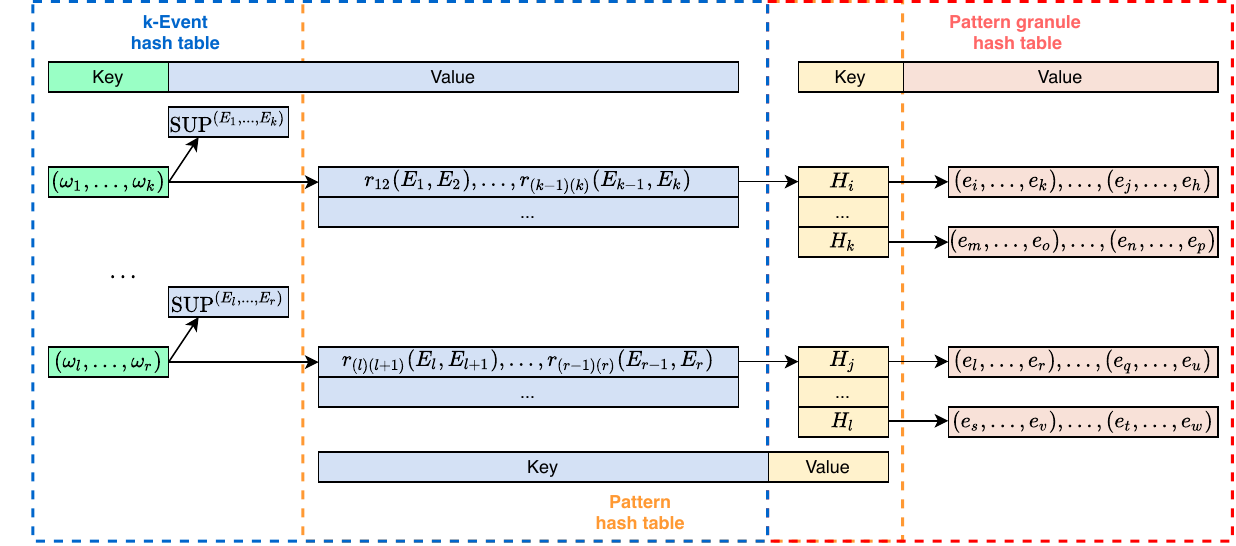}
				\vspace{-0.2in}
				\caption{The $HLH_k (k \geq 2)$ structure}
				\label{fig:hlhk}
			\end{minipage}
		\end{minipage}
	\end{tabularx}
	\vspace{-0.2in}
\end{figure*}

\setlength{\textfloatsep}{0pt}
\SetNlSty{}{}{:} 	
\begin{algorithm}[!t]
	\algsetup{linenosize=\tiny}
	\SetInd{0.5em}{0.5em}
	\small
	\DontPrintSemicolon
	\caption{\mbox{Frequent Seasonal Temporal Pattern Mining}}
	\label{algorithmSFTPM}
	
	\KwInput{Temporal sequence database $\mathcal{D_{\text{SEQ}}}$, the thresholds: $\textit{maxPeriod}$, $\textit{minDensity}$, $\textit{distInterval}$, $\textit{minSeason}$}
	\KwOutput{The set of frequent seasonal temporal patterns $\mathcal{P}$}
	
	\nonl // Step 2.1: Mining frequent seasonal single events \;
	\ForEach{\textit{event} $E_i \in \mathcal{D_{\text{SEQ}}}$}{
		Find $SUP^{E_i}$ and compute $\textit{maxSeason}(E_i)$ ;\;
		\If{$\textit{maxSeason}(E_i) \geq \textit{minSeason}$}{
			Insert $E_i$ into $\textit{Candidate1Event}$ ;\;
		}
	}
	
	\ForEach{\textit{candidate} $E_i \in \textit{Candidate1Event}$}{
		Find $\text{NearSUP}^{E_i}$ that satisfies $\textit{maxPeriod}$ and $\textit{minDensity}$ ;\;
		Find $PS^{E_i}$ that adheres $\textit{distInterval}$ ;\;
		\If{$|PS^{E_i}| \geq \textit{minSeason}$}{
			Insert $E_i$ into $\mathcal{P}$;  //$E_i$ is a frequent seasonal event
		}
	}
	
	\nonl // Step 2.2: Mining frequent seasonal k-event patterns, $k \geq 2$ \;
	FilteredF1 $\leftarrow$ Transitivity\_Filtering($F_1$); \;
	kEventGroups $\leftarrow$ Cartesian(FilteredF1, $F_{k-1}$);\;  
	CandidatekEvent $\leftarrow$ maxSeason\_Filtering(kEventGroups);\;
	\ForEach{\textit{kEvent} in CandidatekEvent}{
		(k-1)-event\_patterns $\leftarrow$ Retrieve\_Relations($PH_{k-1}$);\;  
		\mbox{k-event\_patterns $\leftarrow$ Iterative\_Check((k-1)-event\_patterns, $E_k$);\;}
		
		\ForEach{P in k-event\_patterns}{
			\If{$\textit{maxSeason}(P) \geq minSeason$}{
				Insert $P$ into $\textit{CandidatekPatterns}$;\;
			}
		}
	}
	\ForEach{\textit{candidate} $P \in \textit{CandidatekPatterns}$}{
		Find $\text{NearSUP}^{P}$ satisfying $\textit{maxPeriod}$ and $\textit{minDensity}$;\;
		Identify $\mathcal{PS}^{P}$ adhering to $\textit{distInterval}$ ;\;
		\If{$|\mathcal{PS}^{P}| \geq \textit{minSeason}$}{
			Insert $P$ into $\mathcal{P}$;  //$P$ is a frequent seasonal pattern
		}
	}
\end{algorithm}

The first step in STPM is to mine frequent seasonal single events (Alg. \ref{algorithmSFTPM}, lines 1-9) that satisfy the constraints of $\textit{maxPeriod}$, $\textit{minDensity}$, $\textit{distInterval}$ and $\textit{minSeason}$. To do that, we first look for the candidate single events defined in Section \ref{sec:Candidate}, and then use only the found candidates to mine frequent seasonal events.

The candidate single events are found by first scanning $\mathcal{D}_{\text{SEQ}}$ to identify the support set $SUP^{E_i}$ for each event $E_i$, from which we compute the maximum seasonal occurrence $\textit{maxSeason}(E_i)$. If $\textit{maxSeason}(E_i) \geq \textit{minSeason}$, then $E_i$ is a candidate seasonal single event. Otherwise, $E_i$ is not a candidate and is removed from the search space. Note that we only need to scan $\mathcal{D}_{\text{SEQ}}$ once to find all candidate events. 

To mine frequent seasonal events, for each candidate event $E_i$, we iterate through the support set $SUP^{E_i}$, and calculate the period $pr_{ij}$ between every two consecutive granules in $SUP^{E_i}$, and determine the near support sets $\text{NearSUP}^{E_i}$ that satisfy $\textit{maxPeriod}$ and $\textit{minDensity}$. Next, the set of seasons $\mathcal{PS}^{E_i}$ is identified by selecting the near support sets that adhere to the $\textit{distInterval}$ constraint. Finally, the frequent seasonal events are determined by comparing the number of seasons of $E_i$ to $\textit{minSeason}$, selecting only those that have $\textit{seasons}(E_i) = \mid \mathcal{PS}^{E_i} \mid$ $\geq \textit{minSeason}$. 

We use a \textit{hierarchical lookup hash structure} $HLH_1$ to store the candidate seasonal single events. This data structure enables fast search when mining seasonal k-events patterns ($k \geq 2$). Note that we maintain the candidate events in $HLH_1$ instead of the frequent seasonal events, as the $\textit{maxSeason}$ of candidate events upholds the anti-monotonicity property, and can thus be used for pruning. We illustrate $HLH_1$ in Fig. \ref{fig:hlh1}, and describe the data structure below.
 
\textbf{Hierarchical lookup hash structure $HLH_1$:} The $HLH_1$ is a hierarchical data structure that consists of two hash tables: the \textit{single event hash table} $EH$, 
and the \textit{event granule hash table} $GH$. 
Each hash table has a list of $<$key, value$>$ pairs. 
In $EH$, the key is the event symbol $\omega \in \Sigma_X$ representing the candidate $E_i$, and the value is the list of granules $<H_i,...,H_k>$ in $SUP^{E_i}$. In $GH$, the key is the list of granules shared in the value field of $EH$, while the value stores event instances of $E_i$ that appear at the corresponding granule in $\mathcal{D}_{\text{SEQ}}$.
The $HLH_1$ structure enables fast retrieval of event granules and instances when mining candidate seasonal k-event patterns in the next step of STPM. 

\begin{figure}
	\centering
	\captionsetup{justification=centering, font=small}
	\begin{minipage}{\linewidth}
		\includegraphics[width=\textwidth,height=4.1cm]{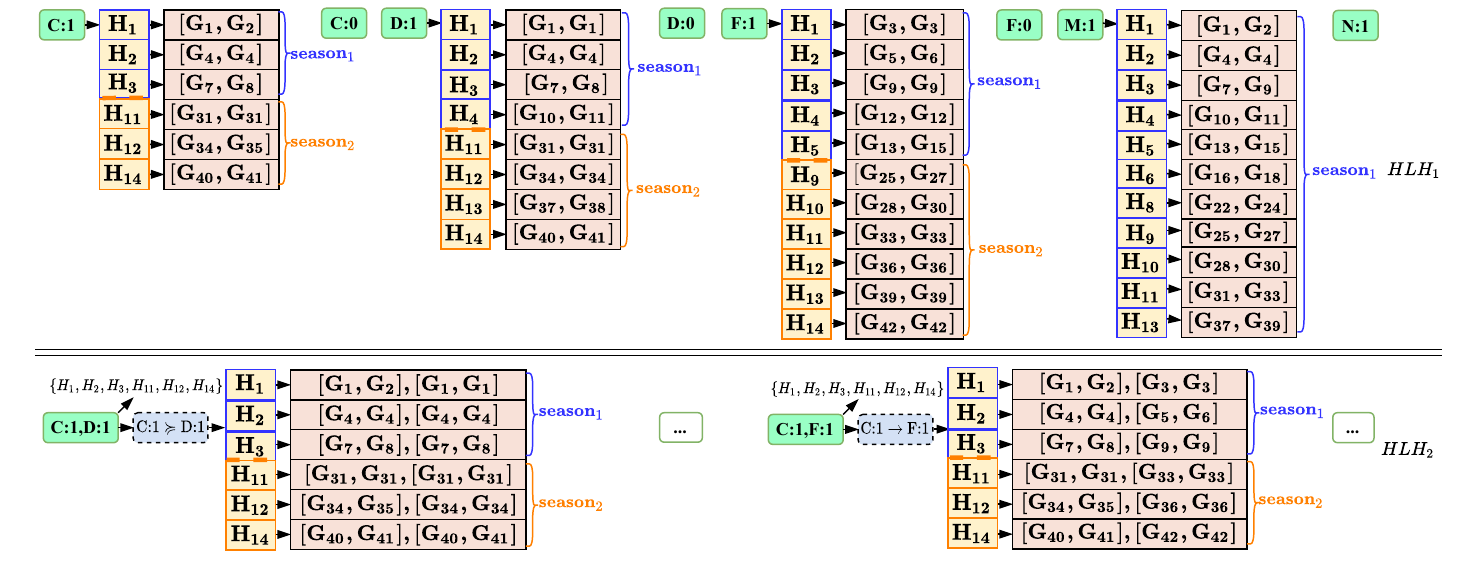}
		\vspace{-0.3in}
		\caption{A hierarchical lookup hash tables for the running example}
		\label{fig:patternTree}
	\end{minipage}
\end{figure}

We provide an example of $HLH_1$ in Fig. \ref{fig:patternTree} using data in Table \ref{tbl:SequenceDatabase} with \textit{maxPeriod} = 2, \textit{minDensity} = 3, \textit{distInterval} = [4, 10], and \textit{minSeason} = 2. Here, out of $10$ events in $\mathcal{D}_{\text{SEQ}}$, we have eight candidate seasonal single events stored in $HLH_1$: C:1, C:0, D:1, D:0, F:1, F:0, M:1, and N:1. Due to space limitations, we only provide the detailed internal structure of four candidate events. Among the eight candidates, the event M:1 does not satisfy the $\textit{minSeason}$ threshold since $\textit{season}$(M:1) = 1, and thus, is not a frequent seasonal event. However, M:1 is still present in $HLH_1$ as M:1 might create frequent seasonal k-event patterns. In contrast, N:0 and M:0 are not the candidate seasonal events because they do not satisfy the $\textit{maxSeason}$ constraint, and are omitted from $HLH_1$.

\textbf{Complexity:} The complexity of finding frequent seasonal events is $O(n \cdot | \mathcal{D}_{\text{SEQ}} |)$, where $n$ is the number of events.
\vspace{-0.07in}
\begin{proof}
	(Sketch - Full proof in \cite{ho2022seasonal}). Computing $\textit{maxSeason}$ for $n$ events takes $O(n \cdot | \mathcal{D}_{\text{SEQ}} |)$. Identifying the set of seasons $\mathcal{PS}$ of all candidate events $E_i$ takes $O(n \cdot |SUP^{E_i}|)$. The overall complexity is thus: $O(n \cdot | \mathcal{D}_{\text{SEQ}} | + n \cdot |SUP^{E_i}|) \sim O(n \cdot | \mathcal{D}_{\text{SEQ}} |)$.
\end{proof}

\vspace{-0.1in}
\subsection{Mining Seasonal k-event Patterns} \label{sec:kpatternmining}
\textbf{Search space of STPM.} The next step of STPM is to mine frequent seasonal k-event patterns ($k \ge 2$). A straightforward approach is to enumerate all possible k-event combinations, and check whether each combination can form frequent seasonal patterns. However, this naive approach is very expensive as it creates a very large search space, approximately of size $O(n^h3^{h^2})$, where $n$ is the number of distinct events in $\mathcal{D}_{\text{SEQ}}$, and $h$ is the maximal length of a temporal pattern, making it computationally prohibitive to mine seasonal patterns.

\begin{proof}
		(Sketch - Full proof in \cite{ho2022seasonal}). 
	The number of seasonal single events is: $N_1=n \sim O(n)$. 
	For mining 2-event groups, the number of 2-event groups is: $N_2\sim O(n^2)$.
	Each 2-event group in $N_2$ can form $3$ different temporal relations, and thus, the total number of seasonal 2-event patterns is: $N_2\times 3^1$ $\sim O(n^23^1)$. Similarly, the number of seasonal h-event patterns is $ O(n^h3^{h^2})$. Therefore, the total number of seasonal temporal patterns is $O(n)+O(n^23^1)+...+O(n^h3^{h^2}) \sim O(n^h3^{h^2})$.
\end{proof}

The problem of a large search space is thus alleviated by using an iterative mining process that first finds candidate seasonal k-event groups, and then mines frequent seasonal k-event patterns only from the candidates. Below, we first introduce the data structure used in this mining step. 

\textbf{The hierarchical lookup hash structure $HLH_k$:} We use the \textit{hierarchical lookup hash structure} $HLH_k$ $(k \geq 2)$ to maintain candidate seasonal k-event groups and patterns, as illustrated in Fig. \ref{fig:hlhk}. 
The $HLH_k$ contains three hash tables: the \textit{k-event hash table} $EH_k$, the \textit{pattern hash table} $PH_k$, and the \textit{pattern granule hash table} $GH_k$. 
For each {\em $<$key, value$>$} pair of $EH_k$, {\em key} is the list of symbols $(\omega_1 ..., \omega_k)$ representing the candidate k-event group $(E_1,...,E_k)$, and {\em value} is an \textit{object} which consists of two components: (1) the support set $SUP^{(E_1,...,E_k)}$, and (2) a list of candidate seasonal k-event temporal patterns.  
In $PH_k$, {\em key} is the candidate pattern $P$ which indeed takes the {\em value} component of $EH_k$, while {\em value} is the list of granules that contain $P$. 
In $GH_k$, {\em key} is the list of granules containing $P$ which indeed takes the {\em value} component of $PH_k$, while {\em value} is the list of event instances from which the temporal relations in $P$ are formed. 
The $HLH_k$ hash structure helps speed up the candidate seasonal k-event group mining through the use of the support set in $EH_k$, and enables fast search for temporal relations between $k$ events using the information in $PH_k$ and $GH_k$.

\textbf{4.1 Mining candidate seasonal k-event groups.} We first find candidate seasonal k-event groups (Alg. \ref{algorithmSFTPM}, lines 10-12).
 
Let $F_{k-1}$ be the set of candidate seasonal (k-1)-event groups found in $HLH_{k-1}$, and $F_1$ be the set of candidate seasonal single events in $HLH_1$. We first generate all possible k-event groups by computing the Cartesian product $F_{k-1} \times F_1$. Next, for each k-event group $(E_1, ..., E_k)$, we compute the support set $SUP^{(E_1, ..., E_k)}$ by taking the \textit{intersection} between $SUP^{(E_1, ..., E_{k-1})}$ in $EH_{k-1}$ and $SUP^{E_k}$ in $EH$. 
We then compute $\textit{maxSeason}(E_1, ..., E_k)$, and evaluate whether $(E_1, ..., E_k)$ is a candidate k-event group, i.e., $\textit{maxSeason}(E_1, ..., E_k) \geq \textit{minSeason}$. If $(E_1, ..., E_k)$ is a candidate, it is kept in $EH_k$ of $HLH_k$.

\textbf{4.2 Mining frequent seasonal k-event patterns.} We use the found candidate k-event groups to mine frequent seasonal k-event patterns (Alg. \ref{algorithmSFTPM}, lines 13-23).  
We first discuss the case of 2-event patterns, and then generalize to k-event patterns.

\textit{4.2.1 Mining frequent seasonal 2-event patterns:} 
For each candidate 2-event group $(E_i,E_j)$, we use the support set $SUP^{(E_i,E_j)}$ to retrieve the temporal sequences $\mathcal{S}$ that contain $(E_i,E_j)$. Next, for each sequence $S \in \mathcal{S}$, we extract their event instances $(e_i,e_j)$, and verify the relation between them. We then compute the $\textit{maxSeason}$ of the 2-event pattern $P$ and determine if $P$ is a candidate pattern, i.e., $\textit{maxSeason}(P) \geq \textit{minSeason}$. Finally, the candidate seasonal 2-event patterns are stored in $PH_2$, while their event instances are stored in $GH_2$.

Based on the set of candidate seasonal 2-event patterns $P$, we determine whether $P$ is a frequent seasonal 2-event pattern by checking the constraints of $\textit{maxPeriod}$, $\textit{minDensity}$, $\textit{distInterval}$ and $\textit{minSeason}$ as in the case of single events, using the support set $SUP^P$ retrieved from the value of $PH_2$. 

\textit{4.2.2 Mining frequent seasonal k-event patterns:} 
Let $N_{k-1}=(E_1,...,E_{k-1})$ be a candidate (k-1)-event group in $HLH_{k-1}$, $N_1=(E_k)$ be a candidate single event in $HLH_1$, and $N_k=N_{k-1} \cup N_1 = (E_1,...,E_k)$ be a candidate k-event in $HLH_k$. 
To find k-event patterns for $N_k$, we first retrieve the set of candidate (k-1)-event patterns $\mathcal{P}_{k-1}$ by accessing the $EH_{k-1}$ table. Each $P_{k-1} \in \mathcal{P}_{k-1}$ is a list of $\frac{1}{2}(k-1)(k-2)$ triples: $\{(r_{12}$, $E_{1}$, $E_{2})$,...,$(r_{(k-2)(k-1)}$, $E_{k-2}$, $E_{k-1})\}$. We iteratively verify the possibility of $P_{k-1}$ forming a k-event pattern $P_k$ with $E_k$ as follows.  

We first start with the triple $(r_{(k-1)k}$, $E_{k-1}$, $E_{k})$. If $(r_{(k-1)k}$, $E_{k-1}$, $E_{k})$ does not exist in $HLH_2$, then $P_k$ is not a candidate k-event pattern, and the verification stops immediately. Otherwise, we continue the similar verification on the triple $(r_{(k-2)k}$, $E_{k-2}$, $E_k)$, until it reaches $(r_{1k}$, $E_{1}$, $E_{k})$. Next, we compute $\textit{maxSeason}(P_k)$ to determine whether $P_k$ is a candidate k-event pattern, i.e., $\textit{maxSeason}(P_k)$ $\geq$ $\textit{minSeason}$. The candidate k-event patterns are maintained in $PH_k$ and $GH_k$. 
Finally, we mine frequent seasonal k-event patterns from the found candidates, similar to 2-event patterns.

\textbf{Using transitivity property to optimize candidate k-event groups:} In Section 4.1, when mining candidate k-event groups, we perform the Cartesian product between $F_{k-1}$ and $F_1$. However, using the candidate single events in $F_1$ to generate k-event groups can create redundancy, since events in $F_1$ when combined with $F_{k-1}$ might not form any frequent seasonal k-event patterns. For example, consider the event F:0 in $HLH_1$ in Fig. \ref{fig:patternTree}. Here, F:0 is a candidate single event, and thus, can be combined with 2-event groups in $HLH_2$ such as (C:1, D:1) to create a 3-event group (C:1, D:1, F:0). However, (C:1, D:1, F:0) cannot form any candidate seasonal 3-event patterns, since F:0 is not present in any candidate 2-event patterns in $HLH_2$. To reduce such redundancy and further optimize the mining, we use the \textit{transitivity property} of temporal relations to identify such event groups.
\begin{lem} \label{lem:transitivity}
Let $\scalemath{0.9}{S=<e_1}$,..., $\scalemath{0.9}{e_{k-1}>}$ be a temporal sequence, $\scalemath{0.9}{P=<(r_{12}, E_{1_{\triangleright e_1}}, E_{2_{\triangleright e_2}}),...,(r_{(k-2)(k-1)}, E_{{k-2}_{\triangleright e_{k-2}}}, E_{{k-1}_{\triangleright e_{k-1}}})>}$ be a (k-1)-event pattern that occurs in $S$, $e_k$ be a new event instance added to $S$ to create the temporal sequence $\scalemath{0.9}{S^{'}=<e_1}$,..., $\scalemath{0.9}{e_{k}>}$. 
	The set of temporal relations $\Re$ is transitive on $S^{'}$: $\forall e_i \in S^{'}$, $i < k$, $\exists r \in \Re$ s.t. $r(E_{i_{\triangleright e_i}}$,$E_{k_{\triangleright e_k}})$ hold.
	\vspace{-0.05in}
\end{lem}

Lemma \ref{lem:transitivity} states the temporal transitivity property between temporal events, and is used to prove the following lemma.

\begin{lem} \label{lem:filter}
	Let $N_{k-1}=(E_1,...,E_{k-1})$ be a candidate seasonal (\textit{k-1})-event group, and $E_k$ be a candidate seasonal single event. The group $N_k= N_{k-1} \cup E_k$ can form candidate seasonal k-event temporal patterns if $\forall E_i \in N_{k-1}$, $\exists r \in \Re$ s.t. $r(E_i,E_k)$ is a candidate seasonal temporal relation. 
	\vspace{-0.05in}
\end{lem}
From Lemma \ref{lem:filter}, only single events in $HLH_1$ that occur in $HLH_{k-1}$ should be used to create k-event groups. We identify these single events by filtering $F_1$, and creating the set \textit{FilteredF1}. Then, the Cartesian product $F_{k-1}$ $\times$ $F_1$ is replaced by $F_{k-1}$ $\times$ \textit{FilteredF1} to generate k-event groups. 

\textbf{Complexity:} 
Let $n$ be the number of single events in $HLH_1$, $i$ be the average number of instances of each event, $r$ be the number of (k-1)-event patterns in $HLH_{k-1}$, and $u$ be the average number of granules of each event/ temporal relation. The complexity of frequent seasonal k-event pattern mining is $O(n^2 i^2 u^2)$ + $O(|F_1|$ $\cdot$ $|F_{k-1}|$ $\cdot$ $r$ $\cdot$ $k^2 \cdot u$$)$.
\begin{proof}
	(Sketch - Full proof in \cite{ho2022seasonal}). Computing $\textit{maxSeason}$ of 2-event patterns takes $O(n^2 i^2 u^2)$. Identifying the set of seasons $\mathcal{PS}$ of candidate 2-event patterns takes $O(n^2  u)$. The complexity of frequent seasonal 2-event pattern mining is: $O($ $n^2 i^2 u^2 $ + $n^2  u$ $) \sim O(n^2 i^2 u^2)$.
	Computing $\textit{maxSeason}$ of k-event patterns $(k > 2)$ takes $O(|F_1|$ $\cdot$ $|F_{k-1}|$ $\cdot$ $r$ $\cdot$ $k^2 \cdot u$$)$. Identifying the set of seasons $\mathcal{PS}$ of candidate k-event patterns takes $O(|F_1| \cdot |F_{k-1}| \cdot r \cdot u)$. The complexity of frequent seasonal k-event pattern mining is: $O(|F_1|$ $\cdot$ $|F_{k-1}|$ $\cdot$ $r$ $\cdot$ $k^2 \cdot u$ + $|F_1| \cdot |F_{k-1}| \cdot r \cdot u) $ $\sim O(|F_1|$ $\cdot$ $|F_{k-1}|$ $\cdot$ $r$ $\cdot$ $k^2 \cdot u$$)$.
	Thus, the total time complexity is $\scalemath{0.95}{O(n^2 i^2 u^2)$ + $O(|F_1|$ $\cdot$ $|F_{k-1}|$ $\cdot$ $r$ $\cdot$ $k^2 \cdot u$$)}$.
\end{proof}

\textbf{STPM overall complexity:} The space complexity of STPM is $O(n^h3^{h^2})$. The time complexity of STPM depends on the size of the search space $O(n^h3^{h^2})$, i.e., STPM scales exponentially with quadratic exponent in the pattern length $h$, and on the complexity of the mining process itself, i.e., $O(n \cdot | \mathcal{D}_{\text{SEQ}} |)$ $+$ $O(n^2 i^2 u^2)$ $+$ $O($$\mid$$F_1$$\mid$ $\cdot$ $\mid$$F_{k-1}$$\mid$ $\cdot$ $r$ $\cdot$ $k^2 \cdot u)$. While the parameters $n$, $h$, $i$, $r$ and $k$ depend on the number of time series, others such as $\mid$${F_1}$$\mid$, $\mid$$F_{k-1}$$\mid$ and $u$ depend on the number of temporal sequences. Thus, STPM space and time complexities are driven by two main factors: the number of time series and the number of temporal sequences.

\section{Approximate STPM}\label{sec:MI}\vspace{-0.02in}
\subsection{Correlated Symbolic Time Series}\vspace{-0.02in}
Let $X_S$ and $Y_S$ be the symbolic series representing the time series $X$ and $Y$, and $\Sigma_X$, $\Sigma_Y$ be their symbolic alphabets.

\hspace{-0.15in}\textbf{Definition 5.1} (Entropy) The \textit{entropy} of $X_S$, denoted as $H(X_S)$, is defined as\vspace{-0.1in}
\begin{equation}
\scalemath{0.9}{H(X_S)= - \sum_{x \in \Sigma_X} p(x) \cdot \log p(x)}
\vspace{-0.1in}
\end{equation}	
where $p(x)$ is the probability of $X_S$. Intuitively, the entropy measures the uncertainty of the possible outcomes of $X_S$ \cite{thomas}. 

The \textit{conditional entropy} $H(X_S\vert Y_S)$ is defined as\vspace{-0.1in}
\begin{equation}
\scalemath{0.9}{
H(X_S\vert Y_S) =  - \sum_{x \in \Sigma_X} \sum_{y \in \Sigma_Y} p(x,y) \cdot \log \frac{p(x,y)}{p(y)}
}
\vspace{-0.1in}
\end{equation}
where $p(x,y)$ is the joint probability of $(X_S,Y_S)$, and $p(y)$ is the probability of $Y_S$.
 
\hspace{-0.15in}\textbf{Definition 5.2} (Mutual information) The \textit{mutual information} (MI) of two symbolic series $X_S$ and $Y_S$, denoted as $I(X_S;Y_S)$, \mbox{is defined as} \vspace{-0.15in}
\begin{equation}	
\scalemath{0.9}{
I(X_S;Y_S)=\sum_{x \in \Sigma_X} \sum_{y \in \Sigma_Y} p(x,y) \cdot \log \frac{p(x,y)}{p(x) \cdot p(y)} 
}
\label{eq:MI}
\vspace{-0.1in}
\end{equation}
The MI represents the reduction of uncertainty of one variable (e.g., $X_S$), given the knowledge of another variable (e.g., $Y_S$). The larger $I(X_S;Y_S)$, the more information is shared between $X_S$ and $Y_S$.  
Since $0 \le I(X_S;Y_S) \leq \min\{H(X_S), H(Y_S)\}$ \cite{thomas}, the MI value has no upper bound. To scale it into the range $[0-1]$, we normalize the MI as defined below. 

\hspace{-0.15in}\textbf{Definition 5.3} (Normalized mutual information) The \textit{normalized mutual information} (NMI) of two symbolic time series $X_S$ and $Y_S$, denoted as $\widetilde{I}(X_S;Y_S)$, is defined as\vspace{-0.05in}
\begin{equation}	
\scalemath{0.9}{
\widetilde{I}(X_S;Y_S)= \frac{I(X_S;Y_S)}{H(X_S)} =1-\frac{H(X_S|Y_S)}{H(X_S)}
}
\label{eq:NMI}
\vspace{-0.1in}
\end{equation}
$\widetilde{I}(X_S;Y_S)$ represents the reduction (in percentage) of the uncertainty of $X_S$ due to knowing $Y_S$. Based on Eq. \eqref{eq:NMI}, a pair of variables $(X_S,Y_S)$ has a mutual dependency if $\widetilde{I}(X_S;Y_S) > 0$. Moreover, Eq. \eqref{eq:NMI} also shows that NMI is not symmetric, i.e., $\widetilde{I}(X_S;Y_S)  \neq \widetilde{I}(Y_S;X_S)$.

\hspace{-0.15in}\textbf{Definition 5.4} (Correlated symbolic time series) Let $\mu$ ($0 < \mu \le 1$) be the mutual information threshold. We say that $X_S$ and $Y_S$ are \textit{correlated} iff $\min \lbrace \widetilde{I}(X_S; Y_S), \widetilde{I}(Y_S; X_S) \rbrace \geq \mu$, and \textit{uncorrelated} otherwise.
 
\vspace{-0.08in}
\subsection{Lower Bound of the maxSeason}\vspace{-0.02in}
Consider two symbolic series $X_S$ and $Y_S$. Let $X_1$ be a temporal event in $X_S$, $Y_1$ be a temporal event in $Y_S$, $\mathcal{D}_{\text{SYB}}$ and $\mathcal{D}_{\text{SEQ}}$ be the symbolic and the sequence databases created from $X_S$ and $Y_S$, respectively. 
We have the following relation between $\widetilde{I}(X_S;Y_S)$ in $\mathcal{D}_{\text{SYB}}$, and \textit{maxSeason}$(X_1,Y_1)$ in $\mathcal{D}_{\text{SEQ}}$. 
\begin{theorem}
(Lower bound of the maximum seasonal occurrence) Let $\mu$ be the mutual information threshold. If the NMI\hspace{0.04in} $\widetilde{I}(X_S$;$Y_S) \ge \mu$, then the maximum seasonal occurrence of $(X_1,Y_1)$ in $\mathcal{D}_{\text{SEQ}}$ has a lower bound: \vspace{-0.1in}
\begin{align} \vspace{-0.15in}
	\scalemath{0.9}{
	\textit{maxSeason}(X_1,Y_1) \geq \frac{\lambda_2 \cdot |\mathcal{D}_{\text{SEQ}}|}{\textit{minDensity}} \cdot e^{W\left( \frac{\log{\lambda_{1}^{1-\mu}} \cdot ln2}{\lambda_2} \right)}
}
	\label{eq:lowerbound}
\end{align}
where: $\lambda_1 = \min \lbrace p(X_i), \forall X_i \in X_S\rbrace$ is the minimum probability of $X_i \in X_S$, and $\lambda_2=p(Y_1)$ is the probability of $Y_1 \in Y_S$, and $W$ is the Lambert function \cite{corless1996lambertw}.
\label{theorem:lowerbound}
\end{theorem}
\vspace{-0.1in}
\begin{proof} (Sketch - Full proof in \cite{ho2022seasonal}). From Eq. \eqref{eq:NMI}, we have:\vspace{-0.05in}
	\begin{align}
		\scalemath{0.9}{\widetilde{I}(X_S;Y_S)= 1 - \frac{H(X_S \vert Y_S)}{H(X_S)} \ge \mu} 
	\end{align}
	\vspace{-0.2in}
	\begin{align}		
		\scalemath{0.9}{\Rightarrow \frac{H(X_S \vert Y_S)}{H(X_S)}} &\scalemath{0.9}{= \frac{p(X_1, Y_1) \cdot \log p(X_1 \vert Y_1) }   {\sum_{i} p(X_i) \cdot \log p(X_i)} \nonumber} \\ &\hspace{-0.4in}\scalemath{0.9}{+ \frac{\sum_{i \neq 1 \& j \neq 1} p(X_i, Y_j) \cdot \log \frac{p(X_i , Y_j)} {p(Y_j)}}{\sum_{i} p(X_i) \cdot \log p(X_i)}
		\le  1-\mu}
		\label{eq:sketchsupp1}
	\end{align}
	Let: $\scalemath{0.9}{\lambda_1 = \min \lbrace p(X_i), \forall i \rbrace}$,
	 $\lambda_2=p(Y_1)$. 
	\vspace{-0.05in}
	\begin{align}
		\scalemath{0.9}{\frac{H(X_S \vert Y_S)}{H(X_S)} \geq  \frac{p(X_1, Y_1) \cdot \log \frac{p(X_1 , Y_1)} {\lambda_2} }{\log \lambda_1} }	
		\label{eq:sketchsupp2}
	\end{align}	
From Eqs. \eqref{eq:sketchsupp1}, \eqref{eq:sketchsupp2}, we derive: $\scalemath{0.9}{
	p(X_1,Y_1) \geq  \lambda_2 \cdot e^{W\left(\frac{\log \lambda_{1}^{1-\mu} \cdot \ln 2}{\lambda_2}  \right)} 
}$
	\vspace{0.05in}
	Since: \vspace{-0.2in}
	\begin{align}
		\scalemath{0.9}{
		\frac{\left| SUP^{(X_1,Y_1)} \right|} { |\mathcal{D}_{\text{SEQ}}| } \geq p(X_1,Y_1) \geq \lambda_2 \cdot e^{W\left(\frac{\log \lambda_{1}^{1-\mu} \cdot \ln 2}{\lambda_2}  \right)} \nonumber }
	\end{align}
	Thus:
	\vspace{-0.325in} 
		\begin{align}
		\hspace{0.2in}
		\scalemath{0.9}{
		\textit{maxSeason}(X_1,Y_1) \geq \frac{\lambda_2 \cdot |\mathcal{D}_{\text{SEQ}}|}{\textit{minDensity}} \cdot e^{W\left(\frac{\log \lambda_{1}^{1-\mu} \cdot \ln 2}{\lambda_2}  \right)} 
	}
		\end{align}
\end{proof}

\textit{\bf Setting the parameters:} To compute the lower bound of \textit{maxSeason}$(X_1, Y_1)$ in Eq. \eqref{eq:lowerbound}, several parameters need to be defined: $\lambda_1$, $\lambda_2$, and $\mu$. Given $\mathcal{D}_{\text{SYB}}$, $\lambda_1$ and $\lambda_2$ can easily be determined since $\lambda_1$ is the minimum probability among all events $X_i \in X_S$, and $\lambda_2$ is the probability of $Y_1 \in Y_S$.  
To set the value of $\mu$, we use the lower bound of \textit{maxSeason} in Theorem \ref{theorem:lowerbound} to derive $\mu$ as follows.

\begin{corollary} 
	\label{corollary:lowerbound} 
	The maximum seasonal occurrence of an event pair $(X_1,Y_1) \in (X_S,Y_S)$ in $\mathcal{D}_{\text{SEQ}}$ is at least \textit{minSeason} if \hspace{0.01in} $\widetilde{I}(X_S;Y_S)$ is at least $\mu$, where: \vspace{-0.07in}
	\begin{footnotesize}
		\begin{align}
			\scalemath{0.9}{
			\mu \geq \begin{cases}
				1- \frac{\lambda_2}{e \cdot \ln 2 \cdot \log \frac{1}{\lambda_1} },&\hspace{-0.1in} \text{if} \hspace{0.05in}0 \leq \rho \leq \frac{1}{e}\\
				1 - \frac{\rho \cdot \lambda_2 \cdot \log \rho }{\ln 2 \cdot \log \lambda_1},&\hspace{-0.1in}\text{otherwise}
			\end{cases}
		}
			, \scalemath{0.9}{ \text{where }  \rho = \frac{\textit{minSeason} \cdot \textit{minDensity}}{\lambda_2 \cdot |\mathcal{D}_{\text{SEQ}}|} }
			\label{eq:muSupportSetting}
		\end{align}
	\end{footnotesize}
\end{corollary}

Note that $\mu$ in Eq. \eqref{eq:muSupportSetting} only ensures that the \textit{maxSeason} of the pair $(X_1,Y_1)$ is at least \textit{minSeason}. Thus, given $(X_S,Y_S)$, $\mu$ has to be computed for each event pair in $(X_S,Y_S)$. The final chosen $\mu$ value to be compared against $\widetilde{I}(X_S;Y_S)$ is the minimum $\mu$ value among all the event pairs in $(X_S,Y_S)$. 

\textbf{Interpretation of the lower bound of the maximum seasonal occurrence:} 
Theorem \ref{theorem:lowerbound} says that, given an MI threshold $\mu$, if the two symbolic series $X_S$ and $Y_S$ are correlated, i.e., $\widetilde{I}(X_S;Y_S) \ge \mu$, then the maximum seasonal occurrence of an event pair in ($X_S$,$Y_S$) is at least the lower bound in Eq. \eqref{eq:lowerbound}. 
Combining Theorem \ref{theorem:lowerbound} and Lemma \ref{lemeventpattern}, we can conclude that given a pair of symbolic series ($X_S$,$Y_S$), if its event pair ($X_1$, $Y_1$) has a maximum seasonal occurrence less than the lower bound in Eq. \eqref{eq:lowerbound}, then any 2-event pattern $P$ formed by that event pair also has a maximum seasonal occurrence less than that lower bound. This allows us to construct the approximate STPM algorithm, discussed in the next section. 
\vspace{-0.05in}
\subsection{Using the Bound to Approximate STPM}  \label{sec:approximateHSTPM}

\textit{Approximate STPM:} 
We construct an approximate version of STPM using Theorem \ref{theorem:lowerbound}.
Specifically, using the STPM thresholds \textit{minSeason} and \textit{minDensity}, we derive $\mu$ (Eq. \ref{eq:muSupportSetting}) and use it to identify \textit{correlated symbolic series} (defined in Def. 5.4). Next, the approximate STPM performs the mining only on \textit{the set of correlated symbolic series} $\mathcal{X}_C \subseteq \mathcal{X}$. Algorithm \ref{algorithmMIHSTPM} outlines the approximate STPM.

\SetNlSty{}{}{:} 
\begin{algorithm}[!t] 
	\algsetup{linenosize=\tiny}
	\SetInd{0.5em}{0.5em}
	\small
	\DontPrintSemicolon
	\caption{Approximate STPM using MI}
	\label{algorithmMIHSTPM}
	\KwInput{A set of time series $\mathcal{X}$, the thresholds: $maxPeriod$, $minDensity$, $distInterval$, $minSeason$} 
	\KwOutput{The set of frequent seasonal temporal patterns $\mathcal{P}$}
	
	\nonl // Find the correlated symbolic series \;
	\ForEach{\textit{pair of symbolic time series $(X_S,Y_S)$} $\in \mathcal{D}_{\text{SYB}}$}{
		$minNMI \leftarrow \min \lbrace \widetilde{I}(X_S; Y_S), \widetilde{I}(Y_S; X_S) \rbrace$;\; 
		Compute $\mu$ using Eq. (\ref{eq:muSupportSetting});\;
		\If{$minNMI \geq \mu$}{
			Insert $X_S$ and $Y_S$ into $\mathcal{X}_C$;\;
		} 	
	}
	Mine frequent seasonal single events from $\mathcal{X}_C$;\;
	\ForEach{ $(X_S,Y_S) \in \mathcal{X}_C$}{
		Mine frequent seasonal 2-event patterns from $(X_S,Y_S)$;\;
	}
	\If{$k \ge 3$}{
		Perform STPM using $HLH_1$ and $HLH_{k-1}$;
	}
\end{algorithm}

First, NMI and $\mu$ are computed for each pair of symbolic series $(X_S,Y_S)$ in $\mathcal{D_{\text{SYB}}}$ (lines 2-3). Then, only pairs whose $\min \lbrace \widetilde{I}(X_S; Y_S), \widetilde{I}(Y_S; X_S) \rbrace$ is at least $\mu$ are inserted into $\mathcal{X}_C$.
Next, only the correlated symbolic series in $\mathcal{X}_C$ are used to mine frequent seasonal single events (line 6). For frequent seasonal 2-event patterns, we mine frequent seasonal patterns only from event pairs in $\mathcal{X}_C$ (lines 7-8). For frequent seasonal k-event patterns ($k \ge 3$), the exact STPM is used (lines 9-10).

\textbf{Complexity analysis of approximate STPM:} 
The approximate STPM differs from STPM in two mining steps, the seasonal single events at $HLH_1$ and the seasonal 2-event patterns at $HLH_2$ by mining those only from correlated time series. To compute NMI and $\mu$, the approximate STPM only need to scan $\mathcal{D}_{\text{SYB}}$ once to calculate the probability for each single event and event pairs. Thus, the cost of NMI and $\mu$ computations is $O(\mid$$\mathcal{D}_{\text{SYB}}$$\mid)$. In contrast, the complexities of the exact STPM at $HLH_1$ and $HLH_2$ are $O(n \cdot | \mathcal{D}_{\text{SEQ}} |)$ $+$ $O(n^2 i^2 u^2)$ (Sections \ref{sec:1patternmining} and \ref{sec:kpatternmining}). Thus, the more time series are pruned, the faster and less memory usage of the approximate STPM. However, overall, the approximate STPM still scales exponentially with quadratic exponent in the pattern length $h$ as in STPM.

\section{Experimental Evaluation}\label{sec:experiment}
Due to space limitations, we only present here the most important results, and discuss other findings in \cite{ho2022seasonal}. 
\vspace{-0.05in}\subsection{Experimental Setup} \label{sec:ExperimentalSetup} \vspace{-0.02in}
\hspace{-0.15in}\textbf{Datasets:}
We use three real-world datasets from three application domains: renewable energy, smart city, and health. For \textit{renewable energy} (RE), we use energy data \cite{renewableEnergy} and weather data \cite{openweather} from Spain. 
For \textit{smart city} (SC), we use traffic and weather datasets \cite{smartcity} from New York City. For \textit{health}, we combine the \textit{influenza} (INF) and \textit{hand-foot-mouth} (HFM) datasets \cite{diseasedata} and weather data \cite{openweather} from Kawasaki, Japan. Besides real-world datasets, we also generate synthetic data for the scalability evaluation. Specifically, starting from each real-world dataset, we generate $1,000$ \textit{times more sequences} and $10,000$ \textit{synthetic time series} for each of them. Table \ref{tbl:datasetCharacteristic} summarizes the dataset characteristics. 
 
\hspace{-0.15in}\textbf{Baseline method:}  
Our exact method is referred to as E-STPM, and the approximate one as A-STPM. 
Since our work is the first that studies frequent seasonal temporal pattern mining, there does not exist an exact baseline to compare against STPM. However, we adapt the state-of-the-art method for recurring itemset mining PS-growth \cite{kiran2019finding} to find seasonal temporal patterns. Specifically, the adaptation is done through 2-phase process: (1) PS-growth is applied to find frequent recurring events, and (2), mine temporal patterns from extracted events. The adapted PS-growth is referred to as APS-growth. 
  
\hspace{-0.15in}\textbf{Infrastructure:}
We use a virtual machine with 32 AMD EPYC cores (2GHz), 512 GB RAM, and 1 TB storage. 
  
\hspace{-0.15in}\textbf{Parameters:} Table \ref{tbl:params} lists the parameters and their values used in our experiments, where \textit{maxPeriod} and \textit{minDensity} are expressed as the percentage of $\mathcal{D}_{\text{SEQ}}$. While the four parameters in Table \ref{tbl:params} are user-defined, we also provide the intuition of how to set them. \textit{maxPeriod} determines how close the patterns should occur within the same season. The smaller the \textit{maxPeriod}, the closer the occurred patterns should be and vice versa. \textit{minDensity} decides how dense a season should be. Combining these two, a small \textit{maxPeriod} and a large \textit{minDensity} will find dense seasons with close-by pattern occurrences. In contrast, a large \textit{maxPeriod} and a small \textit{minDensity} will find sparse seasons. On the other hand, \textit{minSeason} and \textit{distInterval} values often depend on the granularity of $\mathcal{D_{\text{SEQ}}}$. For example, if $\mathcal{D_{\text{SEQ}}}$ has month granularity, we then can look for patterns with yearly seasonality. Thus, \textit{distInterval} is often between $3$ and $9$ months, and \textit{minSeason} is the minimum number of years the patterns should have occurred seasonally.

\begin{table}[!t]
	\begin{minipage}{\columnwidth}
		\caption{Characteristics of the Datasets}
		\vspace{-0.1in}
		\centering
		\resizebox{0.9\columnwidth}{1.1cm}{
				\begin{tabular}{ |c|c|c|c|c|} 
				\hline {\bfseries Datasets} & {\bfseries \#seq.} & {\bfseries \#time series} & {\bfseries \#events} & {\bfseries \#ins./seq.} \\ 
				\hline
				RE (real)  &  1,460 &  21 &  102 &   93\\
				\hline 
				\centering SC (real) &  1,249 &  14 &  56 &  55\\
				\hline 
				\centering INF (real) &  608 &  25 &  124 &  48\\
				\hline 
				\centering HFM (real) &  730 &  24 &  115 &  40\\
				\hline
				\centering RE (syn.) &  1,460 $\times 10^3$ & $10^4$ &  48,500 &  38,012\\
				\hline
				\centering SC (syn.) &  1,249 $\times 10^3$ & $10^4$ &  40,020 &  37,106\\
				\hline
				\centering INF (syn.) &  608 $\times 10^3$ & $10^4$ &  49,600 &  40,623\\
				\hline
				\centering HFM (syn.) &  730 $\times 10^3$ & $10^4$ &  47,825 &  41,241\\
				\hline
			\end{tabular} 
		}
		\label{tbl:datasetCharacteristic}
	\end{minipage}
\end{table}

\subsection{Qualitative Evaluation}
Table \ref{tbl:interestingPatterns} lists some seasonal patterns found in the datasets.
Patterns P1-P3 are extracted from RE, showing that high renewable energy generation and high electricity demand occur seasonally and often at specific \textit{season} throughout the year. 
Patterns P4-P7 are extracted from INF and HFM, showing the detection of seasonal diseases. 
Finally, how weather affects traffic is shown in patterns P8-P11 extracted from SC.

Tables \ref{tbl:numPatternRE} and \ref{tbl:numPatternINF} list the number of seasonal patterns found in the RE and INF datasets. 
It can be seen that high \textit{minSeason} leads to less generated patterns, as many have few seasonal occurrences. Moreover, high $\textit{minDensity}$ also generates fewer patterns since only few patterns have high occurrence density. Finally, high $\textit{maxPeriod}$ results in more generated patterns, since high $\textit{maxPeriod}$ allows more temporal relations to be formed, thus increasing the number of patterns. 
	
\begin{table*}[!t]	
	\centering
	\begin{minipage}{.25\linewidth}
		\begin{minipage}{\linewidth}
			\captionsetup{justification=centering, font=small}
			\caption{\small Parameters and values}
			\vspace{-0.1in}
			\small
			\resizebox{\columnwidth}{1.03cm}{
				\begin{tabular}{ |l|c| }
					\hline &\\[0.01em]  \thead{Params} & {\bfseries Values (User-defined)}  \\ &\\[0.01em]
					\hline 
					$\textit{maxPeriod}$ & 0.2\%, 0.4\%, 0.6\%, 0.8\%, 1.0\% \\
					\hline
					$\textit{minDensity}$ & 0.5\%, 0.75\%, 1.0\%, 1.25\%, 1.5\% \\
					\hline
					$\textit{minSeason}$ & 4, 8, 12, 16, 20 \\
					\hline
					$\textit{distInterval}$ &  
					[90, 270] (RE, SC), [30, 90] (INF, HFM) \\
					\hline
				\end{tabular}
			}			
			\label{tbl:params}
		\end{minipage}
		\\
		\\
		\\
		\begin{minipage}{\linewidth}
			\captionsetup{justification=centering, font=small}
			\caption{\small \mbox{A-STPM} Accuracy}
			\vspace{-0.1in}
			\resizebox{1\columnwidth}{1.03cm}{
				\small
				\begin{tabular}{|c|c|c|c|c|c|c|}
					\hline 
					\multirow{3}{*}{\bfseries \# minSeason} & \multicolumn{6}{c|}{\bfseries minDensity (\%)} \\  
					\cline{2-7}  
					& \multicolumn{3}{c|}{\bfseries RE (real)} & \multicolumn{3}{c|}{\bfseries INF (real)}
					\\  \cline{2-7}  
					& {\bfseries 0.5} & {\bfseries 0.75} & {\bfseries 1}  & {\bfseries 0.5} & {\bfseries 0.75} & {\bfseries 1}  \\
					\hline
					8 & 81  & 82   & 86 & 81 & 83  & 87  \\  \hline			
					12 & 84  & 86   & 92  & 88  & 90  & 93  \\  \hline				
					16 & 94  & 95   & 100  & 95   & 96  & 100   \\  \hline					
					20 & 97  & 100   & 100  & 100 & 100  & 100   \\  \hline								
				\end{tabular}
			}			
			\label{tbl:accuracyRealdata}
		\end{minipage}
	\end{minipage}
	\begin{minipage}{.74\linewidth}
		\captionsetup{justification=centering, font=small}
		\caption{\small Summary of Interesting Seasonal Patterns}
		\vspace{-0.1in}
		\centering
		\resizebox{\textwidth}{2.3cm}{
			\begin{tabular}{ |l|c|c|c|c|} 
				\hline &&&&\\[-1em]  {\bfseries \;\;\;\;\;\;\;\;\;\;\;\;\;\;\;\;\;\;\;\;\;\;\;\;\;\;\;\;\;\;\;\;\;\;\;\;\;\;\;\;\;\;\;\;\;\;\;\;\;\;\;\;\;\;\;\;\;\;\;\;\;\;\;\;\;\;\;\;\;\;\;\;\;\; \small Patterns} & {\bfseries \small minDensity (\%)} & {\bfseries \small  maxPeriod (\%)} & {\bfseries \small   \# minSeason} & {\bfseries \small  Seasonal occurrence} \\
				\hline &&&&\\[-1em]
				\normalsize (P1) Strong Wind $\succcurlyeq$ High Wind Power Generation  & \normalsize 0.5 & \normalsize 0.4 & \normalsize 12 & \normalsize December, January, February \\
				\hline  &&&&\\[-1em]
				\normalsize (P2) Low Temperature $\succcurlyeq$ High Energy Consumption & \normalsize 0.5 & \normalsize 0.4 & \normalsize 12 & \normalsize December, January, February \\			
				\hline   &&&&\\[-1em]
				\normalsize (P3) Very Few Clouds $\succcurlyeq$ Very High Temperature $\between$ High Solar Power Generation & \normalsize 0.75 & \normalsize  0.6 & \normalsize  8 & \normalsize  July, August \\
				\specialrule{1.5pt}{1pt}{1pt}
				\normalsize  (P4) High Humidity $\between$ Very Low Temperature $\rightarrow$ Very High Influenza Cases  & \normalsize  0.5 & \normalsize  0.4 & \normalsize  12 & \normalsize  January, February \\
				\hline   &&&&\\[-1em]
				\normalsize  (P5) Strong Wind $\succcurlyeq$ Heavy Rain $\succcurlyeq$ High Influenza Cases & \normalsize  0.5 & \normalsize  0.4 & \normalsize 12 & \normalsize  January, February \\
				\specialrule{1.5pt}{1pt}{1pt} 
				\normalsize  (P6) Low Humidity $\succcurlyeq$ High Temperature $\succcurlyeq$ Very High Hand-Foot-Mouth Disease Cases & \normalsize 1.0 & \normalsize 0.6 & \normalsize 12 & \normalsize  May, June \\
				\hline   &&&&\\[-1em]
				\normalsize  (P7) Very High Temperature $\succcurlyeq$ High Wind $\succcurlyeq$  High Hand-Foot-Mouth Disease Cases  & \normalsize 1.0 & \normalsize 0.6 & \normalsize 12 & \normalsize  May, June \\
				\specialrule{1.5pt}{1pt}{1pt}  
				\normalsize  (P8) High Temperature $\succcurlyeq$ Strong Wind $\rightarrow$ High Congestion  & \normalsize 0.5 & \normalsize 0.6 & \normalsize 8 & \normalsize  July, August \\
				\hline  &&&&\\[-1em]
				\normalsize  (P9) Strong Wind $\succcurlyeq$ Unclear Visibility  $\succcurlyeq$ High Congestion  & \normalsize 0.5 & \normalsize 0.6 & \normalsize 8 & \normalsize  July, August \\
				\hline  &&&&\\[-1em]
				\normalsize  (P10) Heavy Rain $\succcurlyeq$ Unclear Visibility  $\succcurlyeq$ High Lane-Blocked  & \normalsize 0.4 & \normalsize 0.8 & \normalsize 8 & \normalsize  July, August \\
				\hline  &&&&\\[-1em]
				\normalsize  (P11) Heavy Rain $\succcurlyeq$ Strong Wind $\succcurlyeq$ High Flow-Incident & \normalsize 0.4 & \normalsize 0.8 & \normalsize 8 & \normalsize  July, August \\
				\hline     
			\end{tabular} 
		}
		\label{tbl:interestingPatterns}
	\end{minipage}
\end{table*}

\vspace{-0.05in}
\subsection{Quantitative Evaluation}

\subsubsection{Baseline comparison on real-world datasets}\label{sec:baselinecomparison}

\begin{table*}[!t]
	\centering
	\begin{minipage}{0.495\linewidth}
		\captionsetup{justification=centering, font=small}
		\caption{\small The Number of Seasonal Patterns on RE}
		\vspace{-0.1in}
		\resizebox{\columnwidth}{0.9cm}{
			\begin{tabular}{|c|c|c|c|c|c|c|c|c|c|} 
				\hline 
				\multirow{2}{*}{\bfseries maxPeriod (\%)} & \multicolumn{9}{c|}{\bfseries minSeason (\#) - minDensity (\%)} 
				\\  \cline{2-10}  
				& {\bfseries 8-0.5} & {\bfseries 8-0.75} & {\bfseries 8-1.0} & {\bfseries 12-0.5} & {\bfseries 12-0.75} & {\bfseries 12-1.0} & {\bfseries 16-0.5} & {\bfseries 16-0.75} & {\bfseries 16-1.0}
				\\ \hline 
				0.2  &  35626 &  20427 &  11339 &  21309 &  12941 &  6935 &  8045 &  4218 &  3018\\
				\hline 
				0.4  &  41462 &  29729 &  14281 &  25207 &  17381 &  7294 &  10261 &  7480 &  5483\\
				\hline 
				0.6  &  48651 &  35018 &  16247 &  31860 &  24627 &  9826 &  14061 &  9738 &  7409\\
				\hline 
			\end{tabular} 
		}
		\label{tbl:numPatternRE}
	\end{minipage}
	\begin{minipage}{0.495\linewidth}
		\captionsetup{justification=centering, font=small}
		\caption{The Number of Seasonal Patterns on INF}
		\vspace{-0.1in}
		\resizebox{\columnwidth}{0.9cm}{
			\begin{tabular}{|c|c|c|c|c|c|c|c|c|c|} 
				\hline 
				\multirow{2}{*}{\bfseries maxPeriod (\%)} & \multicolumn{9}{c|}{\bfseries minSeason (\#) - minDensity (\%)} 
				\\  \cline{2-10}  
				& {\bfseries 8-0.5} & {\bfseries 8-0.75} & {\bfseries 8-1.0} & {\bfseries 12-0.5} & {\bfseries 12-0.75} & {\bfseries 12-1.0} & {\bfseries 16-0.5} & {\bfseries 16-0.75} & {\bfseries 16-1.0}
				\\ \hline 
				0.2  &  7812 &  5704 &  4285 &  5159 &  3163 &  2157 &  3521 &  2105 &  1284\\
				\hline 
				0.4  &  10581 &  8294 &  6535 &  7952 &  5863 &  4068 &  5293 &  4618 &  2690\\
				\hline 
				0.6  &  12084 &  9618 &  8260 &  11850 &  8591 &  6028 &  6809 &  5073 &  3529\\
				\hline 
			\end{tabular} 
		}
		\label{tbl:numPatternINF}
	\end{minipage}
\end{table*}
\begin{table*}[!t]
	\centering
	\begin{minipage}{.56\linewidth}
		\captionsetup{justification=centering, font=small}
		\caption{Pruned Time Series and Events from A-STPM}
		\vspace{-0.1in}
		\resizebox{\columnwidth}{1.2cm}{
			\begin{tabular}{|c|c|c|c|c|c|c|c|c|c|c|c|c|}
				\hline 
				\multirow{3}{*}{\# Attr.} & \multicolumn{6}{c|}{\bfseries RE} & \multicolumn{6}{c|}{\bfseries INF}
				\\  \cline{2-13}  
				& \multicolumn{3}{c|}{\bfseries Pruned Time Series (\%)} & \multicolumn{3}{c|}{\bfseries Pruned Events (\%)} & \multicolumn{3}{c|}{\bfseries Pruned Time Series (\%)}  & \multicolumn{3}{c|}{\bfseries Pruned Events (\%)}    
				\\  \cline{2-13} 
				& {\bfseries 12-0.5\%} & {\bfseries 16-0.75\%} & {\bfseries 20-1.0\%}  & {\bfseries 12-0.5\%} & {\bfseries 16-0.75\%} & {\bfseries 20-1.0\%} & {\bfseries 12-0.5\%} & {\bfseries 16-0.75\%} & {\bfseries 20-1.0\%}  & {\bfseries 12-0.5\%} & {\bfseries 16-0.75\%} & {\bfseries 20-1.0\%} \\
				\hline
				2000 & 35.20 & 32.10 & 26.80 & 27.22 & 23.53 & 19.03 & 42.60 &	36.75 & 29.70 & 28.63 & 26.12 & 22.10 \\  \hline		
				
				4000 & 33.05 & 29.15 & 22.05 & 25.24 & 22.41 & 17.95 & 35.70 &	31.03 & 24.80 & 27.35 & 25.77 & 22.01 \\  \hline	
				
				6000 & 30.25 & 26.32 & 19.55 & 24.75 & 21.60 & 17.28 & 33.22 &	28.78 & 22.13 & 26.98 & 25.29 & 20.81   \\  \hline		
				
				8000 & 29.48 & 25.38 & 19.15 & 24.70 & 21.12 & 16.96 & 31.75 &	28.51 & 21.58 & 26.74 & 24.52 & 20.74  \\  \hline	
				
				10000 & 28.59 & 24.87 & 18.91 & 24.50 & 21.07 & 16.69 & 31.06 &	26.48 & 21.15 & 26.61 & 24.36 & 20.27  \\  \hline					
			\end{tabular}
		}
		\label{tbl:prunedAttributesEventsPercent}
	\end{minipage}
	\begin{minipage}{.42\linewidth}
		\captionsetup{justification=centering, font=small}
		\caption{The Accuracy of A-STPM on Syn. Data}
		\vspace{-0.1in}
		\resizebox{\columnwidth}{1.2cm}{
		\begin{tabular}{|c|c|c|c|c|c|c|}
			\hline 
			\multirow{3}{*}{\# Attr.} & \multicolumn{3}{c|}{\bfseries RE} & \multicolumn{3}{c|}{\bfseries INF}
			\\  \cline{2-7}  
			& \multicolumn{3}{c|}{\bfseries Accuracy (\%)} & \multicolumn{3}{c|}{\bfseries Accuracy (\%)}  
			\\  \cline{2-7}
			& {\bfseries 12-0.5\%} & {\bfseries 16-0.75\%} & {\bfseries 20-1.0\%}  & {\bfseries 12-0.5\%} & {\bfseries 16-0.75\%} & {\bfseries 20-1.0\%}  \\
			\hline
			2000 & 85 & 96 & 100 & 89 & 96 &	100 \\  \hline		
			
			4000 & 86 &	96 & 100 & 90 & 98 & 100 \\  \hline	
			
			6000 & 86 &	96 & 100 & 91 & 98 & 100  \\  \hline		
			
			8000 & 88 &	97 & 100 & 93 & 98 & 100  \\  \hline	
			
			10000 & 89 & 98 & 100 & 93 & 98 & 100  \\  \hline								
		\end{tabular}
	}
		\label{tbl:accuracyScalePercent}
	\end{minipage}
\end{table*}
\begin{figure*}[!t]
	\vspace{-0.1in}
	\hspace{2.72in}
	\ref{legendcomparison}
	\clearpage
	\vspace{-0.15in}
	\begin{minipage}[t]{1\columnwidth} 
		\centering
		\begin{subfigure}{0.32\columnwidth}
			\centering
			\resizebox{\linewidth}{!}{
				\begin{tikzpicture}[scale=0.2]
					\begin{axis}[
						compat=newest,
						xlabel={\# minSeason},
						ylabel={Runtime (sec)}, 
						label style={font=\Huge},
						ticklabel style = {font=\Huge},
						xmin=4, xmax=20,
						ymin=0, ymax=10000,
						xtick={4,8,12,16,20},
						ytick ={0,2000,4000,6000,8000,10000},
						legend columns=-1,
						legend entries = {A-STPM, E-STPM, APS-growth},
						legend image post style={scale=0.6}, 
						legend style={nodes={scale=0.5,  transform shape}, font=\Large},
						legend to name={legendcomparison},,
						ymajorgrids=true,
						grid style=dashed,
						line width=2pt
						]
						\addplot[
						color=blue,
						mark=x,
						mark size=4pt,
						]
						coordinates {
							(4,1053)(8,1002)(12,894)(16,735)(20,599)
						};						
						\addplot[
						color=teal,
						mark=diamond,
						mark size=4pt,
						]
						coordinates {
							(4,3021)(8,2572)(12,2265)(16,1870)(20,1427)
						};
						\addplot[
						color=red,
						mark=triangle,
						mark size=4pt,
						]
						coordinates {
							(4,8273)(8,7183)(12,6070)(16,4496)(20,3530)
						};						
					\end{axis}
				\end{tikzpicture}
			}
		\captionsetup{justification=centering, font=scriptsize}
			\caption{\scriptsize Varying minSeason}
		\end{subfigure}
		\begin{subfigure}{0.32\columnwidth}
			\centering
			\resizebox{\linewidth}{!}{
				\begin{tikzpicture}[scale=0.2]
					\begin{axis}[
						compat=newest,
						xlabel={minDensity (\%)},
						ylabel={Runtime (sec)}, 
						label style={font=\Huge},
						ticklabel style = {font=\Huge},
						xticklabel style = {xshift=3mm}, 
						xmin=0.5, xmax=1.5,
						ymin=0, ymax=10000,
						xtick={0.5,0.75,1,1.25,1.5},
						ytick ={0,2000,4000,6000,8000,10000},
						legend columns=-1,
						legend entries = {A-STPM, E-STPM, APS-growth},
						legend image post style={scale=0.6}, 
						legend style={nodes={scale=0.5,  transform shape}, font=\Large},
						legend to name={legendcomparison},
						ymajorgrids=true,
						grid style=dashed,
						line width=2pt
						]
						\addplot[
						color=blue,
						mark=x,
						mark size=4pt,
						]
						coordinates {
							(0.5,1002)(0.75,683)(1,372)(1.25,306)(1.5,251)
						};						
						\addplot[
						color=teal,
						mark=diamond,
						mark size=4pt,
						]
						coordinates {
							(0.5,2572)(0.75,1826)(1,1153)(1.25,943)(1.5,726)
						};						
						\addplot[
						color=red,
						mark=triangle,
						mark size=4pt,
						]
						coordinates {
							(0.5,7183)(0.75,5259)(1,3824)(1.25,2742)(1.5,2139)
						};
					\end{axis}
				\end{tikzpicture}
			}
		\captionsetup{justification=centering, font=scriptsize}
			\caption{\scriptsize Varying minDensity}
		\end{subfigure}
		\begin{subfigure}{0.32\columnwidth}
			\centering
			\resizebox{\linewidth}{!}{
				\begin{tikzpicture}[scale=0.2]
					\begin{axis}[
						compat=newest,
						xlabel={maxPeriod (\%)},
						ylabel={Runtime (sec)}, 
						label style={font=\Huge},
						ticklabel style = {font=\Huge},
						xticklabel style = {xshift=3mm}, 
						xmin=0.2, xmax=1,
						ymin=0, ymax=25000,
						xtick={0.2,0.4,0.6,0.8,1},
						ytick={0,5000,10000,15000,20000,25000},
						legend columns=-1,
						legend entries = {A-STPM, E-STPM, APS-growth},
						legend image post style={scale=0.6}, 
						legend style={nodes={scale=0.5,   transform shape}, font=\Large},
						legend to name={legendcomparison},
						ymajorgrids=true,
						grid style=dashed,
						line width=2pt
						]
						\addplot[
						color=blue,
						mark=x,
						mark size=4pt,
						]
						coordinates {
							(0.2,831)(0.4,1002)(0.6,1362)(0.8,1583)(1,2019)
						};						
						\addplot[
						color=teal,
						mark=diamond,
						mark size=4pt,
						]
						coordinates {
							(0.2,2035)(0.4,2572)(0.6,3077)(0.8,3515)(1,4984)
						};
						\addplot[
						color=red,
						mark=triangle,
						mark size=4pt,
						]
						coordinates {
							(0.2,6381)(0.4,7183)(0.6,10382)(0.8,13819)(1,21392)
						};
					\end{axis}
				\end{tikzpicture}
			}
		\captionsetup{justification=centering, font=scriptsize}
			\caption{\scriptsize Varying maxPeriod}
		\end{subfigure}
		\vspace{-0.08in}
		\captionsetup{justification=centering, font=small}
		\caption{Runtime Comparison on RE (real-world)}
		\label{fig:runtimebaselineRE}
	\end{minipage}
	\hspace{0.2in}  
	\begin{minipage}[t]{1\columnwidth} 
		\centering
		\begin{subfigure}{0.32\columnwidth}
			\centering
			\resizebox{\linewidth}{!}{
				\begin{tikzpicture}[scale=0.2]
					\begin{axis}[
						compat=newest,
						xlabel={\# minSeason},
						ylabel={Runtime (sec)}, 
						label style={font=\Huge},
						ticklabel style = {font=\Huge},
						xmin=4, xmax=20,
						ymin=0, ymax=5000,
						xtick={4,8,12,16,20},
						ytick={0,1000,2000,3000,4000,5000},
						scaled y ticks=base 10:-4,
						legend columns=-1,
						legend entries = {A-STPM, E-STPM, APS-growth},
						legend image post style={scale=0.6}, 
						legend style={nodes={scale=0.5,  transform shape}, font=\Large},
						legend to name={legendcomparison},
						ymajorgrids=true,
						grid style=dashed,
						line width=2pt
						]
						\addplot[
						color=blue,
						mark=x,
						mark size=4pt,
						]
						coordinates {
							(4,664)(8,621)(12,454)(16,345)(20,236)
						};
						\addplot[
						color=teal,
						mark=diamond,
						mark size=4pt,
						]
						coordinates {
							(4,1181)(8,939)(12,765)(16,500)(20,326)
						};
						\addplot[
						color=red,
						mark=triangle,
						mark size=4pt,
						]
						coordinates {
							(4,4910)(8,3930)(12,2483)(16,1706)(20,983)
						};						
					\end{axis}
				\end{tikzpicture}
			}
			\captionsetup{justification=centering, font=scriptsize}
			\caption{\scriptsize Varying minSeason}
		\end{subfigure}
		\begin{subfigure}{0.32\columnwidth}
			\centering
			\resizebox{\linewidth}{!}{
				\begin{tikzpicture}[scale=0.2]
					\begin{axis}[
						compat=newest,
						xlabel={minDensity (\%)},
						ylabel={Runtime (sec)}, 
						label style={font=\Huge},
						ticklabel style = {font=\Huge},
						xmin=0.5, xmax=1.5,
						ymin=0, ymax=5000,
						xticklabel style = {xshift=3mm}, 
						xtick={0.5,0.75,1,1.25,1.5},
						ytick={0,1000,2000,3000,4000,5000},
						scaled y ticks=base 10:-4,
						legend columns=-1,
						legend entries = {A-STPM, E-STPM, APS-growth},
						legend image post style={scale=0.6}, 
						legend style={nodes={scale=0.5,  transform shape}, font=\Large},
						legend to name={legendcomparison},
						ymajorgrids=true,
						grid style=dashed,
						line width=2pt
						]
						\addplot[
						color=blue,
						mark=x,
						mark size=4pt,
						]
						coordinates {
							(0.5,621)(0.75,462)(1,267)(1.25,127)(1.5,101)
						};
						\addplot[
						color=teal,
						mark=diamond,
						mark size=4pt,
						]
						coordinates {
							(0.5,939)(0.75,765)(1,464)(1.25,202)(1.5,150)
						};
						\addplot[
						color=red,
						mark=triangle,
						mark size=4pt,
						]
						coordinates {
							(0.5,3930)(0.75,3091)(1,2612)(1.25,938)(1.5,642)
						};						
					\end{axis}
				\end{tikzpicture}
			}
			\captionsetup{justification=centering, font=scriptsize}
			\caption{\scriptsize Varying minDensity}
		\end{subfigure}
		\begin{subfigure}{0.32\columnwidth}
			\centering
			\resizebox{\linewidth}{!}{
				\begin{tikzpicture}[scale=0.2]
					\begin{axis}[
						compat=newest,
						xlabel={maxPeriod (\%)},
						ylabel={Runtime (sec)}, 
						label style={font=\Huge},
						ticklabel style = {font=\Huge},
						xmin=0.2, xmax=1,
						ymin=0, ymax=6000,
						xticklabel style = {xshift=3mm}, 
						xtick={0.2,0.4,0.6,0.8,1},
						ytick={0,1000,2000,3000,4000,5000,6000},
						scaled y ticks=base 10:-4,
						legend columns=-1,
						legend entries = {A-STPM, E-STPM, APS-growth},
						legend image post style={scale=0.6},  
						legend style={nodes={scale=0.5,   transform shape}, font=\Large},
						legend to name={legendcomparison},
						ymajorgrids=true,
						grid style=dashed,
						line width=2pt
						]
						\addplot[
						color=blue,
						mark=x,
						mark size=4pt,
						]
						coordinates {
							(0.2,483)(0.4,621)(0.6,783)(0.8,982)(1,1003)
						};						
						\addplot[
						color=teal,
						mark=diamond,
						mark size=4pt,
						]
						coordinates {
							(0.2,706)(0.4,939)(0.6,1095)(0.8,1264)(1,1469)
						};
						\addplot[
						color=red,
						mark=triangle,
						mark size=4pt,
						]
						coordinates {
							(0.2,3001)(0.4,3930)(0.6,4421)(0.8,5005)(1,5631)
						};						
					\end{axis}
				\end{tikzpicture}
			}
			\captionsetup{justification=centering, font=scriptsize}
			\caption{\scriptsize Varying maxPeriod}
		\end{subfigure}
		\vspace{-0.08in}
		\captionsetup{justification=centering, font=small}
		\caption{Runtime Comparison on INF (real-world)}
		\label{fig:runtimebaselineINF}
	\end{minipage}
	\vspace{-0.02in}  
\end{figure*}  

\begin{figure*}[!t]
	\vspace{-0.1in}
	\begin{minipage}[t]{1\columnwidth} 
		\centering
		\begin{subfigure}{0.32\columnwidth}
			\centering
			\resizebox{\linewidth}{!}{
				\begin{tikzpicture}[scale=0.2]
					\begin{axis}[
						compat=newest,
						xlabel={\# minSeason},
						ylabel={Memory Usage (MB)}, 
						label style={font=\Huge},
						ticklabel style = {font=\Huge},
						xmin=4, xmax=20,
						ymin=0, ymax=15000,
						xtick={4,8,12,16,20},
						ytick ={0, 5000,10000,15000},
						legend columns=-1,
						legend entries = {A-STPM, E-STPM, APS-growth},
						legend style={nodes={scale=0.5,  transform shape}, font=\Large},
						legend to name={legendcomparison1},
						ymajorgrids=true,
						grid style=dashed,
						line width=2pt
						]
						\addplot[
						color=blue,
						mark=x,
						mark size=4pt,
						]
						coordinates {
							(4,5305)(8,4501)(12,3895)(16,3449)(20,2258)
						};
						\addplot[
						color=teal,
						mark=diamond,
						mark size=4pt,
						]
						coordinates {
							(4,9251)(8,8091)(12,7206)(16,6858)(20,4970)
						};
						\addplot[
						color=red,
						mark=triangle,
						mark size=4pt,
						]
						coordinates {
							(4,14201)(8,12982)(12,11916)(16,11281)(20,9430)
						};						
					\end{axis}
				\end{tikzpicture}
			}
			\captionsetup{justification=centering, font=scriptsize}
			\caption{\scriptsize Varying minSeason}
		\end{subfigure}
		\begin{subfigure}{0.32\columnwidth}
			\centering
			\resizebox{\linewidth}{!}{
				\begin{tikzpicture}[scale=0.2]
					\begin{axis}[
						compat=newest,
						xlabel={minDensity (\%)},
						ylabel={Memory Usage (MB)}, 
						label style={font=\Huge},
						ticklabel style = {font=\Huge},
						xmin=0.5, xmax=1.5,
						ymin=0, ymax=15000,
						xtick={0.5,0.75,1,1.25,1.5},
						ytick ={0, 5000,10000,15000},
						xticklabel style = {xshift=3mm}, 
						legend columns=-1,
						legend entries = {A-STPM, E-STPM, APS-growth},
						legend style={nodes={scale=0.5,  transform shape}, font=\Large},
						legend to name={legendcomparison1},
						ymajorgrids=true,
						grid style=dashed,
						line width=2pt
						]
						\addplot[
						color=blue,
						mark=x,
						mark size=4pt,
						]
						coordinates {
							(0.5,4501)(0.75,3675)(1,1862)(1.25,1684)(1.5,1432)
						};
						\addplot[
						color=teal,
						mark=diamond,
						mark size=4pt,
						]
						coordinates {
							(0.5,8091)(0.75,6500)(1,4106)(1.25,3274)(1.5,2842)
						};
						\addplot[
						color=red,
						mark=triangle,
						mark size=4pt,
						]
						coordinates {
							(0.5,12982)(0.75,10595)(1,9261)(1.25,7693)(1.5,7075)
						};						
					\end{axis}
				\end{tikzpicture}
			}
			\captionsetup{justification=centering, font=scriptsize}
			\caption{\scriptsize Varying minDensity}
		\end{subfigure}
		\begin{subfigure}{0.32\columnwidth}
			\centering
			\resizebox{\linewidth}{!}{
				\begin{tikzpicture}[scale=0.2]
					\begin{axis}[
						compat=newest,
						xlabel={maxPeriod (\%)},
						ylabel={Memory Usage (MB)}, 
						label style={font=\Huge},
						ticklabel style = {font=\Huge},
						xmin=0.2, xmax=1,
						ymin=0, ymax=20000,
						xtick={0.2,0.4,0.6,0.8,1},
						ytick ={0, 5000,10000,15000,20000},
						xticklabel style = {xshift=3mm}, 
						legend columns=-1,
						legend entries = {A-STPM, E-STPM, APS-growth},
						legend style={nodes={scale=0.5,   transform shape}, font=\Large},
						legend to name={legendcomparison1},
						ymajorgrids=true,
						grid style=dashed,
						line width=2pt
						]
						\addplot[
						color=blue,
						mark=x,
						mark size=4pt,
						]
						coordinates {
							(0.2,3761)(0.4,4501)(0.6,4704)(0.8,4983)(1,5903)
						};
						\addplot[
						color=teal,
						mark=diamond,
						mark size=4pt,
						]
						coordinates {
							(0.2,7001)(0.4,8091)(0.6,8437)(0.8,9223)(1,10531)
						};
						\addplot[
						color=red,
						mark=triangle,
						mark size=4pt,
						]
						coordinates {
							(0.2,12074)(0.4,12982)(0.6,15031)(0.8,16392)(1,19834)
						};						
					\end{axis}
				\end{tikzpicture}
			}
			\captionsetup{justification=centering, font=scriptsize}
			\caption{\scriptsize Varying maxPeriod}
		\end{subfigure}
		\vspace{-0.08in}
		\captionsetup{justification=centering, font=small}
		\caption{Memory Usage Comparison on RE (real-world)}
		\label{fig:memorybaselineRE}
	\end{minipage}    
	\hspace{0.2in}  
	\begin{minipage}[t]{1\columnwidth} 
		\centering
		\begin{subfigure}{0.32\columnwidth}
			\centering
			\resizebox{\linewidth}{!}{
				\begin{tikzpicture}[scale=0.2]
					\begin{axis}[
						compat=newest,
						xlabel={\# minSeason},
						ylabel={Memory Usage (MB)}, 
						label style={font=\Huge},
						ticklabel style = {font=\Huge},
						xmin=4, xmax=20,
						ymin=0, ymax=10000,
						xtick={4,8,12,16,20},
						ytick ={0,3000,6000,9000},
						legend columns=-1,
						legend entries = {A-STPM, E-STPM, APS-growth},
						legend style={nodes={scale=0.5,  transform shape}, font=\Large},
						legend to name={legendcomparison1},
						ymajorgrids=true,
						grid style=dashed,
						line width=2pt
						]
						\addplot[
						color=blue,
						mark=x,
						mark size=4pt,
						]
						coordinates {
							(4,2732)(8,2401)(12,2071)(16,1743)(20,869)
						};
						\addplot[
						color=teal,
						mark=diamond,
						mark size=4pt,
						]
						coordinates {
							(4,4083)(8,3671)(12,3334)(16,2820)(20,1847)
						};
						\addplot[
						color=red,
						mark=triangle,
						mark size=4pt,
						]
						coordinates {
							(4,9083)(8,8335)(12,7469)(16,6240)(20,4505)
						};						
					\end{axis}
				\end{tikzpicture}
			}
			\captionsetup{justification=centering, font=scriptsize}
			\caption{\scriptsize Varying minSeason}
		\end{subfigure}
		\begin{subfigure}{0.32\columnwidth}
			\centering
			\resizebox{\linewidth}{!}{
				\begin{tikzpicture}[scale=0.2]
					\begin{axis}[
						compat=newest,
						xlabel={minDensity (\%)},
						ylabel={Memory Usage (MB)}, 
						label style={font=\Huge},
						ticklabel style = {font=\Huge},
						xmin=0.5, xmax=1.5,
						ymin=0, ymax=10000,
						xtick={0.5,0.75,1,1.25,1.5},
						ytick ={0,3000,6000,9000},
						xticklabel style = {xshift=3mm}, 
						legend columns=-1,
						legend entries = {A-STPM, E-STPM, APS-growth},
						legend style={nodes={scale=0.5,  transform shape}, font=\Large},
						legend to name={legendcomparison1},
						ymajorgrids=true,
						grid style=dashed,
						line width=2pt
						]
						\addplot[
						color=blue,
						mark=x,
						mark size=4pt,
						]
						coordinates {
							(0.5,2401)(0.75,2107)(1,1802)(1.25,703)(1.5,506)
						};
						\addplot[
						color=teal,
						mark=diamond,
						mark size=4pt,
						]
						coordinates {
							(0.5,3671)(0.75,3334)(1,2605)(1.25,1432)(1.5,942)
						};
						\addplot[
						color=red,
						mark=triangle,
						mark size=4pt,
						]
						coordinates {
							(0.5,8335)(0.75,7928)(1,7502)(1.25,5400)(1.5,3736)
						};						
					\end{axis}
				\end{tikzpicture}
			}
			\captionsetup{justification=centering, font=scriptsize}
			\caption{\scriptsize Varying minDensity}
		\end{subfigure}
		\begin{subfigure}{0.32\columnwidth}
			\centering
			\resizebox{\linewidth}{!}{
				\begin{tikzpicture}[scale=0.2]
					\begin{axis}[
						compat=newest,
						xlabel={maxPeriod (\%)},
						ylabel={Memory Usage (MB)}, 
						label style={font=\Huge},
						ticklabel style = {font=\Huge},
						xmin=0.2, xmax=1,
						ymin=0, ymax=15000,
						xtick={0.2,0.4,0.6,0.8,1},
						ytick ={0,3000,6000,9000,12000},
						xticklabel style = {xshift=3mm}, 
						legend columns=-1,
						legend entries = {A-STPM, E-STPM, APS-growth},
						legend style={nodes={scale=0.5,   transform shape}, font=\Large},
						legend to name={legendcomparison1},
						ymajorgrids=true,
						grid style=dashed,
						line width=2pt
						]
						\addplot[
						color=blue,
						mark=x,
						mark size=4pt,
						]
						coordinates {
							(0.2,1994)(0.4,2401)(0.6,2793)(0.8,2983)(1,3106)
						};
						\addplot[
						color=teal,
						mark=diamond,
						mark size=4pt,
						]
						coordinates {
							(0.2,3001)(0.4,3671)(0.6,3903)(0.8,4406)(1,4901)
						};
						\addplot[
						color=red,
						mark=triangle,
						mark size=4pt,
						]
						coordinates {
							(0.2,7882)(0.4,8635)(0.6,9863)(0.8,10982)(1,11936)
						};						
					\end{axis}
				\end{tikzpicture}
			}
			\captionsetup{justification=centering, font=scriptsize}
			\caption{\scriptsize Varying maxPeriod}
		\end{subfigure}
		\vspace{-0.08in}
		\captionsetup{justification=centering, font=small}
		\caption{Memory Usage Comparison on INF (real-world)}
		\label{fig:memorybaselineINF}
	\end{minipage}      
	\vspace{-0.04in}  
\end{figure*}  
\begin{figure*}[!t]
	\begin{minipage}{1\columnwidth}
		\centering
		\begin{subfigure}[t]{0.32\columnwidth}
			\centering
			\captionsetup{justification=centering}
			\resizebox{\linewidth}{!}{
				\begin{tikzpicture}[scale=0.2]
					\begin{axis}[
						xlabel={\# Sequences (\%)},
						ylabel={Runtime (sec)}, 
						label style={font=\Huge},
						ticklabel style = {font=\Huge},
						xticklabel style = {xshift=3mm}, 
						xmin=20, xmax=100,
						ymin=0, ymax=200000,
						xtick={20,40,60,80,100},
						legend columns=-1,
						legend entries = {A-STPM, E-STPM, APS-growth},
						legend image post style={scale=0.6},  
						legend style={nodes={scale=0.5, transform shape}, font=\Large},
						legend to name={legendcomparison2},
						legend pos= north west,			
						ymajorgrids=true,
						grid style=dashed,
						line width=2pt
						]
						\addplot[
						color=blue,
						mark=x,
						mark size=4pt,
						]
						coordinates {
							(20,7251.64)(40,16329.05)(60,32384.32)(80,59023.68)(100,98320.17)
						};
						\addplot[
						color=teal,
						mark=diamond,
						mark size=4pt,
						]
						coordinates {
							(20,15549.82)(40,31296.73)(60,67261.38)(80,110036.84)(100,150631.58)
						};
						\addplot[
						color=red,
						mark=triangle,
						mark size=4pt,
						]
						coordinates {
							(20,43038.36)(40,91657.08)(60,NaN)(80,NaN)(100,NaN)
						};
					\end{axis}
				\end{tikzpicture}
			}
			\captionsetup{justification=centering, font=scriptsize}
			\caption{\scriptsize minSeason=12,\\\;\;\;\;\;\;\;minDensity=0.5\%}
			\label{fig:scaleSequence_RE_8}
		\end{subfigure}
		\begin{subfigure}[t]{0.32\columnwidth}
			\centering
			\captionsetup{justification=centering}
			\resizebox{\linewidth}{!}{
				\begin{tikzpicture}[scale=0.2]
					\begin{axis}[
						xlabel={\# Sequences (\%)},
						ylabel={Runtime (sec)}, 
						label style={font=\Huge},
						ticklabel style = {font=\Huge},
						xticklabel style = {xshift=3mm}, 
						xmin=20, xmax=100,
						ymin=0, ymax=200000,
						xtick={20,40,60,80,100},
						legend columns=-1,
						legend entries = {A-STPM, E-STPM, APS-growth},
						legend image post style={scale=0.6}, 
						legend style={nodes={scale=0.5, transform shape}, font=\Large},
						legend to name={legendcomparison2},
						legend pos= north west,			
						ymajorgrids=true,
						grid style=dashed,
						line width=2pt
						]
						\addplot[
						color=blue,
						mark=x,
						mark size=4pt,
						]
						coordinates {
							(20,5104.18)(40,10142.92)(60,18339.13)(80,31910.93)(100,45163.82)
						};
						\addplot[
						color=teal,
						mark=diamond,
						mark size=4pt,
						]
						coordinates {
							(20,9728.72)(40,24068.29)(60,45082.73)(80,75013.42)(100,98291.94)
						};
						\addplot[
						color=red,
						mark=triangle,
						mark size=4pt,
						]
						coordinates {
							(20,28258.39)(40,55931.03)(60,93913.92)(80,140120.29)(100,180132.93)
						};
						
					\end{axis}
				\end{tikzpicture}
			}
			\captionsetup{justification=centering, font=scriptsize}
			\caption{\scriptsize minSeason=16,\\\;\;\;\;\;\;\;minDensity=0.75\%}
		\end{subfigure}
		\begin{subfigure}[t]{0.32\columnwidth}
			\centering
			\captionsetup{justification=centering}			
			\resizebox{\linewidth}{!}{
				\begin{tikzpicture}[scale=0.2]
					\begin{axis}[
						xlabel={\# Sequences (\%)},
						ylabel={Runtime (sec)}, 
						label style={font=\Huge},
						ticklabel style = {font=\Huge},
						xticklabel style = {xshift=3mm}, 
						xmin=20, xmax=100,
						ymin=0, ymax=150000,
						xtick={20,40,60,80,100},
						legend columns=-1,
						legend entries = {A-STPM, E-STPM, APS-growth},
						legend image post style={scale=0.6}, 
						legend style={nodes={scale=0.5, transform shape}, font=\Large},
						legend to name={legendcomparison2},
						legend pos= north west,			
						ymajorgrids=true,
						grid style=dashed,
						line width=2pt
						]
						\addplot[
						color=blue,
						mark=x,
						mark size=4pt,
						]
						coordinates {
							(20,3251.43)(40,7095.94)(60,13941.71)(80,19302.85)(100,29843.95)					
						};
						\addplot[
						color=teal,
						mark=diamond,
						mark size=4pt,
						]
						coordinates {
							(20,7001.84)(40,14132.93)(60,31382.84)(80,48094.82)(100,73294.19)
						};
						\addplot[
						color=red,
						mark=triangle,
						mark size=4pt,
						]
						coordinates {
							(20,18304.92)(40,35901.42)(60,58104.82)(80,97019.39)(100,117019.39)
						};
					\end{axis}
				\end{tikzpicture}
			}
			\captionsetup{justification=centering, font=scriptsize}
			\caption{\scriptsize minSeason=20,\\\;\;\;\;\;\;\;minDensity=1.0\%}
		\end{subfigure}
		\vspace{-0.08in}
		\captionsetup{justification=centering, font=small}
		\caption{Scalability: Varying \#Sequences on RE (synthetic)}
		\label{fig:scaleSequence_RE}
	\end{minipage}
	\hspace{0.2in}
	\begin{minipage}{1\columnwidth}
		\centering
		\begin{subfigure}[t]{0.32\columnwidth}
			\centering
			\captionsetup{justification=centering}
			\resizebox{\linewidth}{!}{
				\begin{tikzpicture}[scale=0.2]
					\begin{axis}[
						xlabel={\# Sequences (\%)},
						ylabel={Runtime (sec)}, 
						label style={font=\Huge},
						ticklabel style = {font=\Huge},
						xticklabel style = {xshift=3mm},
						xmin=20, xmax=100,
						ymin=0, ymax=200000,
						xtick={20,40,60,80,100},
						legend columns=-1,
						legend entries = {A-STPM, E-STPM, APS-growth},
						legend image post style={scale=0.6}, 
						legend style={nodes={scale=0.5, transform shape}, font=\Large},
						legend to name={legendcomparison2},
						legend pos= north west,			
						ymajorgrids=true,
						grid style=dashed,
						line width=2pt
						]
						\addplot[
						color=blue,
						mark=x,
						mark size=4pt,
						]
						coordinates {
							(20,5148.29)(40,8261.37)(60,22375.49)(80,48378.46)(100,62748.34)
						};
						\addplot[
						color=teal,
						mark=diamond,
						mark size=4pt,
						]
						coordinates {
							(20,6861.27)(40,14025.98)(60,52077.02)(80,70207.38)(100,102358.47)
						};
						\addplot[
						color=red,
						mark=triangle,
						mark size=4pt,
						]
						coordinates {
							(20,12856.41)(40,40367.18)(60,120317.54)(80,180342.08)(100,NaN)
						};
					\end{axis}
				\end{tikzpicture}
			}
			\captionsetup{justification=centering, font=scriptsize}
			\caption{\scriptsize minSeason=12,\\\;\;\;\;\;\;\;minDensity=0.5\%}
			\label{fig:scaleSequence_Influenza_8}
		\end{subfigure}
		\begin{subfigure}[t]{0.32\columnwidth}
			\centering
			\captionsetup{justification=centering}
			\resizebox{\linewidth}{!}{
				\begin{tikzpicture}[scale=0.2]
					\begin{axis}[
						xlabel={\# Sequences (\%)},
						ylabel={Runtime (sec)}, 
						label style={font=\Huge},
						ticklabel style = {font=\Huge},
						xticklabel style = {xshift=3mm}, 
						xmin=20, xmax=100,
						ymin=0, ymax=150000,
						xtick={20,40,60,80,100},
						legend columns=-1,
						legend entries = {A-STPM, E-STPM, APS-growth},
						legend image post style={scale=0.6}, 
						legend style={nodes={scale=0.5, transform shape}, font=\Large},
						legend to name={legendcomparison2},
						legend pos= north west,			
						ymajorgrids=true,
						grid style=dashed,
						line width=2pt
						]
						\addplot[
						color=blue,
						mark=x,
						mark size=4pt,
						]
						coordinates {
							(20,3636.09)(40,6840.51)(60,15341.38)(80,30250.4)(100,42321.25)
						};
						\addplot[
						color=teal,
						mark=diamond,
						mark size=4pt,
						]
						coordinates {
							(20,6130.47)(40,11281.35)(60,40247.62)(80,54310.19)(100,75671.08)						
						};
						\addplot[
						color=red,
						mark=triangle,
						mark size=4pt,
						]
						coordinates {
							(20,9320.94)(40,33309.38)(60,84842.73)(80,107310.28)(100,126234.18)
						};
						
					\end{axis}
				\end{tikzpicture}
			}
			\captionsetup{justification=centering, font=scriptsize}
			\caption{\scriptsize minSeason=16,\\\;\;\;\;\;\;\;minDensity=0.75\%}
		\end{subfigure}
		\begin{subfigure}[t]{0.32\columnwidth}
			\centering
			\captionsetup{justification=centering}
			\resizebox{\linewidth}{!}{
				\begin{tikzpicture}[scale=0.2]
					\begin{axis}[
						xlabel={\# Sequences (\%)},
						ylabel={Runtime (sec)}, 
						label style={font=\Huge},
						ticklabel style = {font=\Huge},
						xticklabel style = {xshift=3mm},
						xmin=20, xmax=100,
						ymin=0, ymax=100000,
						xtick={20,40,60,80,100},
						legend columns=-1,
						legend entries = {A-STPM, E-STPM, APS-growth},
						legend image post style={scale=0.6}, 
						legend style={nodes={scale=0.5, transform shape}, font=\Large},
						legend to name={legendcomparison2},
						legend pos= north west,			
						ymajorgrids=true,
						grid style=dashed,
						line width=2pt
						]
						\addplot[
						color=blue,
						mark=x,
						mark size=4pt,
						]
						coordinates {
							(20,2403.64)(40,5031.82)(60,10281.39)(80,21623.17)(100,27520.26)
						};
						\addplot[
						color=teal,
						mark=diamond,
						mark size=4pt,
						]
						coordinates {
							(20,3730.24)(40,9010.26)(60,30103.51)(80,41207.38)(100,58173.04)
						};
						\addplot[
						color=red,
						mark=triangle,
						mark size=4pt,
						]
						coordinates {
							(20,8401.06)(40,23036.47)(60,55631.85)(80,82681.27)(100,90127.69)
						};
					\end{axis}
				\end{tikzpicture}
			}
			\captionsetup{justification=centering, font=scriptsize}
			\caption{\scriptsize minSeason=20,\\\;\;\;\;\;\;\;minDensity=1.0\%}
		\end{subfigure}
		\vspace{-0.08in}
		\captionsetup{justification=centering, font=small}
		\caption{Scalability: Varying \#Sequences on INF (synthetic)}
		\label{fig:scaleSequence_Influenza}
	\end{minipage}
	\vspace{-0.04in}
\end{figure*}
\begin{figure*}[!t]
	\begin{minipage}{1\columnwidth}
		\centering
		\captionsetup{justification=centering}
		\begin{subfigure}[t]{0.32\columnwidth}
			\centering
			\resizebox{\linewidth}{!}{
				\begin{tikzpicture}[scale=0.2]
					\begin{axis}[
						ybar stacked,    
						bar width=15pt,
						enlarge x limits=0.035,
						label style={font=\Huge},
						ticklabel style = {font=\Huge},
						ymin=0, ymax=200000,
						ytick=\empty,
						xmin=1, xmax=5,
						xtick={1,2,3,4,5},
						xticklabels = {,,,,},
						legend columns=-1,
						legend entries = {mining time, MI time},
						legend style={nodes={scale=0.5, transform shape}, font=\Large},
						legend to name={legendbarchart},
						legend pos= north west,			
						ymajorgrids=true,
						grid style=dashed,
						]
						\addplot+[ybar] plot coordinates 
						{(1,7827.21)(2,11979.39)(3,20754.69)(4,65847.43)(5,100212.2)};
						\addplot+[ybar, color=red] plot coordinates  {(1,4729.56)(2,5884.19)(3,7859.69)(4,7588.24)(5,9834.99)};
					\end{axis}
					\begin{axis}[
						xlabel={\# Time Series ($\times 10^3$)},
						ylabel={Runtime (sec)}, 
						label style={font=\Huge},
						ticklabel style = {font=\Huge},
						xmin=2, xmax=10,
						ymin=0, ymax=300000,
						xtick={2,4,6,8,10},
						ytick={0,50000,100000,150000,200000,250000,300000},
						legend columns=-1,
						legend entries = {A-STPM, E-STPM, APS-growth},
						legend style={nodes={scale=0.5, transform shape}, font=\Large},
						legend to name={legendcomparison3},
						legend pos= north west,			
						ymajorgrids=true,
						grid style=dashed,
						line width=2pt,
						scaled x ticks=false,
						]
						\addplot[
						color=blue,
						mark=x,
						mark size=4pt,
						]
						coordinates {
							(2,19035)(4,26712)(6,42783)(8,108543)(10,160291)
						};
						\addplot[
						color=teal,
						mark=diamond,
						mark size=4pt,
						]
						coordinates {
							(2,30720)(4,51058)(6,83920)(8,165235)(10,251482)
						};
						\addplot[
						color=red,
						mark=triangle,
						mark size=4pt,
						]
						coordinates {
							(2,51298)(4,120524)(6,NaN)(8,NaN)(10,NaN)
						};
					\end{axis}
				\end{tikzpicture}
			}
			\captionsetup{justification=centering, font=scriptsize}
			\caption{\scriptsize minSeason=12,\\\;\;\;\;\;\;\;minDensity=0.5\%}
			\label{fig:scaleTimeSeries_RE_8}
		\end{subfigure}
		\begin{subfigure}[t]{0.32\columnwidth}
			\centering
			\captionsetup{justification=centering}
			\resizebox{\linewidth}{!}{
				\begin{tikzpicture}[scale=0.2]
					\begin{axis}[
						ybar stacked,    
						bar width=15pt,
						enlarge x limits=0.035,
						label style={font=\Huge},
						ticklabel style = {font=\Huge},
						ymin=0, ymax=300000,
						ytick=\empty,
						xmin=1, xmax=5,
						xtick={1,2,3,4,5},
						xticklabels = {,,,,},
						legend columns=-1,
						legend entries = {mining time, MI time},
						legend style={nodes={scale=0.5, transform shape}, font=\Large},
						legend to name={legendbarchart},
						legend pos= north west,			
						ymajorgrids=true,
						grid style=dashed,
						]
						\addplot+[ybar] plot coordinates 
						{(1,6227.21)(2,11979.39)(3,23754.69)(4,48847.43)(5,55212.2)};
						\addplot+[ybar, color=red] plot coordinates  {(1,6029.56)(2,10984.19)(3,9259.69)(4,10588.24)(5,12834.99)};
					\end{axis}
					\begin{axis}[
						xlabel={\# Time Series ($\times 10^3$)},
						ylabel={Runtime (sec)}, 
						label style={font=\Huge},
						ticklabel style = {font=\Huge},
						xmin=2, xmax=10,
						ymin=0, ymax=300000,
						xtick={2,4,6,8,10},
						ytick={0,50000,100000,150000,200000,250000,300000},
						legend columns=-1,
						legend entries = {A-STPM, E-STPM, APS-growth},
						legend image post style={scale=0.6}, 
						legend style={nodes={scale=0.5, transform shape}, font=\large},
						legend to name={legendcomparison3},
						legend pos= north west,			
						ymajorgrids=true,
						grid style=dashed,
						line width=2pt,
						scaled x ticks=false,
						]
						\addplot[
						color=blue,
						mark=x,
						mark size=4pt,
						]
						coordinates {
							(2,12573)(4,16386)(6,32614)(8,60368)(10,70427)
						};
						\addplot[
						color=teal,
						mark=diamond,
						mark size=4pt,
						]
						coordinates {
							(2,24043)(4,40068)(6,79873)(8,100610)(10,120763)
						};
						\addplot[
						color=red,
						mark=triangle,
						mark size=4pt,
						]
						coordinates {
							(2,38741)(4,100372)(6,202873)(8,250968)(10,291860)
						};
						
					\end{axis}
				\end{tikzpicture}
			}
			\captionsetup{justification=centering, font=scriptsize}
			\caption{\scriptsize minSeason=16,\\\;\;\;\;\;\;\;minDensity=0.75\%}
			\label{fig:scaleTimeSeries_RE_12}
		\end{subfigure}
		\begin{subfigure}[t]{0.32\columnwidth}
			\centering
			\captionsetup{justification=centering}
			\resizebox{\linewidth}{!}{
				\begin{tikzpicture}[scale=0.2]
					\begin{axis}[
						ybar stacked,    
						bar width=15pt,
						enlarge x limits=0.035,
						label style={font=\Huge},
						ticklabel style = {font=\Huge},
						ymin=0, ymax=300000,
						ytick=\empty,
						xmin=1, xmax=5,
						xtick={1,2,3,4,5},
						xticklabels = {,,,,},
						legend columns=-1,
						legend entries = {mining time, MI time},
						legend style={nodes={scale=0.5, transform shape}, font=\Large},
						legend to name={legendbarchart},
						legend pos= north west,			
						ymajorgrids=true,
						grid style=dashed,
						]
						\addplot+[ybar] plot coordinates 
						{(1,6227.21)(2,8979.39)(3,13754.69)(4,16847.43)(5,30212.2)};
						\addplot+[ybar, color=red] plot coordinates  {(1,6029.56)(2,8984.19)(3,12259.69)(4,12788.24)(5,12834.99)};
					\end{axis}
					\begin{axis}[
						xlabel={\# Time Series ($\times 10^3$)},
						ylabel={Runtime (sec)}, 
						label style={font=\Huge},
						ticklabel style = {font=\Huge},
						xmin=2, xmax=10,
						ymin=0, ymax=300000,
						xtick={2,4,6,8,10},
						ytick={0,50000,100000,150000,200000,250000,300000},
						legend columns=-1,
						legend entries = {A-STPM, E-STPM, APS-growth},
						legend image post style={scale=0.6}, 
						legend style={nodes={scale=0.5, transform shape}, font=\Large},
						legend to name={legendcomparison3},
						legend pos= north west,			
						ymajorgrids=true,
						grid style=dashed,
						line width=2pt,
						scaled x ticks=false,
						]
						\addplot[
						color=blue,
						mark=x,
						mark size=4pt,
						]
						coordinates {
							(2,5463)(4,11464)(6,20729)(8,25630)(10,40638)
						};
						\addplot[
						color=teal,
						mark=diamond,
						mark size=4pt,
						]
						coordinates {
							(2,16523)(4,28117)(6,55238)(8,86729)(10,112011)
						};
						\addplot[
						color=red,
						mark=triangle,
						mark size=4pt,
						]
						coordinates {
							(2,28461)(4,64618)(6,144189)(8,228982)(10,245841)
						};
					\end{axis}
				\end{tikzpicture}
			}
			\captionsetup{justification=centering, font=scriptsize}
			\caption{\scriptsize minSeason=20,\\\;\;\;\;\;\;\;minDensity=1.0\%}
			\label{fig:scaleTimeSeries_RE_16}
		\end{subfigure}
		\vspace{-0.08in}
		\captionsetup{justification=centering, font=small}
		\caption{Scalability: Varying \#TimeSeries on RE (synthetic)}
		\label{fig:scaleTimeSeries_RE}
	\end{minipage}
	\hspace{0.2in}
	\begin{minipage}{1\columnwidth}
		\centering
		\begin{subfigure}[t]{0.32\columnwidth}
			\centering
			\captionsetup{justification=centering}
			\resizebox{\linewidth}{!}{
				\begin{tikzpicture}[scale=0.2]
					\begin{axis}[
						ybar stacked,    
						bar width=15pt,
						enlarge x limits=0.035,
						label style={font=\Huge},
						ticklabel style = {font=\Huge},
						ymin=0, ymax=200000,
						ytick=\empty,
						xmin=1, xmax=5,
						xtick={1,2,3,4,5},
						xticklabels = {,,,,},
						legend columns=-1,
						legend entries = {mining time, MI time},
						legend style={nodes={scale=0.5, transform shape}, font=\Large},
						legend to name={legendbarchart},
						legend pos= north west,			
						ymajorgrids=true,
						grid style=dashed,
						]
						\addplot+[ybar] plot coordinates 
						{(1,5827.21)(2,9579.39)(3,28754.69)(4,45847.43)(5,92212.2)};
						\addplot+[ybar, color=red] plot coordinates  {(1,5829.56)(2,5684.19)(3,7859.69)(4,9988.24)(5,10834.99)};
					\end{axis}
					\begin{axis}[
						xlabel={\# Time Series ($\times 10^3$)},
						ylabel={Runtime (sec)}, 
						label style={font=\Huge},
						ticklabel style = {font=\Huge},
						xmin=2, xmax=10,
						ymin=0, ymax=200000,
						xtick={2,4,6,8,10},
						legend columns=-1,
						legend entries = {A-STPM, E-STPM, APS-growth},
						legend image post style={scale=0.6}, 
						legend style={nodes={scale=0.5, transform shape}, font=\Large},
						legend to name={legendcomparison3},
						legend pos= north west,			
						ymajorgrids=true,
						grid style=dashed,
						line width=2pt,
						scaled x ticks=false
						]
						\addplot[
						color=blue,
						mark=x,
						mark size=4pt,
						]
						coordinates {
							(2,9183)(4,14642)(6,36641)(8,53267)(10,100648)
						};
						\addplot[
						color=teal,
						mark=diamond,
						mark size=4pt,
						]
						coordinates {
							(2,19570)(4,35811)(6,70379)(8,90421)(10,182417)
						};
						\addplot[
						color=red,
						mark=triangle,
						mark size=4pt,
						]
						coordinates {
							(2,36603)(4,82317)(6,152650)(8,NaN)(10,NaN)
						};
					\end{axis}
				\end{tikzpicture}
			}
			\captionsetup{justification=centering, font=scriptsize}
			\caption{\scriptsize minSeason=12,\\\;\;\;\;\;\;\;minDensity=0.5\%}
			\label{fig:scaleTimeSeries_Influenza_8}
		\end{subfigure}
		\begin{subfigure}[t]{0.32\columnwidth}
			\centering
			\captionsetup{justification=centering}
			\resizebox{\linewidth}{!}{
				\begin{tikzpicture}[scale=0.2]
					\begin{axis}[
						ybar stacked,    
						bar width=15pt,
						enlarge x limits=0.035,
						label style={font=\Huge},
						ticklabel style = {font=\Huge},
						ymin=0, ymax=200000,
						ytick=\empty,
						xmin=1, xmax=5,
						xtick={1,2,3,4,5},
						xticklabels = {,,,,},
						legend columns=-1,
						legend entries = {mining time, MI time},
						legend style={nodes={scale=0.5, transform shape}, font=\Large},
						legend to name={legendbarchart},
						legend pos= north west,			
						ymajorgrids=true,
						grid style=dashed,
						]
						\addplot+[ybar] plot coordinates 
						{(1,3827.21)(2,5579.39)(3,15754.69)(4,27847.43)(5,38212.2)};
						\addplot+[ybar, color=red] plot coordinates  {(1,2829.56)(2,4984.19)(3,6859.69)(4,8988.24)(5,10834.99)};
					\end{axis}
					\begin{axis}[
						xlabel={\# Time Series ($\times 10^3$)},
						ylabel={Runtime (sec)}, 
						label style={font=\Huge},
						ticklabel style = {font=\Huge},
						xmin=2, xmax=10,
						ymin=0, ymax=200000,
						xtick={2,4,6,8,10},
						legend columns=-1,
						legend entries = {A-STPM, E-STPM, APS-growth},
						legend image post style={scale=0.6}, 
						legend style={nodes={scale=0.5, transform shape}, font=\Large},
						legend to name={legendcomparison3},
						legend pos= north west,			
						ymajorgrids=true,
						grid style=dashed,
						line width=2pt,
						scaled x ticks=false
						]
						\addplot[
						color=blue,
						mark=x,
						mark size=4pt,
						]
						coordinates {
							(2,7437)(4,10602)(6,22765)(8,35281)(10,46023)
						};
						\addplot[
						color=teal,
						mark=diamond,
						mark size=4pt,
						]
						coordinates {
							(2,14047)(4,25215)(6,49219)(8,70379)(10,90028)
						};
						\addplot[
						color=red,
						mark=triangle,
						mark size=4pt,
						]
						coordinates {
							(2,24695)(4,61128)(6,124638)(8,166421)(10,194562)
						};
						
					\end{axis}
				\end{tikzpicture}
			}
		\captionsetup{justification=centering, font=scriptsize}
		\caption{\scriptsize minSeason=16,\\\;\;\;\;\;\;\;minDensity=0.75\%}
		\end{subfigure}
		\begin{subfigure}[t]{0.32\columnwidth}
			\centering
			\captionsetup{justification=centering}
			\resizebox{\linewidth}{!}{
				\begin{tikzpicture}[scale=0.2]
					\begin{axis}[
						ybar stacked,    
						bar width=15pt,
						enlarge x limits=0.035,
						label style={font=\Huge},
						ticklabel style = {font=\Huge},
						ymin=0, ymax=200000,
						ytick=\empty,
						xmin=1, xmax=5,
						xtick={1,2,3,4,5},
						xticklabels = {,,,,},
						legend columns=-1,
						legend entries = {mining time, MI time},
						legend style={nodes={scale=0.5, transform shape}, font=\Large},
						legend to name={legendbarchart},
						legend pos= north west,			
						ymajorgrids=true,
						grid style=dashed,
						]
						\addplot+[ybar] plot coordinates 
						{(1,3827.21)(2,5579.39)(3,10754.69)(4,24847.43)(5,30212.2)};
						\addplot+[ybar, color=red] plot coordinates  {(1,3829.56)(2,5984.19)(3,7859.69)(4,7988.24)(5,10834.99)};
					\end{axis}
					\begin{axis}[
						xlabel={\# Time Series ($\times 10^3$)},
						ylabel={Runtime (sec)}, 
						label style={font=\Huge},
						ticklabel style = {font=\Huge},
						xmin=2, xmax=10,
						ymin=0, ymax=200000,
						xtick={2,4,6,8,10},
						legend columns=-1,
						legend entries = {A-STPM, E-STPM, APS-growth},
						legend image post style={scale=0.6}, 
						legend style={nodes={scale=0.5, transform shape}, font=\Large},
						legend to name={legendcomparison3},
						legend pos= north west,			
						ymajorgrids=true,
						grid style=dashed,
						line width=2pt,
						scaled x ticks=false
						]
						\addplot[
						color=blue,
						mark=x,
						mark size=4pt,
						]
						coordinates {
							(2,4699)(4,6349)(6,14629)(8,32603)(10,38603)
						};
						\addplot[
						color=teal,
						mark=diamond,
						mark size=4pt,
						]
						coordinates {
							(2,10986)(4,18221)(6,32614)(8,59685)(10,65923)
						};
						\addplot[
						color=red,
						mark=triangle,
						mark size=4pt,
						]
						coordinates {
							(2,16421)(4,52728)(6,90642)(8,130283)(10,147823)
						};
					\end{axis}
				\end{tikzpicture}
			}
			\captionsetup{justification=centering, font=scriptsize}
			\caption{\scriptsize minSeason=20,\\\;\;\;\;\;\;\;minDensity=1.0\%}
		\end{subfigure}
		\vspace{-0.08in}
		\captionsetup{justification=centering, font=small}
		\caption{Scalability: Varying \#TimeSeries on INF (synthetic)}
		\label{fig:scaleTimeSeries_Influenza}
	\end{minipage}
	\vspace{-0.1in}
\end{figure*}
\begin{figure*}[!t]
	\vspace{-0.04in}
	\begin{minipage}[t]{1\columnwidth} 
		\centering
		\begin{subfigure}{0.32\columnwidth}
			\centering
			\resizebox{\linewidth}{!}{
				\begin{tikzpicture}[scale=0.2]
					\begin{axis}[
						compat=newest,
						xlabel={\# minSeason},
						ylabel={Runtime (sec)}, 
						label style={font=\Huge},
						ticklabel style = {font=\Huge},
						xmin=4, xmax=20,
						ymin=0, ymax=7500,
						xtick={4,8,12,16,20},
						ytick={0,1500,3000,4500,6000,7500},
						scaled y ticks=base 10:-3,
						legend columns=-1,
						legend entries = {NoPrune, Apriori, Trans, All},
						legend image post style={scale=0.6}, 
						legend style={nodes={scale=0.5,  transform shape}, font=\Large},
						legend to name={legendpruningExact},
						ymajorgrids=true,
						grid style=dashed,
						line width=2pt
						]
						\addplot[
						color=blue,
						mark=diamond,
						mark size=4pt,
						] 
						coordinates {
							(4,6731.46)(8,6120.18)(12,5430.13)(16,4007.19)(20,2942.61)
						};
						
						\addplot[
						color=black,
						mark=triangle,
						mark size=4pt,
						] 
						coordinates {
							(4,5192.37)(8,4831.08)(12,3861.3)(16,3217.48)(20,2346.14)
						};
						
						\addplot[
						color=teal,
						mark=square*,
						mark size=4pt,
						]	
						coordinates {
							(4,4162.18)(8,3829.47)(12,3258.39)(16,2783.46)(20,2164.28)
						};
						
						\addplot[
						color=red,
						mark=*,
						mark size=4pt,
						] 
						coordinates {
							(4,3021)(8,2572)(12,2265)(16,1870)(20,1627)
						};
					\end{axis}
				\end{tikzpicture}
			}
			\captionsetup{justification=centering, font=scriptsize}
			\caption{\scriptsize Varying minSeason}
		\end{subfigure}
		\begin{subfigure}{0.32\columnwidth}
			\centering
			\resizebox{\linewidth}{!}{
				\begin{tikzpicture}[scale=0.2]
					\begin{axis}[
						compat=newest,
						xlabel={minDensity (\%)},
						ylabel={Runtime (sec)}, 
						label style={font=\Huge},
						ticklabel style = {font=\Huge},
						xticklabel style = {xshift=3mm},
						xmin=0.5, xmax=1.5,
						ymin=0, ymax=5000,
						xtick={0.5,0.75,1,1.25,1.5},
						ytick={0,1000,2000,3000,4000,5000},
						scaled y ticks=base 10:-3,
						legend columns=-1,
						legend entries = {NoPrune, Apriori, Trans, All},
						legend image post style={scale=0.6},  
						legend style={nodes={scale=0.5,  transform shape}, font=\Large},
						legend to name={legendpruningExact},
						ymajorgrids=true,
						grid style=dashed,
						line width=2pt
						]
						\addplot[
						color=blue,
						mark=diamond,
						mark size=4pt,
						] 	
						coordinates {
							(0.5,4281.39)(0.75,3652.84)(1,2493.57)(1.25,2016.32)(1.5,1664.21)
						};
						
						\addplot[
						color=black,
						mark=triangle,
						mark size=4pt,
						]	
						coordinates {
							(0.5,3759.28)(0.75,2691.76)(1,1762.83)(1.25,1468.61)(1.5,1143.25)
						};
						
						\addplot[
						color=teal,
						mark=square*,
						mark size=4pt,
						] 
						coordinates {
							(0.5,3429.05)(0.75,2315.62)(1,1528.08)(1.25,1207.35)(1.5,962.38)
						};
						
						\addplot[
						color=red,
						mark=*,
						mark size=4pt,
						] 
						coordinates {
							(0.5,2572)(0.75,1826)(1,1153)(1.25,943)(1.5,726)
						};
					\end{axis}
				\end{tikzpicture}
			}
			\captionsetup{justification=centering, font=scriptsize}
			\caption{\scriptsize Varying minDensity}
		\end{subfigure}
		\begin{subfigure}{0.32\columnwidth}
			\centering
			\resizebox{\linewidth}{!}{
				\begin{tikzpicture}[scale=0.2]
					\begin{axis}[
						compat=newest,
						xlabel={maxPeriod (\%)},
						ylabel={Runtime (sec)}, 
						label style={font=\Huge},
						ticklabel style = {font=\Huge},
						xmin=0.2, xmax=1,
						ymin=0, ymax=6000,
						xtick={0.2,0.4,0.6,0.8,1},
						ytick={0,1500,3000,4500,6000},
						xticklabel style = {xshift=3mm},
						scaled y ticks=base 10:-3,
						legend columns=-1,
						legend entries = {NoPrune, Apriori, Trans, All},
						legend image post style={scale=0.6}, 
						legend style={nodes={scale=0.5,   transform shape}, font=\Large},
						legend to name={legendpruningExact},
						ymajorgrids=true,
						grid style=dashed,
						line width=2pt
						]
						\addplot[
						color=blue,
						mark=diamond,
						mark size=4pt,
						] 	
						coordinates {
								(0.2,2683.46)(0.4,3761.28)(0.6,4160.83)(0.8,4830.57)(1,5489.37)
						};
						
						\addplot[
						color=black,
						mark=triangle,
						mark size=4pt,
						]	
						coordinates {
							(0.2,2058.17)(0.4,3491.26)(0.6,3741.07)(0.8,4201.85)(1,4561.35)
						};
						
						\addplot[
						color=teal,
						mark=square*,
						mark size=4pt,
						] 
						coordinates {
							(0.2,1536.25)(0.4,2954.08)(0.6,3361.83)(0.8,3694.27)(1,4001.62)
						};
						
						\addplot[
						color=red,
						mark=*,
						mark size=4pt,
						] 
						coordinates {
							(0.2,1235)(0.4,2472)(0.6,3077)(0.8,3315)(1,3684)
						};
					\end{axis}
				\end{tikzpicture}
			}
			\captionsetup{justification=centering, font=scriptsize}
			\caption{\scriptsize Varying maxPeriod}
		\end{subfigure}
		\ref{legendpruningExact}
		\vspace{-0.06in}
		\captionsetup{justification=centering, font=small}
		\caption{Pruning Techniques of E-STPM on RE (real-world)}
		\label{fig:performance}
	\end{minipage}
	\hspace{0.2in}  
	\begin{minipage}[t]{1\columnwidth} 
		\centering
		\begin{subfigure}{0.32\columnwidth}
			\centering
			\resizebox{\linewidth}{!}{
				\begin{tikzpicture}[scale=0.2]
					\begin{axis}[
						compat=newest,
						xlabel={\# minSeason},
						ylabel={Runtime (sec)}, 
						label style={font=\Huge},
						ticklabel style = {font=\Huge},
						xmin=4, xmax=20,
						ymin=0, ymax=3500,
						xtick={4,8,12,16,20},
						ytick={0,500,1000,1500,2000,2500,3000,3500},
						scaled y ticks=base 10:-3,
						legend columns=-1,
						legend entries = {NoPrune, Apriori, Trans, All},
						legend image post style={scale=0.6}, 
						legend style={nodes={scale=0.5,  transform shape}, font=\Large},
						legend to name={legendpruningExact},
						ymajorgrids=true,
						grid style=dashed,
						line width=2pt
						]
						\addplot[
						color=blue,
						mark=diamond,
						mark size=4pt,
						] 
						coordinates {
							(4,3036.46)(8,2561.32)(12,1830.08)(16,1236.18)(20,730.06)
						};
						
						\addplot[
						color=black,
						mark=triangle,
						mark size=4pt,
						] 
						coordinates {
							(4,2102.34)(8,1569.27)(12,1268.73)(16,863.13)(20,453.15)
						};
						
						\addplot[
						color=teal,
						mark=square*,
						mark size=4pt,
						]	
						coordinates {
							(4,1501.25)(8,1217.73)(12,1027.82)(16,525.43)(20,326.38)
						};
						
						\addplot[
						color=red,
						mark=*,
						mark size=4pt,
						] 
						coordinates {
							(4,1181)(8,939)(12,765)(16,400)(20,256)
						};
					\end{axis}
				\end{tikzpicture}
			}
			\captionsetup{justification=centering, font=scriptsize}
			\caption{\scriptsize Varying minSeason}
		\end{subfigure}
		\begin{subfigure}{0.32\columnwidth}
			\centering
			\resizebox{\linewidth}{!}{
				\begin{tikzpicture}[scale=0.2]
					\begin{axis}[
						compat=newest,
						xlabel={minDensity (\%)},
						ylabel={Runtime (sec)}, 
						label style={font=\Huge},
						ticklabel style = {font=\Huge},
						xmin=0.5, xmax=1.5,
						ymin=0, ymax=2500,
						xtick={0.5,0.75,1,1.25,1.5},
						ytick={0,500,1000,1500,2000,2500},
						xticklabel style = {xshift=3mm},
						scaled y ticks=base 10:-3,
						legend columns=-1,
						legend entries = {NoPrune, Apriori, Trans, All},
						legend image post style={scale=0.6}, 
						legend style={nodes={scale=0.5,  transform shape}, font=\Large},
						legend to name={legendpruningExact},
						ymajorgrids=true,
						grid style=dashed,
						line width=2pt
						]
						\addplot[
						color=blue,
						mark=diamond,
						mark size=4pt,
						] 	
						coordinates {
							(0.5,2158.18)(0.75,1530.13)(1,1003.51)(1.25,643.25)(1.5,463.94)
						};
						
						\addplot[
						color=black,
						mark=triangle,
						mark size=4pt,
						]	
						coordinates {
							(0.5,1543.28)(0.75,1223.61)(1,586.34)(1.25,301.59)(1.5,189.36)
						};
						
						\addplot[
						color=teal,
						mark=square*,
						mark size=4pt,
						] 
						coordinates {
							(0.5,1203.29)(0.75,951.39)(1,727.83)(1.25,392.17)(1.50,253.03)
						};
						
						\addplot[
						color=red,
						mark=*,
						mark size=4pt,
						] 
						coordinates {
							(0.5,939)(0.75,765.26)(1,464.03)(1.25,202.37)(1.5,150.38)
						};
					\end{axis}
				\end{tikzpicture}
			}
			\captionsetup{justification=centering, font=scriptsize}
			\caption{\scriptsize Varying minDensity}
		\end{subfigure}
		\begin{subfigure}{0.32\columnwidth}
			\centering
			\resizebox{\linewidth}{!}{
				\begin{tikzpicture}[scale=0.2]
					\begin{axis}[
						compat=newest,
						xlabel={maxPeriod (\%)},
						ylabel={Runtime (sec)}, 
						label style={font=\Huge},
						ticklabel style = {font=\Huge},
						xmin=0.2, xmax=1,
						ymin=0, ymax=2500,
						xtick={0.2,0.4,0.6,0.8,1},
						ytick={0,500,1000,1500,2000,2500,3000},
						xticklabel style = {xshift=3mm}, 
						scaled y ticks=base 10:-3,
						legend columns=-1,
						legend entries = {NoPrune, Apriori, Trans, All},
						legend image post style={scale=0.6}, 
						legend style={nodes={scale=0.5,   transform shape}, font=\Large},
						legend to name={legendpruningExact},
						ymajorgrids=true,
						grid style=dashed,
						line width=2pt
						]
						\addplot[
						color=blue,
						mark=diamond,
						mark size=4pt,
						] 	
						coordinates {
							(0.2,969.31)(0.4,1530.27)(0.6,1836.24)(0.8,2031.04)(1,2483.37)
						};
						
						\addplot[
						color=black,
						mark=triangle,
						mark size=4pt,
						]	
						coordinates {
							(0.2,836.58)(0.4,1203.64)(0.6,1406.95)(0.8,1718.03)(1,1937.49)
						};
						
						\addplot[
						color=teal,
						mark=square*,
						mark size=4pt,
						] 
						coordinates {
							(0.2,626.34)(0.4,1000)(0.6,1206.08)(0.8,1308.61)(1,1608.52)
						};
						
						\addplot[
						color=red,
						mark=*,
						mark size=4pt,
						] 
						coordinates {
							(0.2,406)(0.4,872)(0.6,1000)(0.8,1163)(1,1369)
						};
					\end{axis}
				\end{tikzpicture}
			}
			\captionsetup{justification=centering, font=scriptsize}
			\caption{\scriptsize Varying maxPeriod}
		\end{subfigure}
		\ref{legendpruningExact}
		\vspace{-0.06in}
		\captionsetup{justification=centering, font=small}
		\caption{Pruning Techniques of E-STPM on INF (real-world)}
		\label{fig:performance1}
	\end{minipage}      
	\vspace{-0.15in}  
\end{figure*}

We compare E-STPM and A-STPM with the baseline in terms of the runtime and memory usage. Figs. \ref{fig:runtimebaselineRE}, \ref{fig:runtimebaselineINF}, \ref{fig:memorybaselineRE} and \ref{fig:memorybaselineINF} show the comparison on RE and INF datasets. The results on other datasets are reported in the technical report \cite{ho2022seasonal}. Note that Figs. \ref{fig:runtimebaselineRE}-\ref{fig:scaleTimeSeries_Influenza} use the same legend.

As shown in Figs. \ref{fig:runtimebaselineRE} and \ref{fig:runtimebaselineINF}, A-STPM achieves the best runtime among all methods, and E-STPM has better runtime than the baseline. On the tested datasets, the range and average speedups of A-STPM compared to other methods are: $[1.5$-$4.7]$ and $2.6$ (E-STPM), and $[5.2$-$10.6]$ and $7.1$ (APS-growth). The speedup of E-STPM compared to the baseline is $[3.5$-$7.2]$ and $4.3$ on average. Note that the times to compute MI and $\mu$ for RE and INF in Figs. \ref{fig:runtimebaselineRE} and \ref{fig:runtimebaselineINF} are only $2.6$ and $1.4$ seconds, respectively. Moreover, A-STPM is most efficient, i.e., achieves highest speedup and memory saving, when the $\textit{minSeason}$ threshold is low, e.g., $\textit{minSeason}=4$. This is because there are typically many patterns with few seasonal occurrences. Thus, using A-STPM to prune uncorrelated time series early helps save computational time and resources. However, the speedup comes at the cost of a small loss in accuracy (discussed in Section \ref{sec:accuracy}).

In terms of memory consumption, as shown in Figs. \ref{fig:memorybaselineRE} and \ref{fig:memorybaselineINF}, A-STPM is the most efficient method, while E-STPM is more efficient than the baseline. The range and the average memory consumption of A-STPM compared to other methods are: $[1.4$-$2.7]$ and $1.8$ (E-STPM), and $[2.7$-$7.6]$ and $3.9$ (APS-growth). The memory usage of E-STPM compared to the baseline is $[1.5$-$4.1]$ and $2.3$ on average.
\subsubsection{Scalability evaluation on synthetic datasets}\vspace{-0.02in}
As discussed in Section \ref{sec:FTPMfTSMining}, the complexity of STPM is driven by two main factors: (1) the number of temporal sequences, and (2) the number of time series. 
Thus, to further evaluate STPM scalability, we scale these two factors on synthetic datasets (reported in Table \ref{tbl:datasetCharacteristic}), using two configurations: varying the number of sequences, and varying the number of time series. 

Figs. \ref{fig:scaleSequence_RE} and \ref{fig:scaleSequence_Influenza} show the runtimes of A-STPM, E-STPM and the baseline when the number of sequences changes. We obtain the range and average speedups of A-STPM are: [$1.6$-$3.2$] and $2.2$ (E-STPM), and [$3.1$-$6.4$] and $4.6$ (APS-growth). Similarly, the range and average speedup of E-STPM compared to APS-growth is [$1.9$-$4.3$] and $3.2$. We note that the baseline fails for larger configurations because of memory in this scalability study, i.e., on the synthetic RE at 60\% sequences ($\approx 8\times 10^5$) (Fig. \ref{fig:scaleSequence_RE_8}) and on the synthetic INF at 100\% sequences ($\approx 6\times 10^5$) (Fig. \ref{fig:scaleSequence_Influenza_8}), showing that A-STPM and E-STPM can scale well on big datasets while the baseline cannot.

Figs. \ref{fig:scaleTimeSeries_RE} and \ref{fig:scaleTimeSeries_Influenza} compare the runtimes of A-STPM, E-STPM and APS-growth when changing the number of time series. We obtain the range and average speedups of A-STPM are: [$1.7$-$3.5$] and $2.3$ (E-STPM), and [$3.8$-$9.5$] and $5.3$ (APS-growth), and of E-STPM is [$2.3$-$4.4$] and $3.6$ (APS-growth). The baseline also fails at large configurations in this study, i.e., when \# Time Series $\ge 6000$ on the synthetic RE (Fig. \ref{fig:scaleTimeSeries_RE_8}), and $\ge 8000$ on the synthetic INF (Fig. \ref{fig:scaleTimeSeries_Influenza_8}).

Furthermore, we provide the computation time of MI and $\mu$ in Figs. \ref{fig:scaleTimeSeries_RE} and \ref{fig:scaleTimeSeries_Influenza} by adding an additional bar chart for A-STPM. Each bar represents the runtime of A-STPM with two separate components: the time to compute MI and $\mu$ (top red), and the mining time (bottom blue). We note that for each dataset, we only need to compute MI once (the computed MIs are used across different $\textit{minSeason}$ and $\textit{minDensity}$ thresholds), while the computation of $\mu$ is negligible (in milliseconds using Eq. \eqref{eq:muSupportSetting}). Thus, the MI and $\mu$ computation times, for example, in Figs. \ref{fig:scaleTimeSeries_RE_8}, \ref{fig:scaleTimeSeries_RE_12}, and \ref{fig:scaleTimeSeries_RE_16}, are added only for comparison and are not all actually used. 

Finally, we provide the percentage of time series and events pruned by A-STPM in the scalability test in Table \ref{tbl:prunedAttributesEventsPercent}.  
Here, we can see that low $\textit{minSeason}$ and $\textit{minDensity}$ lead to more time series (events) to be pruned. This is because $\textit{minSeason}$ and $\textit{minDensity}$ have an inverse relationship with $\mu$, therefore, low $\textit{minSeason}$ and $\textit{minDensity}$ result in higher $\mu$, and thus, more pruned time series. 
\subsubsection{Evaluation of the pruning techniques in E-STPM}\vspace{-0.02in}
To understand how effective the proposed pruning techniques are, we compare different versions of E-STPM:  (1) NoPrune: E-STPM with no pruning, (2) Apriori: E-STPM with Apriori-liked pruning  (Lemmas \ref{lempattern1}, \ref{lemeventpattern}), (3) Trans: E-STPM with transitivity-based pruning (Lemmas \ref{lem:transitivity}, \ref{lem:filter}), and (4) All: E-STPM applied both pruning techniques. 

Figs. \ref{fig:performance}, \ref{fig:performance1} show the results. It can be seen that (All)-E-STPM achieves the best performance among all versions. Its speedup w.r.t. (NoPrune)-E-STPM ranges from $3$ up to $6$ depending on the configurations, showing that the proposed prunings are very effective in improving E-STPM performance. Furthermore, (Trans)-E-STPM delivers larger speedup than (Apriori)-E-STPM. The average speedup is from $2$ to $5$ for (Trans)-E-STPM, and from $1.5$ to $4$ for (Apriori)-E-STPM. However, applying both always yields better speedup than applying either of them. 
\subsubsection{Evaluation of A-STPM}\vspace{-0.02in} \label{sec:accuracy}

We proceed to evaluate the accuracy of A-STPM by comparing the patterns extracted by A-STPM and E-STPM. Table \ref{tbl:accuracyRealdata} shows the accuracies of A-STPM for different $\textit{minSeason}$ and $\textit{minDensity}$ on the real-world datasets. It is seen that, A-STPM obtains high accuracy ($\ge 81\%$) when $\textit{minSeason}$ and $\textit{minDensity}$ are low, e.g., $\textit{minSeason}=8$ and $\textit{minDensity}=0.5\%$, and very high accuracy ($\ge 95\%$) when $\textit{minSeason}$ and $\textit{minDensity}$ are high, e.g., $\textit{minSeason}=16$ and $\textit{minDensity}=0.75\%$. Similarly, Table \ref{tbl:accuracyScalePercent} shows the accuracies of A-STPM on the synthetic datasets: very high accuracy ($\ge 96\%$) when $\textit{minSeason}$ and $\textit{minDensity}$ are high, e.g., $\textit{minSeason}=16$ and $\textit{minDensity}=0.75\%$.

\vspace{-0.05in}\section{Conclusion and Future Work}\vspace{-0.02in}\label{sec:conclusion}
This paper presents our efficient Frequent Seasonal Temporal Pattern
Mining from Time Series (FreqSTPfTS) approach that offers: (1) the first solution for Seasonal Temporal Pattern Mining (STPM), (2) the efficient and exact Seasonal Temporal Pattern Mining (E-STPM) algorithm that employs the efficient data structures and pruning techniques to achieve fast mining, and (3) the approximate A-STPM that uses mutual information to prune unpromising time series and allows STPM to scale on big datasets. Extensive experiments conducted on real-world and synthetic datasets show that both A-STPM and E-STPM outperform the baseline, consume less memory, and scale well to big datasets. Compared to the baseline, the approximate A-STPM delivers up to an order of magnitude speedup. In future work, we plan to extend STPM to prune at the event level to further improve its performance.

\bibliographystyle{IEEEtran}
\bibliography{references}

\begin{appendices}
	\appendix
	\section{Detailed Proofs of Complexities, Lemmas and Theorems}
\subsection{Mutual exclusive property of temporal relations}\label{app:mutualexclusive}
\textbf{Property 1. }\textit{(Mutual exclusive) Consider the set of temporal relations $\Re=$ \{Follows, Contains, Overlaps\}. Let $E_i$ and $E_j$ be two temporal events, and $e_i$ occurring during $[t_{s_i}, t_{e_i}]$, $e_j$ occurring during $[t_{s_j}, t_{e_j}]$ be their corresponding event instances, and $\epsilon$ be the tolerance buffer. The relations in $\Re$ are mutually exclusive on $E_i$ and $E_j$.} 
\begin{proof}
	\textbf{$\ast$ Case 1:} Assume the relation \textbf{Follows($E_{i_{\triangleright e_i}}$, $E_{j_{\triangleright e_j}}$)} holds between $E_i$ and $E_j$. Thus, we have: 
	\begin{equation}
		t_{e_i} \pm \epsilon \le t_{s_j}
		\label{eq:mutualcase11}
	\end{equation}
	and:
	\begin{equation}
		t_{s_j} < t_{e_j} \Rightarrow t_{e_i} \pm \epsilon < t_{e_j}
		\label{eq:mutualcase12}
	\end{equation}
	
	Hence, Contains($E_{i_{\triangleright e_i}}$, $E_{j_{\triangleright e_j}}$) cannot exist between $E_i$ and $E_j$, since Contains($E_{i_{\triangleright e_i}}$, $E_{j_{\triangleright e_j}}$) holds iff ${(t_{s_i} \le t_{s_j})} \wedge$ $(t_{e_i} \pm \epsilon \ge t_{e_j})$ (contradict Eq. \eqref{eq:mutualcase12}). 
	Similarly, Overlaps($E_{i_{\triangleright e_i}}$, $E_{j_{\triangleright e_j}}$) cannot exist between $E_i$ and $E_j$ since Overlaps($E_{i_{\triangleright e_i}}$, $E_{j_{\triangleright e_j}}$) holds iff ${(t_{s_i} < t_{s_j})} \wedge$ $(t_{e_i} \pm \epsilon < t_{e_j})$ $\wedge$ $(t_{e_i}-t_{s_j} \ge d_o \pm \epsilon)$ (contradict Eq. \eqref{eq:mutualcase11}).
	
	In conclusion, if Follows($E_{i_{\triangleright e_i}}$, $E_{j_{\triangleright e_j}}$) holds between $E_i$ and $E_j$, then the two remaining relations cannot exist between $E_i$ and $E_j$.
	
	\textbf{$\ast$ Case 2:} Assume the relation \textbf{Contains($E_{i_{\triangleright e_i}}$, $E_{j_{\triangleright e_j}}$)} holds between $E_i$ and $E_j$. Thus, we have: 
	\begin{equation}
		t_{s_i} \le t_{s_j}
		\label{eq:mutualcase21}
	\end{equation}
	\begin{equation}
		t_{e_i} \pm \epsilon \ge t_{e_j}
		\label{eq:mutualcase22}
	\end{equation}
	
	Hence, Follows($E_{i_{\triangleright e_i}}$, $E_{j_{\triangleright e_j}}$) cannot exist between $E_i$ and $E_j$ since Follows($E_{i_{\triangleright e_i}}$, $E_{j_{\triangleright e_j}}$) holds iff $t_{e_i} \pm \epsilon < t_{e_j}$ (contradict Eq. \eqref{eq:mutualcase22}).
	
	Similarly, Overlaps($E_{i_{\triangleright e_i}}$, $E_{j_{\triangleright e_j}}$) cannot exist between $E_i$ and $E_j$, since Overlaps($E_{i_{\triangleright e_i}}$, $E_{j_{\triangleright e_j}}$) holds iff ${(t_{s_i} < t_{s_j})} \wedge$ $(t_{e_i} \pm \epsilon < t_{e_j})$ $\wedge$ $(t_{e_i}-t_{s_j} \ge d_o \pm \epsilon)$ (contradict Eq. \eqref{eq:mutualcase22}).  
	
	In conclusion, if Contains($E_{i_{\triangleright e_i}}$, $E_{j_{\triangleright e_j}}$) holds between $E_i$ and $E_j$, then the two remaining relations cannot exist between $E_i$ and $E_j$.
	
	\textbf{$\ast$ Case 3:} Assume the relation \textbf{Overlaps($E_{i_{\triangleright e_i}}$, $E_{j_{\triangleright e_j}}$)} holds between $E_i$ and $E_j$. Thus, we have: 
	\begin{equation}
		t_{s_i} < t_{s_j}
		\label{eq:mutualcase31}
	\end{equation}
	\begin{equation}
		t_{e_i} \pm \epsilon < t_{e_j}
		\label{eq:mutualcase32}
	\end{equation}
	\begin{equation}
		t_{e_i}-t_{s_j} \ge d_o \pm \epsilon \Rightarrow t_{s_j} \le t_{e_i} - d_o \pm \epsilon
		\label{eq:mutualcase33}
	\end{equation}
	
	Hence, Follows($E_{i_{\triangleright e_i}}$, $E_{j_{\triangleright e_j}}$) cannot exist between $E_i$ and $E_j$, since Follows($E_{i_{\triangleright e_i}}$, $E_{j_{\triangleright e_j}}$) holds iff $t_{e_i} \pm \epsilon < t_{s_j}$ (contradict Eq. \eqref{eq:mutualcase33}).
	
	Similarly, Contains($E_{i_{\triangleright e_i}}$, $E_{j_{\triangleright e_j}}$) cannot exist between $E_i$ and $E_j$, since Contains($E_{i_{\triangleright e_i}}$, $E_{j_{\triangleright e_j}}$) holds iff $t_{e_i} \pm \epsilon \ge t_{e_j}$ (contradict Eq. \eqref{eq:mutualcase32}).
	
	In conclusion, if Overlaps($E_{i_{\triangleright e_i}}$, $E_{j_{\triangleright e_j}}$) holds between $E_i$ and $E_j$, then the two remaining relations cannot exist between $E_i$ and $E_j$.
\end{proof}

\subsection{Lemma \ref{lempattern1}}\label{app:prooflempattern1}
\textbf{Lemma \ref{lempattern1}. }\textit{Let $P$ and $P^{'}$ be two temporal patterns such that $P^{'} \subseteq P$. Then $\textit{maxSeason}(P^{'}) \geq \textit{maxSeason}(P)$. }

\begin{proof}
	We have: 
	
	$\textit{maxSeason}(P^{'}) = \frac { |SUP^{P^{'}}|  }{\textit{minDensity}}$, $ \textit{maxSeason}(P) = \frac { |SUP^P|  }{\textit{minDensity}}$
	
	Since: 
	$|SUP^{P^{'}}| \geq |SUP^P| \text{ (Derived from Def. 3.12)} $
	
	Hence: $\textit{maxSeason}(P^{'}) \geq \textit{maxSeason}(P)$
\end{proof}

\subsection{Lemma \ref{lemeventpattern}}\label{app:prooflemeventpattern}
\textbf{Lemma \ref{lemeventpattern}. }\textit{Let $P$ be a k-event temporal pattern formed by a k-event group $(E_1,...,E_k)$. Then, $\textit{maxSeason}(P) \leq \textit{maxSeason}{(E_1,...,E_k)}$. }
\begin{proof}
	Derived directly from Def. 3.12, and Eq. \eqref{eq:maxSeasonPattern}.
\end{proof}

\subsection{Mining frequent seasonal single event}\label{app:proofcomplexity1event}
\textit{Complexity:} 
The complexity of finding frequent seasonal single events is $O(n \cdot | \mathcal{D}_{\text{SEQ}} |)$, where $n$ is the number of distinct events.
\begin{proof}
	Computing $\textit{maxSeason}$ for each event $E_i$ takes $O(|\mathcal{D}_{\text{SEQ}} |)$. Thus, computing $\textit{maxSeason}$ for $n$ events takes $O(n \cdot | \mathcal{D}_{\text{SEQ}} |)$. Moreover, for each candidate event $E_i$, identifying the set of seasons $\mathcal{PS}$ takes $O(|SUP^{E_i}|)$. We have potentially $n$ events. And thus, it takes $O(n \cdot |SUP^{E_i}|)$. The overal complexity is: $O(n \cdot | \mathcal{D}_{\text{SEQ}} | + n \cdot |SUP^{E_i}|) \sim O(n \cdot | \mathcal{D}_{\text{SEQ}} |)$.
\end{proof}

\subsection{Search space of STPM}
\textit{Complexity:} \textit{The search space of finding seasonal temporal patterns is $O(n^h3^{h^2})$, where $n$ is the number of distinct events in $\mathcal{D}_{\text{SEQ}}$, and $h$ is the maximal length of a temporal pattern.}
\begin{proof}
	 The number of seasonal single events is: $N_1=n \sim O(n)$. For mining 2-event groups, the number of permutations of $n$ distinct events taken $2$ at a time is: $P(n,2)$. However, since the same event can form a pair of events with itself, the total number of 2-event groups is: $N_2=P(n,2)+n$ $\sim O(n^2)$. Each 2-event group in $N_2$ can form $3$ different temporal relations, and thus, the total number of seasonal 2-event patterns is: $N_2\times 3^1$ $\sim O(n^23^1)$. Similarly, the number of 3-event groups is: $N_3= P(n,3)+P(n,2)+n$ $\sim O(n^3)$, and the number of seasonal 3-event patterns is: $N_3 \times 3^3$ $\sim O(n^33^3)$. For mining h-event groups, the number of h-event groups is $O(n^h)$, while the number of seasonal h-event patterns is $O(n^h \times 3^{\frac{1}{2}h(h-1)})$ $\sim O(n^h3^{h^2})$. Therefore, the total number of seasonal temporal patterns is $O(n)+O(n^23^1)+O(n^33^3)+...+O(n^h3^{h^2}) \sim O(n^h3^{h^2})$.
\end{proof}

\subsection{Lemma \ref{lem:transitivity}}\label{app:prooflemtransitivity}
\textbf{Lemma \ref{lem:transitivity}. }\textit{Let $\scalemath{0.9}{S=<e_1}$,..., $\scalemath{0.9}{e_{k-1}>}$ be a temporal sequence, $\scalemath{0.9}{P=<(r_{12}, E_{1_{\triangleright e_1}}, E_{2_{\triangleright e_2}}),...,(r_{(k-2)(k-1)}, E_{{k-2}_{\triangleright e_{k-2}}}, E_{{k-1}_{\triangleright e_{k-1}}})>}$ be a (k-1)-event pattern that occurs in $S$, $e_k$ be a new event instance added to $S$ to create the temporal sequence $\scalemath{0.9}{S^{'}=<e_1}$,..., $\scalemath{0.9}{e_{k}>}$. 
	The set of temporal relations $\Re$ is transitive on $S^{'}$: $\forall e_i \in S^{'}$, $i < k$, $\exists r \in \Re$ s.t. $r(E_{i_{\triangleright e_i}}$,$E_{k_{\triangleright e_k}})$ hold. }
\begin{proof}
	Since $S^{'}=<e_1, ..., e_{n}>$ is a temporal sequence, the event instances in $S^{'}$ are chronologically ordered by their start times. Then, $\forall e_i \in S^{'}, i \neq n$: $t_{s_i} \le t_{s_n}$. We have: 
	\begin{itemize}
		\item If $t_{e_i} \pm \epsilon \le t_{s_n}$, then $E_{i_{\triangleright e_i}}$ $\rightarrow E_{n_{\triangleright e_n}}$.
		\item If ${(t_{s_i} \le t_{s_n})} \wedge$ $(t_{e_i} \pm \epsilon \ge t_{e_n})$, then $E_{i_{\triangleright e_i}} \succcurlyeq E_{n_{\triangleright e_n}}$.
		\item If ${(t_{s_i} < t_{s_n})} \wedge$ $(t_{e_i} \pm \epsilon < t_{e_n})$ $\wedge$ $(t_{e_i}-t_{s_n} \ge d_o \pm \epsilon)$ where $d_o$ is the minimal overlapping duration, then $E_{i_{\triangleright e_i}} \between E_{n_{\triangleright e_n}}$. 
	\end{itemize}
\end{proof}

\subsection{Lemma \ref{lem:filter}}\label{app:prooflemfilter}
\textbf{Lemma \ref{lem:filter}. }\textit{Let $N_{k-1}=(E_1,...,E_{k-1})$ be a candidate seasonal (\textit{k-1})-event group, and $E_k$ be a candidate seasonal single event. The group $N_k= N_{k-1} \cup E_k$ can form candidate seasonal k-event temporal patterns if $\forall E_i \in N_{k-1}$, $\exists r \in \Re$ s.t. $r(E_i,E_k)$ is a candidate seasonal temporal relation. }
\begin{proof}
	Let $p_k$ be any k-event pattern formed by $N_k$. Then $p_k$ is a list of $\frac{1}{2}k(k-1)$ triples $(E_i,r_{ij},E_j)$ where each represents a relation $r(E_i,E_j)$ between two events. In order for $p_k$ to be a candidate seasonal k-event temporal pattern, each of the relations in $p_k$ must be a candidate seasonal temporal pattern (Def. 3.8, Eq. (1), and Lemmas \ref{lempattern1} and  \ref{lem:transitivity}).  
\end{proof}

\subsection{Mining frequent seasonal k-event pattern}\label{app:proofcomplexitykevent}
\textit{Complexity:} Let $n$ be the number of single events in $HLH_1$, $i$ be the average number of event instances of each event, $r$ be the number of (k-1)-event patterns in $HLH_{k-1}$, and $u$ be the average number of granules of each event/temporal relation. The complexity of frequent seasonal k-event pattern mining is $O(n^2 i^2 u^2)$ + $O(|F_1|$ $\cdot$ $|F_{k-1}|$ $\cdot$ $r$ $\cdot$ $k^2 \cdot u$$)$.

\begin{proof}
	\textbf{$\ast$ The complexity of frequent seasonal 2-event pattern mining:}
	The Cartesian product of $n$ events in $HLH_1$ generates $n^2$ 2-event groups. Computing $maxSeason$ of $n^2$ 2-event groups takes $O(n^2 u)$. For each 2-event group, we need to compute $maxSeason$ of their temporal relations, which takes $O(i^2 u^2)$. We have potentially $n^2$ nodes. And thus, it takes $O(n^2 i^2 u^2)$. 
	For each candidate 2-event pattern, identifying the set of season $\mathcal{PS}$ takes $O(u)$. And we have potentially $(3 n^2)$ relations. Thus, finding $\mathcal{PS}$ takes $O(3 n^2   u) \sim O(n^2 u)$.
	The complexity of frequent seasonal 2-event pattern mining is: $O(n^2 u$ $+$ $n^2 i^2 u^2 ) \sim O(n^2 i^2 u^2)$.
	
	\textbf{$\ast$ The complexity of frequent seasonal k-event pattern mining ($k > 2$):}
	For each (k-1)-event pattern, we need to compute the support set of $\frac{1}{2}(k-1)(k-2)$ triples, which takes $O(\frac{1}{2}(k-1)(k-2) u$$) \sim O(k^2 u)$.
	We have $|F_1|$$\times$ $|F_{k-1}|$ events, each has $r$ (k-1)-event patterns.
	Thus, the complexity of computing $maxSeason$ is $O(|F_1|$ $\cdot$ $|F_{k-1}|$ $\cdot$ $r$ $\cdot$ $k^2 \cdot u$$)$. For each candidate pattern, identifying its $\mathcal{PS}$ takes $O(u)$. We have potentially $|F_1|$$\times$ $|F_{k-1}| \times r$ patterns. Thus, it takes $O(|F_1| \cdot |F_{k-1}| \cdot r \cdot u)$. The complexity of frequent seasonal k-event pattern mining ($k > 2$) is: $O(|F_1|$ $\cdot$ $|F_{k-1}|$ $\cdot$ $r$ $\cdot$ $k^2 \cdot u$ + $|F_1| \cdot |F_{k-1}| \cdot r \cdot u) $ $\sim O(|F_1|$ $\cdot$ $|F_{k-1}|$ $\cdot$ $r$ $\cdot$ $k^2 \cdot u$$)$.
	
	Thus, the total complexity is $O(n^2 i^2 u^2)$ + $O(|F_1|$ $\cdot$ $|F_{k-1}|$ $\cdot$ $r$ $\cdot$ $k^2 \cdot u$$)$.
\end{proof}

\subsection{Theorem}\label{app:prooftheorembound}
\textbf{Theorem \ref{theorem:lowerbound}.}   (Lower bound of the maximum seasonal occurrence)
\textit{Let $\mu$ be the mutual information threshold. If the NMI\hspace{0.04in} $\widetilde{I}(X_S$;$Y_S) \ge \mu$, then the maximum seasonal occurrence of $(X_1,Y_1)$ in $\mathcal{D}_{\text{SEQ}}$ has a lower bound: \vspace{-0.1in}
	\begin{align} \vspace{-0.1in}
		\small
		\textit{maxSeason}(X_1,Y_1) \geq \frac{\lambda_2 \cdot |\mathcal{D}_{\text{SEQ}}|}{\textit{minDensity}} \cdot e^{W\left( \frac{\log{\lambda_{1}^{1-\mu}} \cdot ln2}{\lambda_2} \right)}
		\label{eq:lowerbound1}
	\end{align}
	where: $\lambda_1 = \min \lbrace p(X_i), \forall X_i \in X_S\rbrace$ is the minimum probability of $X_i \in X_S$, and $\lambda_2=p(Y_1)$ is the probability of $Y_1 \in Y_S$, and $W$ is the Lambert function \cite{corless1996lambertw}.} 
\begin{proof} 

From Eq. \eqref{eq:NMI}, we have:
\begin{equation}
\small
\widetilde{I}(\mathcal{X}_S;\mathcal{Y}_S)= 1 - \frac{H(\mathcal{X}_S \vert \mathcal{Y}_S)}{H(\mathcal{X}_S)} \ge \mu
\end{equation} 
Hence:
\begin{equation}
\small
\frac{H(\mathcal{X}_S \vert \mathcal{Y}_S)}{H(\mathcal{X}_S)} \le  1-\mu
\label{eq:s30}
\end{equation} 
First, we derive a lower bound for $\frac{H(\mathcal{X}_S \vert \mathcal{Y}_S)}{H(\mathcal{X}_S)}$. We have:
\begin{align}
\small
\frac{H(\mathcal{X}_S \vert \mathcal{Y}_S)}{H(\mathcal{X}_S)} &= \frac{p(X_1, Y_1) \cdot \log \frac{p(X_1 , Y_1)} {p(Y_1)} }   {\sum_{i} p(X_i) \cdot \log p(X_i)} \nonumber \\ &+ \frac{\sum_{i \neq 1 \& j \neq 1} p(X_i, Y_j) \cdot \log \frac{p(X_i , Y_j)} {p(Y_j)}}{\sum_{i} p(X_i) \cdot \log p(X_i)}
\label{eq:s31}
\end{align}		
We first consider the numerator in Eq. \eqref{eq:s31}, we have:
\begin{align}
\small
p(X_1, Y_1) \cdot \log \frac{p(X_1 , Y_1)} {p(Y_1)} &+ \sum_{i \neq 1 \& j \neq 1} p(X_i, Y_j) \cdot \log \frac{p(X_i , Y_j)} {p(Y_j)} \nonumber \\
&\leq p(X_1, Y_1) \cdot \log \frac{p(X_1 , Y_1)} {p(Y_1)} \nonumber \\
&\leq p(X_1, Y_1) \cdot \log \frac{p(X_1 , Y_1)} {\lambda_2}
\label{eq:sNumerator}
\end{align}
where $\lambda_2 = p(Y_1)$.

Next, we consider the denominator in Eq. \eqref{eq:s31}. Suppose that:
\begin{align}
\small
p(X_k) = \min \lbrace p(X_i)\rbrace, \forall X_i \in X_S 
\end{align}
Then we have:
\begin{align}
\small
p(X_i) &\ge p(X_k), \forall X_i \in X_S  \nonumber \\
\Rightarrow \log  p(X_i) &\ge \log p(X_k) \nonumber \\
\Rightarrow p(X_i)\log  p(X_i) &\ge p(X_i)\log p(X_k) \nonumber \\
\Rightarrow \sum_{i} p(X_i)\log  p(X_i) &\ge \sum_{i} p(X_i)\log p(X_k) \nonumber \\
&= \log p(X_k) \sum_{i} p(X_i) \nonumber\\
&= \log p(X_k) \nonumber\\
&= \log \lambda_1
\label{eq:sminmax2}
\end{align}
where $\lambda_1 = p(X_k)$.
\\
Replace Eqs. \eqref{eq:sNumerator} and \eqref{eq:sminmax2} into Eq. \eqref{eq:s31}, we get:		
\begin{align}
\small
\frac{H(\mathcal{X}_S \vert \mathcal{Y}_S)}{H(\mathcal{X}_S)} \geq  \frac{p(X_1, Y_1) \cdot \log \frac{p(X_1 , Y_1)} {\lambda_2} }{\log \lambda_1} 
\label{eq:s33}
\end{align}	

From Eqs. \eqref{eq:s30} and \eqref{eq:s33}, it follows that:
	\begin{align}
	\small
	(1-\mu) \geq \frac{p(X_1, Y_1) \cdot \log \frac{p(X_1 , Y_1)} {\lambda_2} }{\log \lambda_1} \nonumber \\
	\Leftrightarrow p(X_1, Y_1) \cdot \log \frac{p(X_1 , Y_1)} {\lambda_2} \geq \log \lambda_{1}^{1-\mu}
	\label{eq:s34}
	\end{align}
Assign $x=p(X_1,Y_1), b=\log \lambda_{1}^{1-\mu}$. Replace $x$ and $b$ into Eq. \eqref{eq:s34}, we get:
\begin{align}
	\small
	x \cdot \log \frac{x}{\lambda_2} \geq b  \nonumber \\
	\Leftrightarrow \frac{x}{\lambda_2} \cdot \log \frac{x}{\lambda_2} \geq \frac{b}{\lambda_2} \nonumber \\
	\Leftrightarrow \frac{x}{\lambda_2} \cdot \frac{\ln \frac{x}{\lambda_2}}{\ln 2} \geq \frac{b}{\lambda_2} \nonumber \\
	\Leftrightarrow \frac{x}{\lambda_2} \cdot \ln \frac{x}{\lambda_2} \geq \frac{b \cdot \ln 2}{\lambda_2} 
	\label{eq:s35}
\end{align}
Assign $y=\ln \frac{x}{\lambda_2} \Rightarrow \frac{x}{\lambda_2} = e^y$. Replace $y$ and $e^y$ into Eq. \eqref{eq:s35}, we get:
\begin{align}
	\small
	& y \cdot e^y \geq \frac{b \cdot \ln 2}{\lambda_2} \nonumber \\
	\Leftrightarrow & y \geq W\left(\frac{b \cdot \ln 2}{\lambda_2} \right) \text{ , where $W$ is the Lambert function \cite{corless1996lambertw}.} \nonumber \\
	\Leftrightarrow & \ln \frac{x}{\lambda_2} \geq W\left(\frac{\log \lambda_{1}^{1-\mu} \cdot \ln 2}{\lambda_2}  \right) \nonumber \\
	\Leftrightarrow & e^{\ln \frac{x}{\lambda_2} } \geq  e^{W\left(\frac{\log \lambda_{1}^{1-\mu}  \cdot \ln 2}{\lambda_2} \right)} \nonumber \\
	\Leftrightarrow & \frac{x}{\lambda_2} \geq e^{W\left(\frac{\log \lambda_{1}^{1-\mu} \cdot \ln 2}{\lambda_2}  \right)} \nonumber \\
	\Leftrightarrow & x \geq \lambda_2 \cdot e^{W\left(\frac{\log \lambda_{1}^{1-\mu} \cdot \ln 2}{\lambda_2}  \right)} \nonumber \\
	\Leftrightarrow & p(X_1,Y_1) \geq \lambda_2 \cdot e^{W\left(\frac{\log \lambda_{1}^{1-\mu} \cdot \ln 2}{\lambda_2}  \right)} 
	\label{eq:s36_1}
\end{align}

Since the relative support of $(X_1, Y_1)$ in $\mathcal{D}_{\text{SEQ}}$ is greater than or equal to the relative support of $(X_1, Y_1)$ in $\mathcal{D}_{\text{SYB}}$ \cite{ho2020efficient}, hence:
\begin{align}
	\small
	\frac{|SUP^{(X_1,Y_1)}|}{|\mathcal{D}_{\text{SEQ}}|} \geq \frac{|SUP^{(X_1,Y_1)}_{\mathcal{D}_{\text{SYB}}}|}{|\mathcal{D}_{\text{SYB}}|} = p(X_1,Y_1)
	\label{eq:s36_2}
\end{align}
where $\frac{|SUP^{(X_1,Y_1)}_{\mathcal{D}_{\text{SYB}}}|}{|\mathcal{D}_{\text{SYB}}|}$ is the relative support of $(X_1, Y_1)$ in $\mathcal{D}_{\text{SYB}}$.

From Eqs. (\ref{eq:s36_1}) and (\ref{eq:s36_2}), it follows that:
\begin{align}
	\small
	& \frac{|SUP^{(X_1,Y_1)}|}{|\mathcal{D}_{\text{SEQ}}|} \geq \lambda_2 \cdot e^{W\left(\frac{\log \lambda_{1}^{1-\mu} \cdot \ln 2}{\lambda_2}  \right)} \nonumber \\
	\Leftrightarrow & \frac{|SUP^{(X_1,Y_1)}|}{\textit{minDensity}} \geq \frac{\lambda_2 \cdot |\mathcal{D}_{\text{SEQ}}|}{\textit{minDensity}} \cdot e^{W\left(\frac{\log \lambda_{1}^{1-\mu} \cdot \ln 2}{\lambda_2}  \right)} \nonumber \\
\Leftrightarrow	& \scalemath{0.9}{\textit{maxSeason}(X_1,Y_1) \geq \frac{\lambda_2 \cdot |\mathcal{D}_{\text{SEQ}}|}{\textit{minDensity}} \cdot e^{W\left(\frac{\log \lambda_{1}^{1-\mu} \cdot \ln 2}{\lambda_2}  \right)} }
\label{eq:s37}
\end{align}

We can derive a lower bound of $\mu$ from Eq. (\ref{eq:s37}).
\begin{align}
	\small
	\textit{maxSeason}(X_1,Y_1) & \geq \scalemath{0.85}{\frac{\lambda_2 \cdot |\mathcal{D}_{\text{SEQ}}|}{\textit{minDensity}} \cdot e^{W\left(\frac{\log \lambda_{1}^{1-\mu} \cdot \ln 2}{\lambda_2}  \right)} \geq \textit{minSeason} } \nonumber \\	
	\Rightarrow e^{W\left(\frac{\log \lambda_{1}^{1-\mu} \cdot \ln 2}{\lambda_2}  \right)} & \geq \frac{\textit{minSeason} \cdot \textit{minDensity}}{\lambda_2 \cdot |\mathcal{D}_{\text{SEQ}}|} \nonumber \\
	\Leftrightarrow e^{W\left(\frac{\log \lambda_{1}^{1-\mu} \cdot \ln 2}{\lambda_2}  \right)} & \geq \rho
	\label{eq:s38}
\end{align}
where $\rho=\frac{\textit{minSeason} \cdot \textit{minDensity}}{\lambda_2 \cdot |\mathcal{D}_{\text{SEQ}}|}$.

To solve Eq. (\ref{eq:s38}), we consider two cases.

\textbf{$\ast$ Case 1:} $0 \leq \rho \leq \frac{1}{e}$, we have:
\begin{align}
	\small
	& \frac{\log \lambda_{1}^{1-\mu} \cdot \ln 2}{\lambda_2} \geq \frac{-1}{e} \nonumber \\
	\Leftrightarrow & \log \lambda_{1}^{1-\mu} \geq \frac{-\lambda_2}{e \cdot \ln 2} \nonumber \\
	\Leftrightarrow & (1-\mu) \cdot \log \lambda_1 \geq \frac{-\lambda_2}{e \cdot \ln 2} \nonumber \\
	\Leftrightarrow &  1-\mu \leq \frac{-\lambda_2}{e \cdot \ln 2 \cdot \log \lambda_1} \text{ (}\log \lambda_1 < 0 \text{)} \nonumber \\
	\Leftrightarrow & \mu \geq 1 + \frac{\lambda_2}{e \cdot \ln 2 \cdot \log \lambda_1} \nonumber \\
	\Leftrightarrow & \mu \geq 1 - \frac{\lambda_2}{e \cdot \ln 2 \cdot \log \frac{1}{\lambda_1} }
\end{align}

\textbf{$\ast$ Case 2:} $\rho > \frac{1}{e}$, we have:
\begin{align}
	\small
	& \frac{\log \lambda_{1}^{1-\mu} \cdot \ln 2}{\lambda_2} \geq \rho \cdot \log \rho \nonumber \\
	\Leftrightarrow & \log \lambda_{1}^{1-\mu} \geq \rho \cdot \lambda_2 \cdot \frac{\log \rho}{\ln 2} \nonumber \\
	\Leftrightarrow & (1-\mu) \cdot \log \lambda_1 \geq \rho \cdot \lambda_2 \cdot \frac{\log \rho}{\ln 2} \nonumber \\
	\Leftrightarrow &  1-\mu \leq \rho \cdot \lambda_2 \cdot \frac{\log \rho}{\ln 2 \cdot \log \lambda_1} \text{ (}\log \lambda_1 < 0 \text{)} \nonumber \\
	\Leftrightarrow & \mu \geq 1 - \frac{\rho \cdot \lambda_2 \cdot \log \rho }{\ln 2 \cdot \log \lambda_1}
\end{align}
\end{proof}
\section{Additional Experimental Results}
\subsection{Qualitative Evaluation}
Tables \ref{tbl:numPatternSC} and \ref{tbl:numPatternHFM} list the number of seasonal patterns found in the SC and HFM datasets. 
It can be seen that high \textit{minSeason} leads to less generated patterns, as many have few seasonal occurrences. Moreover, high $\textit{minDensity}$ also generates fewer patterns since only few patterns have high occurrence density. Finally, high $\textit{maxPeriod}$ results in more generated patterns, since high $\textit{maxPeriod}$ allows more temporal relations to be formed, thus increasing the number of patterns. 
\begin{table}[!h]
	\centering
	\vspace{-0.05in}
	\begin{minipage}{1\linewidth}
		\captionsetup{justification=centering, font=small}
		\caption{\small The Number of Seasonal Patterns on SC}
		\vspace{-0.1in}
		\resizebox{\columnwidth}{0.9cm}{
			\begin{tabular}{|c|c|c|c|c|c|c|c|c|c|} 
				\hline 
				\multirow{2}{*}{\bfseries maxPeriod (\%)} & \multicolumn{9}{c|}{\bfseries minSeason (\#) - minDensity (\%)} 
				\\  \cline{2-10}  
				& {\bfseries 8-0.5} & {\bfseries 8-0.75} & {\bfseries 8-1.0} & {\bfseries 12-0.5} & {\bfseries 12-0.75} & {\bfseries 12-1.0} & {\bfseries 16-0.5} & {\bfseries 16-0.75} & {\bfseries 16-1.0}
				\\ \hline 
				0.2  &  17241 &  12401 &  8632 &  10973 &  8291 &  3742 &  6207 &  3416 &  2138\\
				\hline 
				0.4  &  24948 &  18293 &  11827 &  16830 &  12726 &  5291 &  8263 &  5084 &  3816\\
				\hline 
				0.6  &  31825 &  26108 &  14039 &  24806 &  19408 &  8032 &  11852 &  8165 &  6010\\
				\hline 
			\end{tabular} 
		}
		\label{tbl:numPatternSC}
	\end{minipage}
\end{table}
\begin{table}[!h]
	\begin{minipage}{1\linewidth}
		\captionsetup{justification=centering, font=small}
		\caption{The Number of Seasonal Patterns on HFM}
		\vspace{-0.1in}
		\resizebox{\columnwidth}{0.9cm}{
			\begin{tabular}{|c|c|c|c|c|c|c|c|c|c|} 
				\hline 
				\multirow{2}{*}{\bfseries maxPeriod (\%)} & \multicolumn{9}{c|}{\bfseries minSeason (\#) - minDensity (\%)} 
				\\  \cline{2-10}  
				& {\bfseries 8-0.5} & {\bfseries 8-0.75} & {\bfseries 8-1.0} & {\bfseries 12-0.5} & {\bfseries 12-0.75} & {\bfseries 12-1.0} & {\bfseries 16-0.5} & {\bfseries 16-0.75} & {\bfseries 16-1.0}
				\\ \hline 
				0.2  &  14763 &  10425 &  7191 &  8014 &  7125 &  2486 &  4452 &  2693 &  1307\\
				\hline 
				0.4  &  19542 &  14018 &  8506 &  11036 &  9082 &  5563 &  6207 &  5261 &  3005\\
				\hline 
				0.6  &  22671 &  17039 &  10617 &  13502 &  10539 &  8035 &  7658 &  7014 &  4092\\
				\hline 
			\end{tabular} 
		}
		\label{tbl:numPatternHFM}
	\end{minipage}
\end{table}
\subsection{Baselines comparison}
Figs. \ref{fig:runtimebaselineSC}, \ref{fig:runtimebaselineHFM}, \ref{fig:memorybaselineSC}, and \ref{fig:memorybaselineHFM} show the experimental results on SC and HFM datasets. Note that Figs. \ref{fig:runtimebaselineSC}-\ref{fig:scaleTimeSeries_HFM} use the same legend.

As shown in Figs. \ref{fig:runtimebaselineSC} and \ref{fig:runtimebaselineHFM}, A-STPM achieves the best runtime among all methods, and E-STPM has better runtime than the baseline. The range and average speedups of A-STPM compared to other methods are: $[1.4$-$3.1]$ and $2.2$ (E-STPM), and $[4.9$-$10.3]$ and $6.9$ (APS-growth). The speedup of E-STPM compared to the baseline is $[2.8$-$5.4]$ and $3.5$ on average.
Note that the time to compute MI and $\mu$ for SC and HFM in Figs. \ref{fig:runtimebaselineSC} and \ref{fig:runtimebaselineHFM} are $0.9$ and $1.2$ seconds, respectively.

In terms of memory consumption, as shown in Figs. \ref{fig:memorybaselineSC} and \ref{fig:memorybaselineHFM}, A-STPM is the most efficient method, while E-STPM is more efficient than the baseline. The range and the average memory consumption of A-STPM compared to other methods are: $[1.4$-$2.6]$ and $1.7$ (E-STPM), and $[2.5$-$6.3]$ and $3.7$ (APS-growth). The memory usage of E-STPM compared to the baseline is $[1.3$-$3.2]$ and $2.1$ on average.

\subsection{Scalability evaluation on synthetic datasets}
Figs. \ref{fig:scaleSequence_SC} and \ref{fig:scaleSequence_HFM} show the runtimes of A-STPM, E-STPM and the baseline when the number of sequences changes. We obtain the range and average speedups of A-STPM are: [$1.4$-$2.7$] and $2.1$ (E-STPM), and [$2.4$-$6.1$] and $4.2$ (APS-growth). Similarly, the range and average speedup of E-STPM compared to APS-growth is [$1.6$-$3.5$] and $2.8$. We note that the baseline fails for larger configurations in this scalability study, i.e., on the synthetic SC at 60\% sequences (Fig. \ref{fig:scaleSequence_SC_8}) and on the synthetic HFM at 100\% sequences (Fig. \ref{fig:scaleSequence_HFM_8}), showing that A-STPM and E-STPM can scale well on big datasets while the baseline cannot.

Figs. \ref{fig:scaleTimeSeries_SC} and \ref{fig:scaleTimeSeries_HFM} compare the runtimes of A-STPM, E-STPM and APS-growth when changing the number of time series. We obtain the range and average speedups of A-STPM are: [$1.4$-$2.9$] and $2.2$ (E-STPM), and [$3.1$-$6.7$] and $4.5$ (APS-growth), and of E-STPM is [$1.7$-$3.8$] and $2.9$ (APS-growth). The baseline also fails at large configurations in this study, i.e., when \# Time Series $\ge 6000$ on the synthetic SC (Fig. \ref{fig:scaleTimeSeries_SC_8}) and the synthetic HFM (Fig. \ref{fig:scaleTimeSeries_HFM_8}).

Finally, we provide the percentage of time series and events pruned by A-STPM in the scalability test in Tables \ref{tbl:prunedAttributesEventsSCAppendix} and \ref{tbl:prunedAttributesEventsHFMAppendix}. 
Here, we can see that low $\textit{minSeason}$ and $\textit{minDensity}$ lead to more time series (events) to be pruned because low $\textit{minSeason}$ and $\textit{minDensity}$ result in higher $\mu$. 
\begin{table}[!h]
	\centering
	\begin{minipage}{1\linewidth}
		\captionsetup{justification=centering, font=small}
		\caption{\small Pruned Time Series and Events from A-STPM on SC}
		\vspace{-0.1in}
		\resizebox{\columnwidth}{!}{
			\begin{tabular}{|c|c|c|c|c|c|c|}
				\hline 
				\multirow{2}{*}{\# Attr.} & \multicolumn{3}{c|}{\bfseries \# Pruned Time Series} & \multicolumn{3}{c|}{\bfseries \# Pruned Events} 
				\\  \cline{2-7} 
				& {\bfseries 12-0.5\%} & {\bfseries 16-0.75\%} & {\bfseries 20-1.0\%}  & {\bfseries 12-0.5\%} & {\bfseries 16-0.75\%} & {\bfseries 20-1.0\%} \\
				\hline
				2000 & 31.60 & 29.20 & 25.30 & 30.23 & 26.26 & 20.04  \\  \hline		
				
				4000 & 30.10 & 26.05 & 18.45 & 29.01 & 25.66 & 19.90  \\  \hline	
				
				6000 & 28.35 & 24.22 & 18.25 & 28.83 & 25.02 & 19.83   \\  \hline		
				
				8000 & 26.78 & 24.05 & 17.80 & 28.79 & 24.64 & 19.49  \\  \hline	
				
				10000 & 26.03 & 23.01 & 17.49 & 25.19 & 22.53 & 18.13  \\  \hline									
			\end{tabular}
		}
		\label{tbl:prunedAttributesEventsSCAppendix}
	\end{minipage}
\end{table}

\begin{table}[!h]
	\centering
	\begin{minipage}{1\linewidth}
		\captionsetup{justification=centering, font=small}
		\caption{Pruned Time Series and Events from \scriptsize{A-STPM} on HFM}
		\vspace{-0.1in}
		\resizebox{\columnwidth}{!}{
			\begin{tabular}{|c|c|c|c|c|c|c|}
				\hline 
				\multirow{2}{*}{\# Attr.} & \multicolumn{3}{c|}{\bfseries \# Pruned Time Series} & \multicolumn{3}{c|}{\bfseries \# Pruned Events}    
				\\  \cline{2-7} 
				& {\bfseries 12-0.5\%} & {\bfseries 16-0.75\%} & {\bfseries 20-1.0\%}  & {\bfseries 12-0.5\%} & {\bfseries 16-0.75\%} & {\bfseries 20-1.0\%} \\
				\hline
				2000 & 38.10 & 31.35 & 28.90 & 27.46 & 24.68 & 20.72 \\  \hline		
				
				4000 & 32.65 & 29.58 & 22.60 & 26.22 & 24.58 & 20.29 \\  \hline	
				
				6000 & 31.12 & 27.85 & 20.28 & 25.41 & 23.78 & 19.84   \\  \hline		
				
				8000 & 29.33 & 27.15 & 20.08 & 24.83 & 23.21 & 19.17  \\  \hline	
				
				10000 & 28.84 & 25.09 & 19.68 & 24.54 & 23.01 & 18.97  \\  \hline									
			\end{tabular}
		}
		\label{tbl:prunedAttributesEventsHFMAppendix}
	\end{minipage}
\end{table}

\subsection{Evaluation of the pruning techniques in E-STPM}
In this section, we report the evaluation results of the proposed pruning techniques in E-STPM on SC and HFM.
We use $3$ different configurations that vary: the mininum season, the minimum density, and the maximum period. Figs. \ref{fig:performanceSC} and  \ref{fig:performanceHFM} 
show the results. It can be seen that All-E-STPM achieves the best performance among all versions. Its speedup w.r.t. NoPrune-E-STPM ranges from $2.5$ up to $4.5$ depending on the configurations, showing that the proposed prunings are very effective in improving E-STPM performance. The average speedup is from $2$ to $4$ for Trans-E-STPM, and from $1.5$ to $3$ for Apriori-E-STPM. However, applying both always yields better speedup than applying either of them.

\subsection{Evaluation of  A-STPM}
Table \ref{tbl:accuracyRealdataAppendix} shows the accuracies of A-STPM for different $\textit{minSeason}$ and $\textit{minDensity}$ on the real world datasets. It is seen that, A-STPM obtains high accuracy ($\ge 80\%$) when $\textit{minSeason}$ and $\textit{minDensity}$ are low, e.g., $\textit{minSeason}=8$ and $\textit{minDensity}=0.5\%$, and very high accuracy ($\ge 95\%$) when $\textit{minSeason}$ and $\textit{minDensity}$ are high, e.g., $\textit{minSeason}=16$ and $\textit{minDensity}=0.75\%$. Similarly, Table \ref{tbl:accuracyScaleAppendix} shows the accuracies of A-STPM on the synthetic datasets: very high accuracy $(\geq 95\%)$ when $\textit{minSeason}$ and $\textit{minDensity}$ are high, e.g.,$\textit{minSeason}$ = 16 and $\textit{minDensity}$ = 0.75\%.

\begin{table}[!h]
	\centering
			\captionsetup{justification=centering, font=small}
		\caption{A-STPM Accuracy}
		\vspace{-0.1in}
		\resizebox{\columnwidth}{1cm}{
			\begin{tabular}{|c|c|c|c|c|c|c|}
				\hline 
				\multirow{3}{*}{\bfseries \# minSeason} & \multicolumn{6}{c|}{\bfseries minDensity (\%)} \\  
				\cline{2-7}  
				& \multicolumn{3}{c|}{\bfseries SC (real)} & \multicolumn{3}{c|}{\bfseries HFM (real)}
				\\  \cline{2-7}  
				& {\bfseries 0.5} & {\bfseries 0.75} & {\bfseries 1}  & {\bfseries 0.5} & {\bfseries 0.75} & {\bfseries 1}  \\
				\hline
				8 & 80  & 81   & 87 & 82 & 84  & 89  \\  \hline			
				12 & 83  & 85   & 93  & 86  & 92  & 94  \\  \hline				
				16 & 92  & 95   & 100  & 96   & 97  & 100   \\  \hline					
				20 & 95  & 99   & 100  & 97 & 100  & 100   \\  \hline								
			\end{tabular}
		}			
		\label{tbl:accuracyRealdataAppendix}
\end{table}
\vspace{-0.25in}
\begin{table}[!h]
	\vspace{0.1in}
	\centering
	\begin{minipage}{1\linewidth}
			\captionsetup{justification=centering, font=small}
		\caption{The Accuracy of A-STPM on Syn. Data}
		\vspace{-0.1in}
		\resizebox{\columnwidth}{1.2cm}{
			\begin{tabular}{|c|c|c|c|c|c|c|}
				\hline 
				\multirow{3}{*}{\# Attr.} & \multicolumn{3}{c|}{\bfseries SC} & \multicolumn{3}{c|}{\bfseries HFM}
				\\  \cline{2-7}  
				& \multicolumn{3}{c|}{\bfseries Accuracy (\%)} & \multicolumn{3}{c|}{\bfseries Accuracy (\%)}  
				\\  \cline{2-7}
				& {\bfseries 12-0.5\%} & {\bfseries 16-0.75\%} & {\bfseries 20-1.0\%}  & {\bfseries 12-0.5\%} & {\bfseries 16-0.75\%} & {\bfseries 20-1.0\%}  \\
				\hline
				2000 & 84 & 95 & 100 & 87 & 98 &	100 \\  \hline		
				
				4000 & 85 &	96 & 100 & 87 & 98 & 100 \\  \hline	
				
				6000 & 85 &	97 & 100 & 90 & 98 & 100  \\  \hline		
				
				8000 & 87 &	97 & 100 & 93 & 98 & 100  \\  \hline	
				
				10000 & 88 & 98 & 100 & 94 & 99 & 100  \\  \hline									
			\end{tabular}
		}
		\label{tbl:accuracyScaleAppendix}
	\end{minipage}
\end{table}

\input{appendix/appendixGraphBaselineComparison}
\input{appendix/scalabilityAppendix}
\input{appendix/runtimePruningEHSTPMappendix}

\subsection{Evaluation of  the tolerance buffer $\epsilon$}
We evaluate the impact of the buffer $\epsilon$ on extracted seasonal patterns. Tables \ref{tbl:epsilon_Appendix1} and \ref{tbl:epsilon_Appendix2} report the number of extracted seasonal patterns for different $\epsilon$ values, and the corresponding percentages of pattern loss compared to $\epsilon$ = 0. For RE and SC datasets, there are no lost patterns with $\epsilon$ = 1 hour. For INF and HFM datasets, there are no lost patterns with $\epsilon$ = 1 day and $\epsilon$ = 2 days. And the losses among other $\epsilon$ values are very low since there is low noise level in the datasets. 
\begin{table*}[!t]
	\centering
	\vspace{-0.05in}
	\begin{minipage}{0.495\linewidth}
		\captionsetup{justification=centering, font=small}
		\caption{\small Number of Extracted Patterns and Percentages of Pattern Loss on RE and SC}
		\vspace{-0.1in}
		\resizebox{\columnwidth}{0.9cm}{
			\begin{tabular}{|c|c|c|c|c|}
				\hline 
				\multirow{2}{*}{$\epsilon$ value} &\multicolumn{2}{c|}{\bfseries RE} &\multicolumn{2}{c|}{\bfseries SC} \\  
				\cline{2-5}  
				& {\bfseries \# Patterns} & {\bfseries Patterns (\%)} & {\bfseries \# Patterns} & {\bfseries Patterns (\%)} \\
				\hline
				1 hour & 35626 & 0.00 & 17241 & 0.00 
				\\  \hline							
				2 hours & 35407  & 0.61   & 16921 & 1.85 
				\\  \hline							
				3 hours & 35192  & 1.21   & 16812 & 2.48  
				\\  \hline							
			\end{tabular}
		}
		\label{tbl:epsilon_Appendix1}
	\end{minipage}
	\begin{minipage}{0.495\linewidth}
		\captionsetup{justification=centering, font=small}
		\caption{Number of Extracted Patterns and Percentages of Pattern Loss on INF and HFM}
		\vspace{-0.1in}
		\resizebox{\columnwidth}{0.9cm}{
			\begin{tabular}{|c|c|c|c|c|}
				\hline 
				\multirow{2}{*}{$\epsilon$ value} &\multicolumn{2}{c|}{\bfseries INF} &\multicolumn{2}{c|}{\bfseries HFM} \\  
				\cline{2-5}  
				& {\bfseries \# Patterns} & {\bfseries Patterns (\%)} & {\bfseries \# Patterns} & {\bfseries Patterns (\%)} \\
				\hline
				1 day & 7812 & 0.00 & 14763 & 0.00 
				\\  \hline							
				2 days & 7812  & 0.00   & 14763 & 0.00 
				\\  \hline							
				3 days & 7803  & 0.11   & 14750 & 0.08  
				\\  \hline							
			\end{tabular}
		}
		\label{tbl:epsilon_Appendix2}
	\end{minipage}
\end{table*}

\end{appendices}

\end{document}